\documentclass[final,3p,times]{elsarticle}

\usepackage{amssymb,bbold}
\usepackage{amsmath}
\usepackage{booktabs}
\usepackage{float}
\usepackage{graphicx}
\usepackage[utf8x]{inputenc}
\usepackage{multirow}
\usepackage{natbib}
\usepackage{hyperref}

\RequirePackage{color}

\hypersetup{
  colorlinks=true,linkcolor=blue,citecolor=blue,
  filecolor=blue,urlcolor=blue,breaklinks=true
}

\biboptions{sort&compress}

\journal{Physics Reports}

\begin{document}

\begin{frontmatter}

\title{Green's function approach for quantum graphs: an overview}

\author[ucl,uepg]{Fabiano M. Andrade}
\ead{f.andrade@ucl.ac.uk,fmandrade@uepg.br}

\author[uff]{A. G. M. Schmidt}

\author[unicentro]{E. Vicentini}

\author[ufpr]{B. K. Cheng}

\author[ufpr]{M. G. E. da Luz}
\ead{luz@fisica.ufpr.br}

\address[ucl]{
  Department of Computer Science and
  Department of Physics and Astronomy,
  University College London,
  WC1E 6BT London, United Kingdom
}

\address[uepg]{
  Departamento de Matem\'{a}tica e Estat\'{i}stica,
  Universidade Estadual de Ponta Grossa,
  84030-900 Ponta Grossa, Paran\'{a}, Brazil}

\address[uff]{
  Departamento de F\'{i}sica,
  Universidade Federal Fluminense,
  27215-350 Volta Redonda, Rio de Janeiro, Brazil}

\address[unicentro]{
  Departamento de F\'{i}sica,
  Universidade Estadual do Centro Oeste,
  85010-990 Guarapuava, Paran\'{a} , Brazil
}

\address[ufpr]{
Departamento de F\'{i}sica,
Universidade Federal do Paran\'{a},
81531-980 Curitiba,  Paran\'{a}, Brazil}

\begin{abstract}
Here we review the many aspects and distinct phenomena associated to 
quantum dynamics on general graph structures.
For so, we discuss such class of systems under the energy domain Green's
function ($G$) framework.
This approach is particularly interesting because $G$ can be written as
a sum over classical-like paths, where local quantum effects are taking
into account through the scattering matrix amplitudes (basically,
transmission and reflection amplitudes) defined on each one of the
graph vertices.
Hence, the {\em exact} $G$ has the functional form of a generalized
semiclassical formula, which through different calculation techniques
(addressed in details here) always can be cast into a closed analytic
expression.
It allows to solve exactly arbitrary large (although finite) graphs in 
a recursive and fast way.
Using the Green's function method, we survey many properties for open
and closed quantum graphs as scattering solutions for the former and
eigenspectrum and eigenstates for the latter, also considering
quasi-bound states.
Concrete examples, like cube, binary trees and Sierpi\'{n}ski-like
topologies are presented.
Along the work, possible distinct applications using the Green's
function methods for quantum graphs are outlined.
\end{abstract}

%%% Local Variables:
%%% mode: latex
%%% TeX-master: "green-qg-pr"
%%% ispell-local-dictionary: "american"
%%% End:


\end{frontmatter}

\tableofcontents

\section{Introduction}
\label{sec:introduction}

A graph can be understood intuitively as a set of elements (the
vertices), attached ones to the others through connections (the edges).
The topological arrangement of a graph is thus completely determined by
the way the vertices are joined by the edges.
The more general concept of a network -- essentially a graph -- has
found  applications in many branches of science and engineering.
Some representative examples include: the analysis of electrical
circuits, verification (in different contexts) of the shortest paths in
grid structures, traffic planning, charge transport in complex chemical
compounds, ecological webs, cybernetics architectures, linguistic
families, and  social connection relations, to cite just a few.
In fact, given that as diverse as the street system of a city, the web
of neurons in the human brain, and the organization of digital database
in distinct storage devices, can all be described as `graphs', we might
be lead to conclude that such idea is one of the most useful and broadly
used abstract mathematical notion in our everyday lives.

Less familiar is which we call quantum graphs\footnote{
Depending on the particular aspect to be studied, quantum graphs
are also named quantum networks or quantum wires.},
or more precisely quantum metric graphs (by associating lengths to the
edges), basically comprising the study of the Helmholtz operator
$\nabla^2 + k^2$ -- when the external potentials for the underlying
Hamiltonian along the edges are null, see later -- on these topological
structures.
Nevertheless, they still attract a lot of attention in the physics and
mathematics specialized literature because their rich behavior and
potential applications \cite{WRCM.2004.14.3,Book.2006.Berkolaiko}, for
instance, regarding wave propagation and diffusive properties
(actually, this latter aspect allowing a possible formal association
between the
Schr\"odinger and the diffusion equations \cite{JPA.2013.46.235202}).

Historically, Linus Pauling seems to be the first to foresee the
usefulness of considering quantum dynamics on graph structures, e.g.,
to model free electrons in organic molecules
\cite{JCP.1936.4.673,HCA.1948.31.1441,JCP.1949.17.484,
JCP.1953.21.1565,PPSA.1954.67.608,JMP.1970.11.635,AoP.1972.73.308}.
Indeed, in a first approximation the molecules can be viewed as
a set of fixed atoms (vertices) connected by chemical bonds
(edges), along which the
electrons obey a 1D Schr\"{o}dinger equation with an effective potential.
Moreover, quantum transport in multiply connected systems
\cite{RMP.1988.60.873}, like electron transport in organic molecules
\cite{Science.1991.252.1285} as proteins and polymers, may be
described by one-dimensional pathways (trajectories through the edges),
changing from one path to another due to scattering at the vertices
centers.
More recently, quantum graphs have also been used to characterize
molecular connectivity \cite{CR.2000.100.3827,CR.2008.108.1127}.

In the realm of condensed matter physics, under certain conditions
\cite{Book.1981.Kao,PRB.2000.62.11473} charge transport in solids is
likewise well described by one-dimensional dynamics in branched
(so network-like) structures, as in polymer films
\cite{APL.2000.77.693,EPL.2007.79.47011}.
Quantum graphs have also been applied in the analysis of disordered
superconductors \cite{PRB.1983.27.1541}, Anderson transition in
disordered wires \cite{PRB.1981.23.4828,PRL.1982.48.823}, quantum Hall
systems \cite{PRL.1997.79.721}, superlattices \cite{PRB.2000.62.16294},
quantum wires \cite{JPA.1996.29.87}, mesoscopic quantum systems
\cite{PRB.1990.42.9009,PRL.2004.92.186801,PRB.2005.72.115327,
JPA.2005.38.3455}, and in connection with laser tomography
technologies \cite{Inproceedings.2003.Bondarenko}.

To understand fundamental aspects of quantum mechanics, graphs are
idealized exactly soluble models to address, e.g., band spectrum
properties of lattices \cite{PRL.1995.74.3503,PRL.2014.112.070406}, the
relation between periodic-orbit theory and Anderson localization
\cite{PRL.2000.84.1427}, general scattering \cite{JMP.2014.55.083524},
chaotic and diffusive scattering
\cite{PRL.2000.85.968,PRE.2001.65.016205,PRB.2009.80.245441}, and
quantum chaos \cite{PRL.2014.112.144102}.
In particular, quantum graphs relevance in grasping distinct features of
quantum chaotic dynamics have been demonstrated in two pioneer papers
\cite{PRL.1997.79.4794,AoP.1999.274.76}.
Through elucidating examples, such works show that the corresponding
spectral statistics follow very closely the predictions of the
random-matrix theory \cite{Book.2004.Mehta}.
They also present an alternative derivation of the trace
formula\footnote{For $G({\bf r}'',{\bf r}';E)$ the energy dependent
Green's function of a quantum system (Sec. \ref{sec:gfqg}), the trace of $G$, or
$g(E) = \int d{\bf r} \, G({\bf r},{\bf r};E)$, is important because
it leads to the problem density of states
$\rho(E) = - (1/\pi) \lim_{\epsilon \rightarrow 0}
\operatorname{Im}[g(E + i \epsilon)]$.
The Gutzwiller trace formula \cite{Book.1990.Gutzwiller} is an elegant
semiclassical approximation for $\rho(E)$, in which $g(E)$ is given in
terms of  sums over classical periodic orbits.},
highlighting the similarities with the famous Gutzwiller's expression
for chaotic Hamiltonian systems
\cite{JMP.1971.12.343,Book.1990.Gutzwiller}.
Actually, a very welcome fact in the area is the possibility to obtain
exact analytic results for quantum graphs even when they present chaotic
behavior
\cite{PRL.2002.88.044101,PRE.2002.65.046222,PE.2001.9.523,
PRE.2001.64.036225}.
Important advances and distinct approaches to spectrum statistics
analysis in quantum graphs, as well as the relation with quantum chaos,
can be found in a nice review in \cite{AP.2006.55.527}.

As a final illustration of the vast applicability of graphs we mention
two issues in the important fields of quantum information and quantum 
computing \cite{Book.2010.Nielsen}.
First, for the metric case (the focus in this review), it has been proposed
that the logic gates necessary to process and operate qubits could be
implemented by tailoring the scattering properties of the vertices
along a quantum graph \cite{FP.2000.48.703,PLA.2004.330.338}.
However, much more common in quantum information is to consider only the
topological features of the graphs \cite{Conference.2008.Raussendorf},
hence not ascribing lengths to the edges.
Such structures are usually referred as discrete or combinatorial graphs
(for a parallel between metric and combinatorial see, e.g.,
\cite{JPA.2005.38.4887}).
They are the basis to construct the so called graph-states
\cite{PRL.2001.86.910,PRL.2001.86.5188,PRA.2003.68.022312,
Inproceedings.2006.Hein,NJP.2014.16.113070}, in which the vertices
are the states themselves (e.g., spins 1/2 constituting the qubits) and
the edges represent the pairwise interactions (for instance, an
Ising-like coupling \cite{PRA.2014.89.052335}) between two vertices
states \cite{Conference.2006.Feder}.
Graph-states are very powerful tools to unveil different aspects of
quantum computation.
For instance, to establish relations between different computational
methods schemes \cite{PRA.2010.82.030303,NJP.2014.16.113070} and to
demonstrate that entanglement can help to outperform the Shannon limit
capacity (of the classical case) in transmitting a message with zero
probability of error throughout a channel presenting noise
\cite{CMP.2012.311.97,PNASU.2012.110.15}.

Second, also relevant in quantum information processing is the concept
of quantum walks, loosely speaking, the quantum version of classical
random walks \cite{PRA.1993.48.1687,CP.2003.44.307,QIP.2012.11.1015}.
Quantum walks are extremely useful either theoretically, as primitives 
of universal quantum computers
\cite{PRL.2009.102.180501,PRA.2010.81.042330,Science.2013.339.791}, or
operationally, as building blocks to quantum algorithms
\cite{PTRSA.2006.364.3407,QIP.2012.11.1015,Incollection.2009.Mosca,
Book.2013.Portugal}.
Thus, since there is a close connection between quantum walks and
quantum graphs
\cite{Inbook.2006.Tanner,IJQI.2006.4.791,JPA.2004.37.6675,
arXiv:math.0404467}, this might open the possibility of extending
different techniques to treat quantum graphs to the study of quantum
walks
\cite{PRA.2009.80.052301,PRA.2011.84.042343,PRA.2012.86.042309,
Phdthesis.2009.Andrade}, therefore helping in the development of quantum
algorithms.

The physical construction of quantum graphs is obviously an essential
matter.
In such regard, an important result is that in
Ref. \cite{PRE.2004.69.056205}.
It shows that quantum graphs can be implemented through microwave
networks due to the formal equivalence between the Schr\"{o}dinger
equation (describing the former) and the telegraph equation (describing
the latter) \cite{PRE.2004.69.056205}.
Currently, these kind of systems are among the most preeminent
experimental realizations of quantum graphs --
as demonstrated by the vast literature on the topic
\cite{PRE.2015.92.022904,PRE.2014.89.032911,PS.2014.2014.014025,
PRE.2013.87.62915,APPA.2013.124.1078,JAP.2013.113.164101,
PS.2013.2013.014041,PRL.2012.109.40402,PS.2012.2012.014018,
Incollection.2011.Lawniczak,PRE.2011.83.66204,PS.2011.2011.014014,
APPA.2011.120.185,CMSIM.2011.1.105,PRE.2010.81.46204,
PS.2009.2009.014050,APPA.2009.116.749,PRE.2008.77.56210,
Incollection.2008.Hul,APPA.2007.112.655,JPA.2005.38.10489}.
Nonetheless, microwave networks are not the only possibility.
In particular, optical lattices
\cite{Book.2009.Goldman,PRA.2010.82.053605,Book.2012.Lewenstein}
and quasi-1D structures of large donor-acceptor molecules (with
quasi-linear optical responses) \cite{JNOPM.2015.24.1550018} might also
constitute very appropriate setups for building quantum graphs.

The implementation of quantum graphs -- of course, alongside with the
concrete applications -- can also be quite helpful in settling relevant
theoretical questions.
As an illustrative example, consider the famous query posed by Mark Kac
in 1966: `can one hear the shape of a drum?' \cite{TAMM.1966.73.1}.
Its modified version in the present context is \cite{JPA.2001.34.6061}:
`can one hear the shape of a graph?'.
It has been proved that for simple graphs (see next Sec.) whose all
edges lengths are incommensurable, the spectrum is uniquely determined
\cite{JPA.2001.34.6061}.
In other words, in this case one should be able to reconstruct the graph
just from its eigenmodes.
But if these assumptions are not verified, then distinct graphs can be
isospectral \cite{JPA.2009.42.175202,JGA.2010.20.439}.
An interesting perspective to the problem arises by adding infinity
leads to originally closed graphs
\cite{AAM.2005.35.58,JPA.2005.38.4901}.
So, we have scattering system which can be analyzed in terms of their
scattering matrices ${\mathcal S}$.
Two metric graphs, ${\Gamma}_A$ and ${\Gamma}_B$, are said isoscattering
either if ${\mathcal S}_A$ and ${\mathcal S}_B$ share the same set of
poles or the phases of $\det[{\mathcal S}_A]$ and
$\det[{\mathcal S}_B]$ are equal \cite{JPA.2010.43.415201}.
Hence, the question is now: can the poles of ${\mathcal S}$ and phases
of  $\det[{\mathcal S}]$ {\em alone} define the graph's shape?
The answer is again negative \cite{AAM.2005.35.58,PRL.2012.109.40402},
as nicely confirmed through microwave networks experiments
\cite{PRL.2012.109.40402} (see also \cite{PRE.2014.89.032911}).
However, by analyzing in more details actual scattering data (e.g.,
in the time instead of frequency domain \cite{PRE.2013.87.62915}) it
does become possible to distinguish isoscattering graphs which are
topologically different.

Quantum graphs as a well posed general mathematical problem requires the
establishment of the underlying self-adjoint operator, i.e., the proper
definition of the wave equation with its correct boundary conditions.
Probably, the first important step along this direction was taken in
1953 in Ref. \cite{JCP.1953.21.1565}.
There, graphs were thought of as idealized web of wires or wave guides,
but for the widths being much smaller than any other spatial scale.
Assuming the lateral size of the wire small enough, any propagating wave
remains in a single transverse mode.
Therefore, instead of the corresponding partial differential
Schr\"{o}dinger equation, one can deal with ordinary differential
operators.
If no external field is applied or no potential $V$ for the wires
is assumed, the one dimensional motion along the edges is free and
anywhere in the graph the wave number reads $k=\sqrt{2\mu E/\hbar^2}$,
with the energy $E$ a constant.
Concerning the nodes, they either can be faced as scattering centers
(thus, conceivably described by local ${\mathcal S}$ matrices) or the
{\em loci} where consistent matching conditions for the partial wave
functions (i.e., the $\psi$'s in the distinct edges) must be imposed
(Sec. \ref{sec:qmgdp}).

In contrast, graphs with non-vanishing potentials -- sometimes referred
to as `dressed' \cite{PRE.2001.63.066201,PRE.2002.65.046222} -- lead
to solutions with spatially dependent $k$'s along the edges.
An important subset of dressed are scaling quantum
graphs{\footnote{Briefly, to each edge $e$ of a scaling quantum graph
one can associate a numerical constant $\gamma_e$.
Then, along $e$ the wave number is $k_e = \gamma_e k_0$, with
$k_0 = \sqrt{2 \mu E / \hbar^2}$ a constant.}}
\cite{PRE.2004.70.046206,JETP.2003.77.530,
PRE.2003.68.055201,PRE.2002.65.046222,PRL.2002.88.044101,
JETP.2002.94.1201}, whose mathematical foundations are discussed
in \cite{arXiv:quant-ph.0203126}.
They are particularly interesting because although their classical limit
is chaotic, the quantum spectrum is exactly obtained through analytic
periodic orbit expansions \cite{PRL.2002.88.044101}.
Another very relevant class of dressed quantum graphs is that described
by magnetic Schr\"odinger operators \cite{CMP.2003.237.161}.
In this case one assumes arbitrary inhomogeneous magnetic fields in the
network \cite{Nanotechnology.2001.12.570}, such that for each edge $e$
there is a corresponding vector potential $A_e$.
So, formally we have to make the traditional momentum operator
substitution in the Schr\"odinger equation:
$d/dx_e \rightarrow d/dx_e - i A_e$.
Recently, quantum graphs with magnetic flux have attracted a lot of
attention due to the many distinct phenomena emerging in these systems
\cite{arXiv:1012.1845,NPCM.2015.6.309,PTRSA.2014.372.20120522,
JPA.2015.48.125302,JMP.2013.54.032104,JMP.2013.54.042103,
MPCPS.2010.148.331,PLA.2013.377.1788}.

Given the discussion so far, it is already clear that a quantum graph
is, after all, just an usual quantum problem.
As such, its solution basically means to determine properties like
wave packets propagation \cite{WRCM.2005.15.101,WRCM.2010.20.260},
eigenstates (either bound and scattering states)
\cite{IJM.2013.193.1,arXiv:1405.5871}, eigenenergies
\cite{PLA.2013.377.439}, etc.
This can be accomplished from, say, a suitable Schr\"{o}dinger equation
and appropriate boundary conditions for each specific graph topology,
Sec. \ref{sec:qmgdp}.
But operationally there are many ways to mathematically deal with these 
systems, so different techniques can be employed.
For instance, we can cite self-adjoint extension approaches
\cite{JPA.1999.32.595}, and the previously mentioned scattering
${\mathcal S}$
matrix methods \cite{PRL.1997.79.4794} and the trace formula based on
classical periodic orbits expansions \cite{AoP.1999.274.76}.

It is well known that the energy Green's function $G$ is a very powerful
tool in quantum mechanics \cite{Book.2006.Economou,Book.1995.Barton}.
Its knowledge allows to determine essentially any relevant quantity
for the problem (e.g., the time evolution can be calculated from the
time-dependent propagator, which is the Fourier transform of $G$).
So, it should be quite natural to consider Green's function approaches
in the study of graph structures.
In fact, one of the first works in this direction
\cite{PRE.2001.65.016205} has employed $G$ to describe transport in open
graphs.
Later, the many possibilities in utilizing Green's functions techniques
for arbitrary quantum graphs have been discussed and exemplified in
\cite{JPA.2003.36.545}, with general and rigorous results further
obtained from such a method in
\cite{JPA.2005.38.4859,Inproceedings.2006.Kostrykin}.
Recently, Green's functions have been used to investigate (always in the
context of quantum graphs): searching algorithms for shortest paths
\cite{NJP.2011.13.013022}, Casimir effects \cite{arXiv:0707.3710},
vacuum energy in quantum field theories
\cite{Inproceedings.2006.Fulling}, and resonances on unbounded
star-shaped networks \cite{RIT.2007.1.1}.
Lastly, but not the least important, the special topological features
of networks make it possible (at least in the undressed
case{\footnote{The Green's function for scaling quantum graphs can also
be calculated exactly. This will be briefly discussed in Sec. 3.}})
to obtain the exact $G$ in a closed analytic form for any finite
(i.e., a large although limited number of nodes and edges) arbitrary
graph.
Certainly, this contrasts with most problems in quantum mechanics,
for which exact analytic solutions are very hard to find
\cite{Book.2005.Takahashi,Book.2011.Blumel}.

Therefore, regarding the purpose of this review, we start observing 
there is a huge literature discussing general features and applications 
of classical graphs.
To cite just one, more physics-oriented, we mention communicability --
so, signal transport -- in classical networks \cite{PR.2012.514.89}.
In the quantum case comprehensive overviews are not so abundant,
notwithstanding particular relevant aspects can be found addressed in
details in some very interesting works
\cite{AoP.1999.274.76,AP.2006.55.527,WRCM.2004.14.3,WRCM.2004.14.107,
JPA.2005.38.4887,Inproceedings.2008.Kuchment} (with also a good source
of a formal and rigorous treatment being \cite{Book.2012.Berkolaiko}).
In this way, our first goal is to survey graphs as ordinary quantum
mechanics problems, but highlighting that their special characteristics 
can give rise to rich quantum phenomena.

The second is to do so by specifically considering one of the most
powerful methods to treat quantum graphs, namely, the Green's function
approach.
For arbitrary graphs, we discuss in an unified manner how to obtain the
exact energy domain $G$ as a general sum over paths
`\textit{a la Feynman}'
\cite{JPA.1998.31.2975,JPA.2001.34.5041,JPA.2003.36.227}.
These paths must be weighted by the proper quantum amplitudes, given by
energy-dependent scattering matrices elements associated to the
vertices.
We examine a schematic way to regroup the multi-scattering contributions
(essentially a factorization method
\cite{JPA.1999.32.595,JMP.2001.42.1563,JGT.1997.24.291,JPA.2000.33.63}),
leading to a final closed analytic expression for $G$.
This particular procedure to construct the exact $G$ is very useful to
interpret many results concerning quantum graphs, like interference
in transport processes
\cite{PRE.2001.65.016205,NL.2010.10.4260,CSR.2015.44.875}.
With the help of illustrative examples, we elaborate on how to extract
from $G$ the graphs quantum properties.

The work is organized as the following.
In Section \ref{sec:qmgdp} we define and discuss general quantum
graphs.
In Section \ref{sec:gfqg} we consider in great detail the Green's
function approach for such systems.
In Section \ref{sec:ogfqggp} we present (with examples) the
factorization protocols which allow to cast $G$ as a closed analytic
formula.
Distinct applications are addressed in the next three Sections.
More specifically, the general determination of bound and scattering
states, analysis of representative graphs (cube, binary trees, and
Sierpi\'{n}ski-like graphs), and quasi-bound states in open structures, are
considered, respectively, in Secs. \ref{sec:cesog}, \ref{sec:rqg}, and
\ref{sec:qbsqg}.
Finally, we drawn our final remarks and conclusion in Section
\ref{sec:conclusion}.

%%% Local Variables:
%%% mode: latex
%%% TeX-master: "green-qg-pr"
%%% ispell-local-dictionary: "american"
%%% End:
\section{Quantum mechanics on graphs: general aspects}
\label{sec:qmgdp}

\subsection{Graphs}
\label{sec:graph}

A finite \textit{graph} $X(V,E)$ is a pair consisting of two sets, of
vertices (or nodes) $V(X)=\{1,2, \ldots, n\}$ and of edges (or bonds)
$E(X)=\{e_{1}, e_{2}, \ldots,e_{m}\}$
\cite{Book.2010.Diestel,Book.2001.Godsil}.
Thus, the total number of vertices and edges is given, respectively, by
$n=|V(X)|$ and $m=|E(X)|$.
If the vertices $i$ and $j$ are linked by the edge $e_s$, then
$e_s \equiv \{i,j\}$ (hereafter $i, j = 1, \ldots, n$ and
$r, s=1,\ldots,m$).
For an undirected graph, any edge $\{i, j\}$ has the same properties
\cite{Book.2008.Bondy} in both $i \rightarrow j$ and $j \rightarrow i$
`directions': $\{i, j\} \equiv \{j, i\}$.
For simple graphs $e_s \neq \{j, j\}$ and $e_{r} = e_{s}$ only if
$r=s$.
Hence, in this case there are no loops or pair of vertices
multiple-connected. 
Finally, for connected graphs the vertices cannot be divided into two
non-empty subsets such that there is no edge joining the two subsets.

The graph topology, i.e., the way the vertices and edges are associated,
can be described in terms of the adjacency matrix $A(X)$ of
dimension $n \times n$.
For simple undirected graphs, the $ij$-th entry of $A(X)$ reads
\begin{equation}
  A_{ij}(X)=
  \begin{cases}
    1, & \;\text{\rm if }\; \{i,j\}\in E(X),\\
    0, & \;\text{\rm otherwise}.
  \end{cases}
\end{equation}
Two vertices are said neighbors whenever they are connected by an edge.
Thus, the set
\begin{equation}
E_i(X)=\left\{j:\{i,j\}\in E(X)\right\}
\end{equation}
is the neighborhood of the vertex $i\in V(X)$ and the degree (or
valence) of $i$ is
\begin{equation}
v_i=|E_i(X)|=\sum_{j=1}^{n} A_{ij}(X).
\label{eq:grau}
\end{equation}
Note that
\begin{equation}
|E(X)| = \frac{1}{2} \sum_{i=1}^{n} |E_i(X)|.
\label{eq:bonds}
\end{equation}

So far, the above definitions refer to \textit{discrete} or
\textit{combinatorial} graphs.
To discuss quantum graphs it is necessary to equip the graphs with a
metric.
Therefore, a \textit{metric graph} $\Gamma(V,E)$ is a graph $X(V, E)$
for  which it is also assigned a length $\ell_{e_{s}}\in(0,+\infty)$ to 
each edge.
If all edges have finite length the metric graph is called
\textit{compact}, otherwise it is \textit{non-compact}.
In this latter case $\Gamma$ has one ore more `leads'.
A lead is a single ended edge $e_r$, which leaves from a vertex and
extends to the semi-infinite (so $\ell_{e_{s}} = + \infty$).

In the quantum description, for each edge $e_s$ (with $e_s$ either
joining two vertices $i$ and $j$ or leaving from vertex $j$ to the
infinite) we assume a coordinate $x_{e_{s}}$, indicating the position
along the edge.
For $e_{s} = \{i,j\}$, to choose at which vertex ($i$ or $j$)
$x_{e_{s}} = 0$ and $x_{e_{s}} = \ell_{e_s}$\footnote{It is an usual practice
in the study of quantum graphs, although not strictly necessary, to
assume $x_{e_s} \geq 0$ (even at the leads, when then 
$0 \leq x_{e_s} < + \infty$). We follow this convention throughout the
present review.}  
is just a matter of convention, 
and can be set according to the convenience in each specific system.
Of course, for $e_{s}$ a lead attached to $j$, a natural choice is
$x_{e_{s}}=0$ at $j$.

In the remaining of this review we will (mainly but not only) focus on 
simple connected graphs, the most studied situation in quantum mechanics
\cite{IJQI.2006.4.791}.
But we stress that the Green's function discussed here is also valid for
non-simple graphs, i.e., for many edges joining the same two vertices
and for the existence of loops:
one just need to consider the proper reflections and transmissions
quantum amplitudes (Sec. \ref{sec:gfqg}) for the propagation along 
these extra edges.
This will be illustrated with certain examples in Sec. \ref{sec:rqg}.

\begin{figure}
  \centering
  \includegraphics*[width=0.5\textwidth]{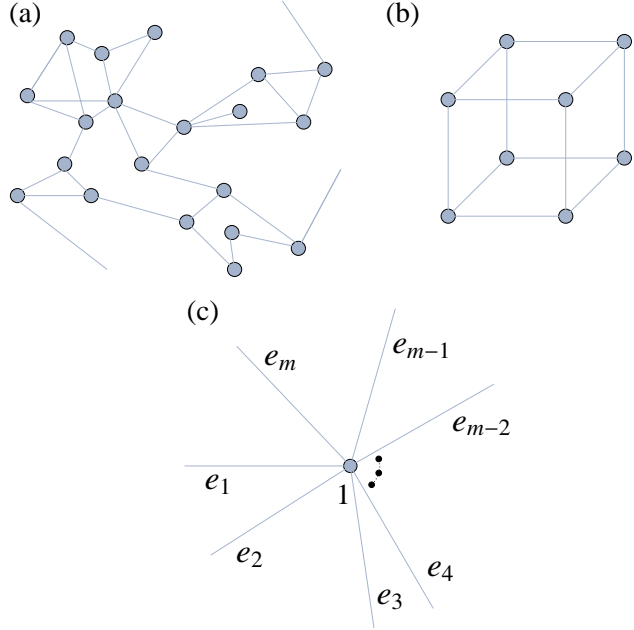}
  \caption{\label{fig:fig1} (Color online).
    Examples of (a) open and (b) closed quantum graphs.
    (c) A open star graph with a single vertex $V(\Gamma)=\{1\}$
    connected to $E(\Gamma)=\{e_{1},\ldots,e_{m}\}$ leads.}
\end{figure}

\subsection{The time-independent Schr\"odinger equation on graphs}
\label{sec:qg}

A \textit{quantum graph} is a metric graph structure $\Gamma(V,E)$, on
which we can define a differential operator $H$ (usually the
Schr\"{o}dinger Hamiltonian) together with proper vertices boundary
conditions \cite{AoP.1999.274.76,AP.2006.55.527}.
In others words, a quantum graph problem is a triple
\begin{equation*}
  \{
  \text{$\Gamma(V,E)$, Hamiltonian operator $H$ on $E(\Gamma)$,
        boundary conditions for $V(\Gamma)$}
  \}.
  \label{eq:tripla}
\end{equation*}
A quantum graph is called \textit{closed} if the respective metric graph
is compact, otherwise it is called \textit{open}.
A schematic representation of quantum graphs \cite{Book.2008.Bondy} is
depicted in Figure \ref{fig:fig1}.

\begin{figure}
  \centering
  \includegraphics*[width=0.25\textwidth]{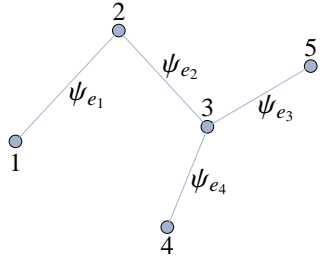}
  \caption{\label{fig:fig2} (Color online).
    A quantum graph $\Gamma(V,E)$ -- with $V(\Gamma)=\{1,2,3,4,5\}$ and
    $E(\Gamma)=\{\{1,2\},\{2,3\},\{3,4\},\{3,5\}\}$ -- and the indication
    of the $\psi_{e_s}$ components of $\Psi$ in each one of the $\Gamma$ 
    edges.
    The wave functions must be matched through the boundary condition at
    each vertex $i \in V(\Gamma)$.
    Specifically:
    at $i=1$: $\psi_{e_1}$;
    at $i=2$: $\psi_{e_1}, \psi_{e_2}$;
    at $i=3$: $\psi_{e_2}, \psi_{e_3}, \psi_{e_4}$;
    at $i=4$: $\psi_{e_4}$;
    at $i=5$: $\psi_{e_3}$.}
\end{figure}

The total wave function $\Psi$ is a vector with $m$ components, written
as
\begin{equation}
\Psi=
\left(
  \begin{array}{c}
    \psi_{e_{1}}(x_{e_1})\\
    \psi_{e_{2}}(x_{e_2})\\
    \vdots\\
    \psi_{{e_m}}(x_{e_{m}})
  \end{array}
\right).
\end{equation}
The Hamiltonian operator on $E(\Gamma)$ consists of the following
unidimensional differential operators defined on each edge $e_{s}$
\cite{PRL.1994.72.896,PRB.1983.27.1541} (the dressed case)
\begin{equation}
  H_{e_{s}}(x_{e_{s}}) = -\frac{\hbar^2}{2 \mu}
\frac{d^2}{dx_{e_{s}}^2}+V_{e_{s}}(x_{e_{s}}).
  \label{eq:schroedingeroperator}
\end{equation}
Here, $V_{e_{s}}(x_{e_{s}})$ is the potential (usually assumed to be non-negative
and smooth) in the interval $0 < x_{e_s} < \ell_{e_{s}}$.
Different works have considered the above Hamiltonian for
non-vanishing potentials (for instance, see
\cite{PRE.2003.68.055201,PRE.2002.65.046222,PRL.2002.88.044101,
JPA.2003.36.545,JNOPM.2013.22.1350041,JPA.2015.48.365201,
arXiv:1503.02253,PRE.93.032204.2016}).
However, in the literature, even in papers discussing quantum chaos
\cite{PRL.1997.79.4794,AoP.1999.274.76,AP.2006.55.527,
PRL.2008.101.264102,PRL.2014.112.144102}, it is usual to
have for any $e_{s}$ that $V_{e_{s}} = 0$ (the case we assume in this
review).
Then, the component $\psi_{e_{s}}(x_{e_{s}})$ of the total wave
function $\Psi$ is the solution of ($k=\sqrt{2 \mu E}/\hbar$)
\begin{equation}
  -\frac{d^2 \psi_{e_{s}}}{dx_{e_{s}}^2} = k^2\psi_{e_{s}}(x_{e_{s}})
  \ \Rightarrow \
  \psi_{e_{s}}(x_{{e}_s}) =
  c_{+, e_{s}}\, \exp[+ i \,k \, x_{e_{s}}] +
  c_{-, e_{s}}\, \exp[- i \, k \, x_{e_{s}}],
 \label{eq:bondschroedinger}
\end{equation}
with the $c$'s constants.
All these wave functions must satisfy appropriate boundary conditions
at the vertices, ensuring continuity, global probability current
conservation, divergence free $\psi$'s and uniqueness.
Technically, the match of the boundary conditions in each vertex is the most
cumbersome step in obtaining the final full $\Psi$
(in Figure \ref{fig:fig2} we illustrate which components must be matched
in which vertices for a particular example of a graph with
$V(\Gamma)=\{1,2,3,4,5\}$ and
$E(\Gamma)=\{\{1,2\},\{2,3\},\{3,4\},\{3,5\}\}$).

Furthermore, the imposition of these boundary conditions
\cite{AoP.1999.274.76,AP.2006.55.527,Book.2009.Teschl}
renders the Hamiltonian operator to be self-adjoint
\footnote{Consider a continuous linear (so bounded) operator
$\mathcal{O}$ of domain  $\mathcal{D}(\mathcal{O})$ in a Hilbert space
${\mathcal H}$.
The adjoint $\mathcal{O}^{\dagger}$ (also bounded) of the operator
$\mathcal{O}$ is such that
$\langle {\mathcal O} \psi | \phi\rangle=
\langle \psi | \mathcal{O}^{\dagger} \phi \rangle$
for $\psi \in {\mathcal D}({\mathcal O})$ and $\varphi \in {\mathcal H}$.
$\mathcal{O}$ is self-adjoint if and only if
$\mathcal{O}=\mathcal{O}^{\dagger}$ and
$\mathcal{D}(\mathcal{O})=\mathcal{D}(\mathcal{O}^{\dagger})$
\cite{Book.2009.Teschl}.}.
In fact, the most general boundary conditions at a vertex of a quantum
graph (consistent with flux conservation \cite{PRL.1995.74.3503}) can be
determined through self-adjoint extension
techniques \cite{PLA.1988.128.493,RMP.1989.28.7}.
Let us denote by \cite{JPA.1999.32.595,JMP.2001.42.1563}
${\Psi_j} = (\psi_{e_{j_1}},\psi_{e_{j_2}}, \ldots, \psi_{e_{j_{v_j}}})^{T}$
and
${\Psi_j'} = (\psi_{e_{j_1}}',\psi_{e_{j_2}}', \ldots, \psi_{e_{j_{v_j}}}')^{T}$,
respectively, the wave functions and their derivatives associated to
the $v_j$ edges attached to the vertex $j$.
Then, the boundary conditions can be specified through $v_j \times v_j$ matrices
${\mathcal A}_j$ and ${\mathcal B}_j$, with
${\mathcal A}_j \Psi_j = {\mathcal B}_j \Psi_j^{'}$ at $j$.
One ensures self-adjointeness of the Hamiltonian operator by imposing
current conservation $\Psi_j^{\dagger}\Psi_j'={\Psi'}_j^{\dagger}\Psi_j$.
As shown in \cite{JPA.1999.32.595,JMP.2001.42.1563},
the general solution for this problem implies that
${\mathcal A}_j {\mathcal B}_j^{\dagger} =
{\mathcal B}_j {\mathcal A}_j^{\dagger}$, resulting in a set of $v_j^{2}$
independent real parameters to characterize the boundary conditions at $j$.
More on this is discussed in the Appendix \ref{app:flux}, but here
we comment that in physical terms, the self-adjointness of the
Hamiltonian implies that the dynamics does not allow the
vertices to behave as sinks or sources.

\subsection{The vertices as zero-range potentials}
\label{sec:vertices-zerorange}

From the previous discussion, in an undressed quantum graph the edges
$e_{s}$ can be viewed as free unidimensional spatial directions of length
$\ell_{e_{s}}$ and the vertices as point structures (0D), whose action is
to impose the proper boundary conditions on the $\psi$'s.
In the usual 1D quantum mechanics, arbitrary zero-range potentials,
also known as point interactions, have exactly such effect
\cite{Book.2004.Albeverio,AJP.2001.69.322}
(see Appendix \ref{appendix:a1}).
A textbook example is the Dirac delta-function potential that simply
determines, at its location, a specific boundary condition to the wave
function \cite{Book.1999.Flugge}.

Hence, to describe the quantum dynamics along a graph we can take the
$j$'s as arbitrary zero-range interactions, an approach fully consistent
with the general boundary conditions treatment described in Sec. 
\ref{sec:qg} (Appendix \ref{app:flux}).
To assume the vertices as potentials brings up two important advantages.
(a) The $j$'s become point scatterers, which are completely characterized 
by their reflections and transmission amplitudes (recall this is exactly 
the case for a delta-function, for which $\psi$ can be obtained without 
considering any boundary conditions).
So, a purely scattering treatment solves the problem -- see, e.g., the 
pedagogical discussion in \cite{Book.1981.Baym}.
(b) General point interactions are very diverse in their scattering
properties.
For instance, the intriguing aspects of transmission and reflection from 
point interactions have been discussed in distinct situations, such as, 
time-dependent potentials \cite{JOB.7.S77.2005}, nonlinear 
Schr\"{o}dinger equation \cite{JPA.2004.37.367} and shredding by sparse 
barriers \cite{PLA.2000.277.1}.
So, the mentioned procedure allows to have all the features of arbitrary 
zero-range potentials also in the context of quantum graphs.

As demonstrated in the Appendix \ref{appendix:a1}, to determine the
boundary conditions that a point interaction in the line (say, at 
$x_0 = 0$) imposes on the the wave function at $x=0$ is entirely 
equivalent to specify the potential scattering ${\mathcal S}$ matrix 
elements.
This also holds true when the vertex, a zero-range potential, instead of
being attached to two edges (the `left' ($-\infty < x < 0$) and
`right' ($0 < x < + \infty$) semi-infinite leads for the
1D line case), is connected to $v_j$ edges, representing $v_j$ 1D
``directions'', see Figure \ref{fig:fig1} (c).
From the Appendix \ref{app:generalization}, we then can define for each
vertex $j$ a matrix ${\mathcal S}_{j}$, of elements
${\mathcal S}_{j}^{(s,s)}(k) = r_{j}^{(s)}(k)$
and ${\mathcal S}_{j}^{(s,r)}(k) = t_{j}^{(s,r)}(k)$
(from now on, we will label edges $e_{j_s}$ and $e_{j_r}$ simply as
$s$ and $r$), such that
\begin{itemize}
\item $t_{j}^{(s,r)}(k)$ is the quantum amplitude for a plane wave,
of wave number $k$, incoming from the edge $r$ towards the vertex $j$
to be transmitted to the edge $s$ outgoing from $j$.
\item $r_{j}^{(s)}(k)$ is the quantum amplitude for a plane wave,
of wave number $k$, incoming from the edge $s$ towards the vertex $j$
to be reflected to the edge $s$ outgoing from $j$.
\end{itemize}
The required conditions for self-adjointeness (i.e., probability flux
conservation) along the whole graph (Appendix \ref{app:general-graph}),
demands that
${{\mathcal S}}(k) {{\mathcal S}}^{\dagger}(k) =
{{\mathcal S}}^{\dagger}(k) {{\mathcal S}}(k) = \mathbf{1}$
and
${{\mathcal S}}(k)={{\mathcal S}}^{\dagger}(-k)$,
so yielding
\begin{equation}
\sum_{l=1}^{v_j} {\mathcal S}_{j}^{(s,l)}(k) \,
               {{\mathcal S}_{j}^{(r,l)}}^{*}(k) =
\sum_{l=1}^{v_j} {\mathcal S}_{j}^{(l,s)}(k) \,
               {{\mathcal S}_{j}^{(l,r)}}^{*}(k) = \delta_{sr},
\ \
\mathcal{S}_{j}^{(s,r)}(k) = {{\mathcal S}_{j}^{(r,s)}}^{*}(-k).
\label{eq:sqg}
\end{equation}

Summarizing, for quantum graphs it is complete equivalent to set either
the boundary conditions for the $\psi$'s at each vertex, as mentioned in
Sec. \ref{sec:qg}, or to specify the scattering properties of the
different $j$'s through the ${\mathcal S}_{j}^{(r,s)}$ matrices obeying
to Eq. \eqref{eq:sqg}.
We also observe that eventually one could have bound states for a
given point interaction potential $j$ depending on the particular
BC imposed to $\psi$ at the vertex location.
In the scattering description, the quantum coefficients
$R$ and $T$ have poles at the upper-half of the complex plane $k$,
corresponding to the possible eigenenergies.
The eigenfunctions can then be obtained from an appropriate extension
of the scattering states to those $k$'s values \cite{Book.1955.Schiff}.
This will be exemplified in Section \ref{sec:rqg}.

%%% Local Variables:
%%% mode: latex
%%% TeX-master: "green-qg-pr"
%%% ispell-local-dictionary: "american"
%%% End:

\section{Energy domain Green's functions for quantum graphs}
\label{sec:gfqg}

\subsection{The basic Green's function definition in 1D}

The Green's function $G(E)$ is an important tool in quantum mechanics
\cite{Book.2006.Economou}.
In the usual 1D case, it is defined by the inhomogeneous differential equation
($H(x)=-(\hbar^{2}/(2 \mu)) \, d^{2}/dx^{2} + V(x)$)
\begin{equation}
  \label{eq:defgf}
  [E-H(x_{f})]G(x_{f},x_{i};E) =
  \delta(x_{f}-x_{i}),
\end{equation}
where $G(x_{f},x_{i};E)$ is also subjected to proper boundary conditions.

Suppose we have a complete set of normalized eigenstates
$\psi_{s}(x)$ ($s=0,1,...$, discrete spectrum) and
$\psi_{\sigma}(x)$ ($\sigma > 0$, continuum spectrum),
with
\begin{equation}
H \, \psi_{s} = E_{s} \, \psi_{s}, \qquad
H \, \psi_{\sigma} = \frac{\hbar^2 \sigma^2}{2 \mu} \, \psi_{\sigma}.
\end{equation}
Then, the solution of Eq. \eqref{eq:defgf} is formally
\begin{equation}
  \label{eq:segf}
G(x_{f},x_{i};E)=
\sum_{s}\frac{\psi_{s}(x_{f}) \, {\psi_{s}}^*(x_{i})}{(E-E_{s})} +
\int_{0}^{\infty} d\sigma \,
\frac{\psi_{\sigma}(x_f) \, {\psi_{\sigma}}^{*}(x_i)}
{(E - \hbar^{2} \sigma^{2}/(2 \mu))}.
\end{equation}
From Eq. (\ref{eq:segf}) we can identify the poles of the Green's
function with the bound states eigenenergies $E_s$ and the residues at
each pole with a tensorial product of the corresponding bound state
eigenfunction.
The  continuous part of the spectrum corresponds to a branch
cut of $G(x_{f},x_{i};E)$ \cite{JMP.1992.33.643,Book.2006.Kleinert}.
Given Eq. \eqref{eq:segf}, the limit
\begin{equation}
\label{eq:limg}
\lim_{E \to E_s} (E-E_s) \, G(x_f,x_i;E) =
\psi_s(x_f) \, {\psi_s}^{*}(x_i)
\end{equation}
can be used to extract the discrete bound states from $G$.

\subsection{The exact Green's function written as a generalized
semiclassical expression}

There are basically three methods for calculating the Green's function
\cite{Book.2006.Economou}:
solving the differential equation in \eqref{eq:defgf};
summing up the spectral representation in \eqref{eq:segf};
or performing the Feynman path integral expansion for the
propagator in the energy representation
\cite{Book.2010.Feynman,Book.2005.Schulman}.
In particular, for contexts similar to the present work (see next), the
latter approach has been used to study scattering by multiple potentials in 1D
\cite{JPA.1998.31.2975,JPA.2001.34.5041}, to calculate the eigenvalues
of multiple well potentials \cite{JPA.2003.36.227}, to study scattering
quantum walks \cite{PRA.2011.84.042343,PRA.2012.86.042309},
and to construct exact Green's function for piecewise constant potentials
\cite{PRA.1993.48.2567,PLA.2014.378.1461}.

The exact Green's function for an arbitrary finite array of potentials of
compact support\footnote{$V_n(x)$ is said to have compact support in
the interval ${\mathcal I}_n \equiv \{x \, | \, a_n < x < b_n \}$ if
$V_n(x)$ identically vanishes for $x \notin {\mathcal I}_n$.
An arbitrary array of $N$ potentials of compact support is given by
$V(x) = \sum_{n=1}^{N} V_n(x)$, for all ${\mathcal I}_n$'s disjoint.}
has been obtained in \cite{JPA.1998.31.2975}, with an extension for more
general cases presented in \cite{JPA.2001.34.5041}.
For the derivations in \cite{JPA.1998.31.2975}, it is necessary for the
$r$'s and $t$'s of each localized potential to satisfy to certain conditions,
which indeed are the ones in the Appendix \ref{app:flux}, Eq.
\eqref{eq:rt-relations} (note that point interactions constitute a particular
class of potentials of compact support \cite{IEOT.2013.75.341}).
Thus, based on \cite{JPA.1998.31.2975} we can calculate the Green's
function for general point interactions by using the corresponding
reflection and transmission coefficients, which are quantities with a
very clear physical interpretation and conceivably amenable to experimental
determination \cite{PRL.1990.64.2215,PRE.2001.64.065201}.

So, for these general array of potentials, according to Refs.
\cite{JPA.1998.31.2975,JPA.2001.34.5041,JPA.2003.36.227} the {\em
exact} (hence in contrast with usual semiclassical approximations,
see footnote 2) Green's function for a fixed energy $E$ (and end points 
$x_i$ and $x_f$) is given by
\begin{equation}
G(x_{f},x_{i}; E) = \frac{\mu}{i\hbar^2 k}
\sum_{\mbox{\scriptsize sp}} W_{\mbox{\scriptsize sp}}
  \exp{[\frac{i}{\hbar} S_{\mbox{\scriptsize sp}}(x_{f},x_{i};k)]}.
  \label{eq:gf}
\end{equation}
The above sum is performed over all scattering paths (sp) starting in
$x_i$ and ending in $x_f$.
A `scattering path' represents a trajectory in which the particle leaves
from $x_{i}$, suffers multiple scattering, and finally arrives at $x_{f}$.
For each sp, $S_{\mbox{\scriptsize sp}}$ is the classical-like action,
i.e., $S_{\mbox{\scriptsize sp}} = k \, L_{\mbox{\scriptsize sp}}$, with
$L_{\mbox{\scriptsize sp}}$ the trajectory length.
The term $W_{\mbox{\scriptsize sp}}$ is the sp quantum amplitude (or weight),
constructed as the following: each time the particle hits a localized
potential $V_n$, quantically it can be reflected or transmitted by the
potential.
In the first case, $W_{\mbox{\scriptsize sp}}$ gets a factor $r_n$ and in the
second, $W_{\mbox{\scriptsize sp}}$ gets a factor $t_n$.
The total $W_{\mbox{\scriptsize sp}}$ is then the product of all quantum
coefficients $r_n$'s and $t_n$'s acquired along the sp.

The direct extension of Eq. (\ref{eq:gf}) -- often called generalized
semiclassical Green's function formula because its functional form --
to quantum graphs is natural.
In fact, the two main ingredients necessary in the rigorous derivation
\cite{JPA.1998.31.2975,JPA.2001.34.5041} of Eq. \eqref{eq:gf}, namely,
unidimensionality and localized potentials, are by construction present
in quantum graphs.
First, since the quantum evolution takes place along the graph edges,
regardless the graph topology, the dynamics is essentially 1D.
Second, the potentials (scatters) are the vertices, which as we have
seen, can be treated as point interactions, so a particular class of
compact support potentials \cite{JPA.2006.39.2493,IEOT.2013.75.341}.

In the Appendix \ref{app:green} we outline the main steps necessary to
prove that the exact Green's function for arbitrary quantum graphs has the
very same form of Eq. \eqref{eq:gf}.
Moreover, as we are going to discuss in length in Sec. 4,
different techniques can be used to identify and sum up all the scattering
paths.
So, for general finite (i.e., $|V(\Gamma)|$ and $|E(\Gamma)|$ both finite)
connected undirected simple metric quantum graphs $\Gamma$, in principle
one always can obtain a closed analytical expression for $G$.
Therefore, given that any information about a quantum system can be extracted
directly from the corresponding Green's function, the results here
constitute a very powerful tool in the analysis of many distinct aspects
of quantum graphs.

As a final observation, we recall that for scaling quantum graphs
\cite{arXiv:quant-ph.0203126}, for each edge $e_s$ we have
$k_{e_s} = \gamma_{e_s} \, k_0$ (see footnote 3).
But this behavior for the wave number also would result from constant 
potentials $V_{e_s}$ along the distinct $e_s$'s.
Moreover, as discussed in \cite{PLA.2014.378.1461}, the correct $G$ for 
these kind of piecewise constant potential systems can too be cast as
above.
Therefore, the exact Green's function for scaling quantum graphs are
likewise given by Eq. (\ref{eq:gf}).

%%% Local Variables:
%%% mode: latex
%%% TeX-master: "green-qg-pr"
%%% ispell-local-dictionary: "american"
%%% End:

\section{Obtaining the Green's function for quantum graphs: general
procedures}
\label{sec:ogfqggp}

The formula in Eq. \eqref{eq:gf} gives the correct Green's function for
arbitrary connected undirected simple quantum graphs.
However, it has no universal practical utility unless we are able to
generally identify all the possible scattering paths and to sum up the 
resulting infinite series regardless the specific system.
So, here we shall describe different protocols to handle Eq. \eqref{eq:gf},
allowing to write the exact $G$ as a closed analytic expression.
To keep the discussion as accessible as possible, we start with few
straightforward illustrative examples.
In the sequence we extend the analysis to more complex situations.

We adopt the following notation:
\begin{itemize}
\item $r_{j}^{(s)}$ and $t_{j}^{(s,r)}$ are the reflection and transmission
amplitudes for the vertex $j$, as described in the end of Sec. 2.
\item $P_{l}$ represents the contribution from an entire infinite family
$l$ of sp to Eq. (\ref{eq:gf}), so that $G = \mu/(i \hbar^2 k) \sum_l P_l$.
\item $G_{s r}(x_f,x_i;k)$ is the Green's function for a particle with
energy  $E = {\hbar^2 k^2}/{2\mu}$, whose initial point $x_i$ lies in the
edge $e_r$ and the final point $x_f$ in the edge $e_s$.
\end{itemize}

Also, whenever there is no room for doubt, for simplicity we represent
edges by $s$ (instead of $e_s$) and vertices by capital letters, $A$, $B$,
etc.

\subsection{Constructing the Green's function: a simple example}
\label{subsec:cuafggq}

\begin{figure}
  \centering
  \includegraphics*[width=0.4\textwidth]{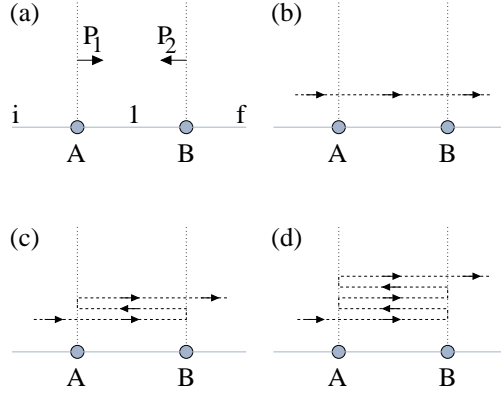}
  \caption{\label{fig:fig3} (Color online).
A simple graph with two vertices, $A$ and $B$, a finite edge labeled 1
(of length $\ell_1$), and left, $i$, and right, $f$, leads.
    (a) The starting positions of two families, $P_1$ and $P_2$, of sp.
    (b)-(d) Schematic examples of individual sp.}
\end{figure}

Consider the open graph shown in Fig. \ref{fig:fig3} (a).
It has two vertices, $A$ and $B$, one finite edge (of length $\ell_1$),
labeled 1, and two semi-infinite edges (leads), labeled $i$ and $f$.
By assuming $0 \leq x_i < +\infty$ ($x_i = 0$ at $A$) in $i$ and 
$0 \leq x_f < +\infty$ ($x_f = 0$ at $B$) in $f$,
the Green's function $G_{f i}(x_f, x_i; k)$ essentially describes the
transmission across the full graph structure, i.e., from the left to the
right leads.
To obtain $G$ we need to sum up all the possible sp for a quantum particle
starting at $x_{i}$, in $i$, going through multiple reflections between the
vertices $A$ and $B$, and finally ending up at $x_f$, in $f$.
As we shall demonstrate, Eq. \eqref{eq:gf} then yields a convergent geometric
series, which therefore can be calculated exactly
\cite{JPA.1992.25.L1043,PhysicaD.1994.72.244,PRA.1995.51.1811,
JPA.1998.31.2975,JPA.2001.34.5041,PRE.2001.64.026201,
PRA.2002.66.062712,JPA.2003.36.227,PRA.2004.69.052708}.

In Fig. \ref{fig:fig3} (b)--(d) it is depicted three examples of sp.
Consider the scattering path in Fig. \ref{fig:fig3} (b), representing
the `direct' propagation from $x_i$ to $x_f$.
The particle starts by leaving $x_i$ towards $A$.
From this first stretch of the trajectory, one gets a factor
$\exp[i k x_i]$ to $G$.
Upon hitting the vertex, the particle is then transmitted through $A$.
This process yields a factor $t_{A}^{(1,i)}$ to $G$.
Next, the particle goes to the vertex $B$ location, leading to a factor
$\exp[i k \ell_1]$.
Once in $B$, the particle is then transmitted through $B$, thus resulting
in $t_{B}^{(f,1)}$.
Finally, from the final trajectory stretch ($B$ to $x_f$), one gets
$\exp[i k x_f]$.
Putting all this together, the sp of Fig. \ref{fig:fig3} (b) contributes
to Eq. \eqref{eq:gf} with $W_{sp} =  t_{A}^{(1,i)} \, t_{B}^{(f,1)}$ and
$L_{sp} = (x_f + x_i) + \ell_1$ (hence the length of this sp).

Following the same type of analysis, for the other two examples in Fig.
\ref{fig:fig3} we have:
\begin{align}
  \mbox{(c) }
  &
    \exp[i k x_{i}] \, t_{A}^{(1,i)} \, \exp[i k \ell_{1}] \,
    r_{B}^{(1)} \, \exp[i k \ell_{1}] \, r_{A}^{(1)} \,
    \exp[i k \ell_{1}] \, t_{B}^{(f,1)} \, \exp[i k x_{f}]:
    \nonumber \\
  &
    W_{sp} =  r_{A}^{(1)} \, r_{B}^{(1)} \, t_{A}^{(1,i)} \, t_{B}^{(f,1)},
    \qquad L_{sp} = (x_f + x_i) + 3 \ell_1; \nonumber \\
  \mbox{(d) }
  &
  \exp[i k x_{i}] \, t_{A}^{(1,i)} \, \exp[i k \ell_{1}] \,
  r_{B}^{(1)} \, \exp[i k \ell_{1}] \, r_{A}^{(1)} \, \exp[i k \ell_{1}]
  \, r_{B}^{(1)} \, \exp[i k \ell_{1}]
  r_{A}^{(1)} \,  \exp[i k \ell_{1}] \, t_{B}^{(f,1)} \,
  \exp[ i k x_{f}]: \nonumber \\
  &
    W_{sp} =  (r_{A}^{(1)})^2 \, (r_{B}^{(1)})^2 \, t_{A}^{(1,i)} \, t_{B}^{(f,1)},
    \qquad L_{sp} = (x_f + x_i) + 5 \ell_1.
\nonumber
\end{align}

Thus, the full Green's function is written as a sum over all the existing
terms of the above form, or
\begin{equation}
  G_{f i}(x_f,x_i;k) =
    \frac{\mu}{i \hbar^2 k} \,
    \exp[i k x_i] \, t_A^{(1,i)} \Big(
    \sum_{n=0}^{\infty}
    [r_A^{(1)}]^{n} \, [r_B^{(1)}]^{n} \, \exp[i k (2n+1) \ell_1]
    \Big) \,  t_B^{(f,1)} \, \exp[i k x_f].
\label{example-1}
\end{equation}
Equation (\ref{example-1}) is in fact a geometric series and since for
the quantum amplitudes we have that $|r_{j}^{(s)}|^{2} \leq 1$ and
$|t_{j}^{(s,r)}|^{2} \leq 1$, the sum in Eq. (\ref{example-1}) always
converges.
So, the Green's function reads
\begin{equation}
G_{f i}(x_f,x_i;k)=\frac{\mu}{i \hbar^2 k} \, T_{f i} \,
\exp[i k (x_f+x_i + \ell_1)],
\label{eq:gexefinal}
\end{equation}
with
\begin{equation}
T_{f i}=\frac{t_A^{(1,i)} \;t_B^{(f,1)}}
{1-r_A^{(1)} r_B^{(1)}\exp{[2ik\ell_1]}}.
\label{transmission-example-1}
\end{equation}

Note that Eq. (\ref{transmission-example-1}) can be recognized as the
transmission amplitude for the whole system \cite{JPA.1998.31.2975}.
This illustrates the fact that by properly regrouping several vertices,
they can be treated as a `single' vertex, effectively contributing with
overall reflection and transmission amplitudes to $G$.
As we discuss in details in Sec. \ref{sec:simplification}, such an
approach strongly simplifies the calculation of the Green's function for
more complicated systems.

For the present example, to identify all the infinite possible sp is
relatively direct.
But when the number of vertices and edges increases, this can become a
very tedious and cumbersome enterprise.
Fortunately, the task can be accomplished by means of a simple diagrammatic
classification scheme, separating the sp into families.

To exemplify it, consider again $G_{f i}$ for the graph of Fig. \ref{fig:fig3}.
For any sp, necessarily at the beginning the particle leaves $x_{i}$, goes
to $A$, and then is transmitted through $A$.
Once tunneling to $x_1 = 0^+$ (always with positive velocity), there are
infinite possibilities to follow (some displayed in Fig. \ref{fig:fig3}
(b)--(d)).
So, schematically we represent all the trajectories headed to the right,
departing from $x_1 = 0^+$, as the family $P_1$, Fig. \ref{fig:fig3} (a).
Now, a sp in $P_1$ initiates traveling from $A$ to $B$.
Then, in $B$ it may either cross the vertex $B$, finally arriving at the
final point $x_f$, or be reflected from $B$, reversing its movement
direction (at $x_1 = \ell_1^{-}$).
For this latter situation, all the subsequent trajectories from
$x_1 = \ell_1^{-}$ can be represented as the family $P_2$, Fig. \ref{fig:fig3}
(a).
But exactly the same reasoning shows that for any sp in $P_2$, the particle
leaves $B$ towards $A$, it is reflected from
$A$\footnote{To be transmitted through $A$ would lead the particle to travel
towards $x_i \rightarrow + \infty$, with no returning (there are no vertices
for $x_i > 0$). So, obviously this sp cannot contribute to $G_{f i}$.},
and then becomes one of the paths in $P_1$.

Hence, the above prescription yields for the Green's function
\begin{equation}
G_{f i}(x_f,x_i;k) =
\frac{\mu}{ik\hbar^2} \, \exp[i k x_i] \, t_A^{(1,i)} \; P_1,
\label{eq:gp1}
\end{equation}
where
\begin{equation}
P_{1} = \exp[i k \ell_1]
 \left\{
    \begin{array}{l}
        r_B^{(1)} \, P_2 \\
        t_B^{(f,1)} \, \exp[i k x_f],
    \end{array}
 \right.
\label{eq:p1p2}
\end{equation}
and
\begin{equation}
P_{2} = \exp[i k \ell_1] \, r_A^{(1)} \, P_1.
\label{eq:p2exe}
\end{equation}
In Eq. \eqref{eq:p1p2}, `$\{$' represents the possible splitting for the
sp in the family $P_1$.
The algebraic equation equivalent to Eq. \eqref{eq:p1p2} is
\begin{equation}
P_{1}=\exp[i k \ell_1] \Big(
r_B^{(1)} \, P_2 + t_B^{(f,1)} \, \exp[i k x_f]\Big).
\label{eq:p1exe}
\end{equation}
Thus, solving Eqs. \eqref{eq:p2exe} and \eqref{eq:p1exe} for $P_1$,
one obtains
\begin{equation}
P_{1} = \frac{t_B^{(f,1)} \, \exp[i k \ell_1] \, \exp[i k x_f]}
{1 - r_A^{(1)} \, r_B^{(1)} \, \exp[2 i k \ell_1]},
\label{eq:p1final}
\end{equation}
which by direct substitution into Eq. \eqref{eq:gp1}, leads to the exact
$G$ in Eq. \eqref{eq:gexefinal}.

In this way, the identification and summation of an infinite number of sp is
reduced to the solution of a simple system of linear algebraic equations.
Such strong recursive nature of the scattering paths in quantum graphs
constitutes a key procedure to solve more complicated problems.

\subsection{Simplification procedures: further details}
\label{sec:simplification}

From the previous example, it is clear that two protocols which drastically
simplify the calculations for $G$ are:
(a) to regroup infinite many scattering paths into finite number of
families of trajectories;
and (b) to divide a large graph into smaller blocks, to solve the
individual blocks, and then to connect the pieces altogether.

Thus, given their importance, here we further elaborate on (a) and (b),
unveiling certain technical aspects which do not arise from a so simple
graph as that in Sec. \ref{subsec:cuafggq}.
Hence, we explicit address two different systems below: a cross shaped
structure, useful to illustrate details about (a), and a tree-like quantum
graph, a system whose solution is considerably facilitated by the block
separation technique (b).

\subsubsection{Regrouping the sp into families: a cross shaped graph
case study}
\label{sec:regroup}

\begin{figure}
  \centering
  \includegraphics*[width=0.5\textwidth]{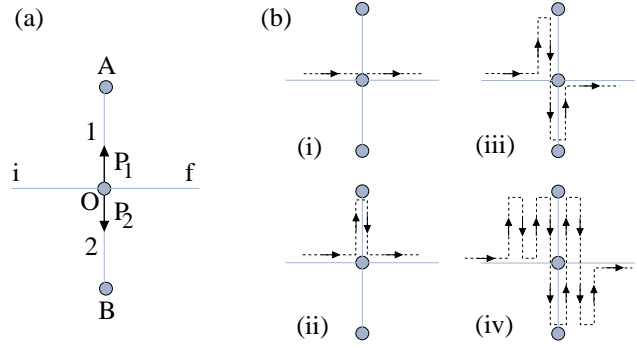}
  \caption{\label{fig:fig4} (Color online).
The cross shaped graph, with two leads, $i$ and $f$ (left and right),
two finite edges, 1 and 2 (up and down), and three vertices, $A$, $O$,
$B$.
(a) The $P_s$'s represent all the trajectories starting at vertex
$O$ along an edge $s$ and finally tunneling $O$, to get to the lead $f$.
(b) Four schematic examples of possible sp.}
\end{figure}

The cross-shaped graph is shown in Fig. \ref{fig:fig4}.
It is composed by three vertices, two edges and two leads.
Observe that the vertex $O$ is the origin (end) of the lead $f$ ($i$).
Let us first discuss the Green's function for the particle leaving $x_i$, 
along the lead $i$, and getting to $x_f$, along the lead $f$.
In the sum Eq. \eqref{eq:gf}, the sp are all the trajectories starting
from $i$, suffering multiple transmissions and reflections between the
edges $1$ and $2$ (of lengths $\ell_1$ and $\ell_2$), and arriving at $f$.
In Fig. \ref{fig:fig4} (b) we show schematic examples of possible
sp:
(i) direct transmission from $i$ to $f$ through the central vertex $O$,
so that $W_{sp} = t_{O}^{(f,i)}$ and $L_{sp} = x_f + x_i$;
(ii) transmission from $i$ to the edge $1$, a reflection at vertex $A$,
and a final transmission from the central vertex to the lead $f$, then
$W_{sp} = t_{O}^{(1,i)} \, r_A^{(1)} \, t_{O}^{(f,1)}$ and
$L_{sp} = x_f + x_i + 2 \ell_1$;
(iii) transmission to edge $1$, a reflection from $A$, then a
transmission to edge $2$, a new reflection, this time from vertex $B$,
and finally at $O$ a transmission to lead $f$, in this way
$W_{sp} = t_{O}^{(1,i)} \, r_A^{(1)} \, t_{O}^{(2,1)} \, r_B^{(2)} \, t_{O}^{(f,2)}$
and $L_{sp} = x_f + x_i + 2 (\ell_1 + \ell_2)$;
(iv) transmission to edge $1$, a double bouncing within edge $1$, then
transmission to edge $2$, a reflection from vertex $B$, a transmission
to edge $1$, a reflection from vertex $A$, another transmission to edge
$2$, a reflection from vertex $B$, and finally a transmission to lead
$f$ from edge $2$ (through vertex $O$), thus
$W_{sp} = t_{O}^{(1,i)} \, [r_A^{(1)}]^3 \, r_{O}^{(1)} \,
[t_{O}^{(2,1)}]^2 \, [r_B^{(2)}]^2 \, t_{O}^{(1,2)} \, t_{O}^{(f,2)}$ and
$L_{sp} = x_f + x_i + 6 \ell_1 + 4 \ell_2$.

Such infinite large proliferation of paths can be factorized in a simple
way.
Indeed, since for any sp we have initially a propagation from $x_i$ to
$O$ along $i$ and finally a propagation from $O$ to $x_f$ along $f$, we
can write
\begin{equation}
G_{f i}(x_f,x_i;k) = \frac{\mu}{i\hbar^2 k} \, T_{f i} \, \exp[i k (x_f + x_i)].
\end{equation}
Here $T_{f i}$ comprises all the contributions resulting from sp in the
region \mbox{$A$---$O$---$B$} of the graph, or
\begin{equation}
  T_{f i} =
  \left\{
    \begin{array}{l}
      t_{O}^{(f,i)}    \\
      t_{O}^{(1,i)} \, P_1 \\
      t_{O}^{(2,i)} \, P_2
    \end{array}
  \right. .
\end{equation}
As before, the symbol `$\{$' represents the trajectories splitting,
which reads
\begin{equation}
  \label{green-cross}
  T_{f i} = t_{O}^{(f,i)} \, + t_{O}^{(1,i)} \, P_1 + t_{O}^{(2,i)} \, P_2.
\end{equation}
The first term is just the amplitude for the direct path, i.e., a simple
tunneling from $i$ to $f$ through $O$.
The second (third) term represents the tunneling from lead $i$ to edge
1 (2) and all the subsequent possible trajectories that the particle can
follow until reaching lead $f$, represented by $P_1$ and $P_2$, Fig.
\ref{fig:fig4} (a).

The reasoning to obtain the two families of infinite trajectories, $P_1$
and $P_2$, is quite simple.
Take, for instance, $P_1$: all such paths start at $x_1 = 0^+$, travel
along edge 1 towards vertex $A$, suffer a reflection at $A$, and then
return to vertex $O$.
This part of the trajectories results in the term
$r_A^{(1)} \, \exp[2 i k \ell_1]$.
Once reaching back vertex $O$ they can either, be reflected from it,
then going into the set of paths $P_1$ again, or to tunnel to edge
$2$, so going into the family of paths $P_2$, or yet to tunnel to lead $f$,
thus terminating the \mbox{$A$---$O$---$B$} part of the sp.
The same type of analysis follows for $P_2$, so
\begin{equation}
  \left\{
    \begin{array}{l}
      P_1 = r_A^{(1)} \exp[2 i k \ell_1]
      \left\{
        \begin{array}{l}
          r_{O}^{(1)} P_1   \\
          t_{O}^{(2,1)} P_2 \\
          t_{O}^{(f,1)}
        \end{array}
      \right.
      \\
      P_2 = r_B^{(2)} \exp[2 i k \ell_2]
      \left\{
        \begin{array}{l}
          r_{O}^{(2)} P_2   \\
          t_{O}^{(1,2)} P_1 \\
          t_{O}^{(f,2)}
        \end{array}
      \right.
    \end{array}
  \right.,
  \label{eq:system-cross-1}
\end{equation}
leading to the algebraic equations
\begin{equation}
  \left\{
    \begin{array}{l}
      P_1 = r_A^{(1)} \exp[2 i k \ell_1] \,
      \Big(
        r_{O}^{(1)} P_1 + t_{O}^{(2,1)} P_2 + t_{O}^{(f,1)}
      \Big)
      \\
      P_2 = r_B^{(2)} \exp[2 i k \ell_2] \,
      \Big(
        r_{O}^{(2)} P_2 + t_{O}^{(1,2)} P_1 + t_{O}^{(f,2)}
      \Big),
    \end{array}
  \right.
  \label{eq:system-cross}
\end{equation}
whose solution reads
\begin{align}
  P_1 = {} & \frac{1}{g} \left\{
    r_A^{(1)} t_{O}^{(f,1)} \exp[2 i k \ell_1] +
    r_A^{(1)} r_B^{(2)} \Big(t_{O}^{(2,1)} t_{O}^{(f,2)}
      - r_{O}^{(2)}  t_{O}^{(f,1)}\Big)
    \exp[2 i k (\ell_1 + \ell_2)] \right\},
  \nonumber \\
  P_2 = & \frac{1}{g} \left\{
    r_B^{(2)} t_{O}^{(f,2)} \exp[2 i k \ell_2] +
    r_A^{(1)} r_B^{(2)} \Big(t_{O}^{(1,2)} t_{O}^{(f,1)}
      - r_{O}^{(1)}  t_{O}^{(f,2)}\Big)
    \exp[2 i k (\ell_1 + \ell_2)] \right\}, \nonumber \\
  \label{eq:ps-cross}
\end{align}
for
\begin{equation}
  g =
  \Big(1 - r_A^{(1)} r_{O}^{(1)} \exp[2 i k \ell_1] \Big)
  \Big(1 - r_B^{(2)} r_{O}^{(2)} \exp[2 i k \ell_2] \Big)
  -r_A^{(1)} r_B^{(2)} t_{O}^{(2,1)} t_{O}^{(1,2)}
  \exp[2 i k (\ell_1 + \ell_2)].
  \label{g-sec4}
\end{equation}

Similarly, we can consider both the initial and end points at the edge
$i$ ($0 \leq x_i, x_f < +\infty \in i$), for which $G_{i i}$ is given by
\begin{equation}
G_{i i}(x_f,x_i;k) =
\frac{\mu}{i\hbar^2k}
\Big\{\exp[i k |x_f-x_i|] + R_{i i} \, \exp[i k (x_f + x_i)]\Big\}.
\end{equation}
In this case, it is not difficult to see that
\begin{equation}
  R_{ii} = r_{O}^{(i)} + t_{O}^{(1,i)} P_1 + t_{O}^{(2,i)} P_2.
\end{equation}
The expressions leading to the correct $P$'s are those in \eqref{eq:ps-cross}
where, however, we must make the obvious substitution of $t_{O}^{(f,s)}$ by
$t_{O}^{(i,s)}$ ($s =1,2$).

Finally, we consider the end point $x_f$ in one of the edges, say edge $1$.
We assume that the origin of the this edge is at vertex $O$, so
$0 < x_f < \ell_1$.
Then, we have that
\begin{equation}
  G_{1 i}(x_f,x_i;k) = \frac{\mu}{i\hbar^2k} \, \exp[i k x_i] \,
  \Big(t_{O}^{(1,i)} P_1 + t_{O}^{(2,i)} P_2 \Big).
  \label{eq:gcross-bra1}
\end{equation}
Of course here we should not take into account any sp for which the
particle tunnels to the edge $f$ or comes back to the edge $i$ 
(for a reason similar to that explained in footnote 7).
Thus, we have for the $P$'s:
\begin{equation}
  \left\{
    \begin{array}{l}
      P_1 = \exp[i k x_f] + r_A^{(1)} \, \exp[2 i k \ell_1] \,
      \Big(
      \exp[-i k x_f]  + r_{O}^{(1)} \, P_1 + t_{O}^{(2,1)} \, P_2
      \Big)
      \\
      P_2 = r_B^{(2)} \, \exp[2 i k \ell_2] \,
      \Big(
      r_{O}^{(2)} \, P_2 + t_{O}^{(1,2)} \, P_1
      \Big).
    \end{array}
  \right.
\end{equation}
By solving the above system and substituting into the expression
\eqref{eq:gcross-bra1}, we get
\begin{align}
  G_{1 i}(x_f,x_i;k) = {}
  &
    \frac{\mu}{i \hbar^2 k} \, \frac{1}{g} \,
    \Big\{
    t_{O}^{(1,i)} + r_B^{(2)}
    \Big(t_{O}^{(2,i)} t_{O}^{(1,2)}
    - r_{O}^{(2)} t_{O}^{(1,i)}
    \Big) \exp[2 i k \ell_2]
    \Big\} \nonumber \\
  &  \times
    \Big\{ \exp[i k (x_f + x_i)]
    + r_A^{(1)} \exp[i k (2 \ell_1 - x_f + x_i)] \Big\},
\end{align}
with $g$ given by Eq. (\ref{g-sec4}).

\subsubsection{Treating a graph in terms of blocks:
a tree-like case study}
\begin{figure}
  \centering
  \includegraphics*[width=0.5\textwidth]{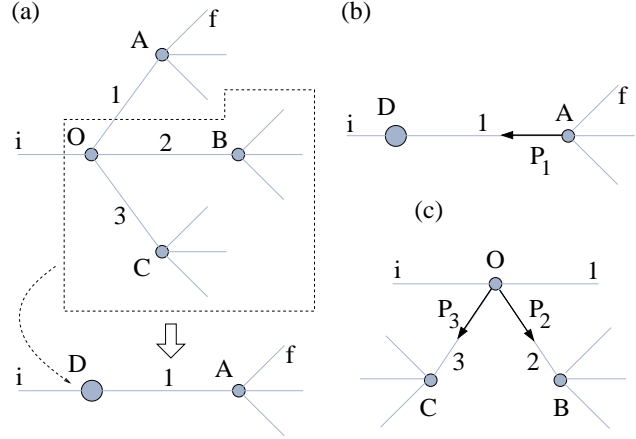}
  \caption{ \label{fig:fig5} (Color online).
A tree-like quantum graph.
(a) By regarding the whole region $C$---$O$---$B$ (including the
leads) as an `unique' effective vertex $D$, the original graph is reduced
as illustrated.
(b) In the reduced graph, $P_1$ represents the family of trajectories which
suffer multiple reflections between $D$ and $A$, and finally tunnel
the vertex $A$ to the lead $f$.
(c) The auxiliary graph (and the corresponding sp families) necessary to
calculate $r_D^{(1)}$ and $t_D^{(i,1)}$.}
\end{figure}

Next we discuss how to shorten the calculations for a large quantum graph by
decomposing it in blocks.
For so, we consider the example shown in Fig. \ref{fig:fig5} (a), a
relatively simple tree-like graph: a lead $i$ is attached to a vertex $O$,
from which emerges three edges 1, 2 and 3, ending, respectively, at vertices
$A$, $B$, and $C$.
Each of these vertices, by their turn, are connected to three leads.

Here we just analyze the Green's function for the initial position
$x_i$ in lead $i$ and the end position $x_f$ in lead
$f$ (this latter lead, $f$, connected to vertex $A$, 
see Fig. \ref{fig:fig5} (a)).
Observe that in this particular situation we do not need to consider any
sp that goes into another lead besides $f$ (because then, it would be 
impossible for the particle to come back to $f$).

The first step to simplify the problem is to treat the whole block
indicated in Fig. \ref{fig:fig5} (a) as a single vertex $D$.
Any information about the inner structure of such region will be contained
in the vertex quantum amplitudes $t_D^{(1,i)}$ and $r_D^{(1)}$.
Thus, we reduce the original graph to the simpler one depicted in Fig.
\ref{fig:fig5} (b).
From Fig. \ref{fig:fig5} (b), we have that the Green's function
can be written as
$G_{f i}(x_f,x_i;k) = \mu/(i \hbar^2 k) \, T_{f i} \, \exp[i k (x_f + x_i)]$,
with $T_{f i} =
t_D^{(1,i)} \exp[i k \ell_1] \Big(r_A^{(1)} P_1 + t_A^{(f,1)}\Big)$.
Then, based on our previous discussions, one quickly realizes that the
infinite family of trajectories $P_1$ is given by
$P_1 = r_D^{(1)} \exp[2 i k \ell_1] \Big(r_A^{(1)} P_1 + t_A^{(f,1)}\Big)$, or
\begin{equation}
  P_1 = \frac{r_D^{(1)} t_A^{(f,1)} \exp[2 i k \ell_1]}
  {1 - r_D^{(1)} r_A^{(1)} \, \exp[2 i k \ell_1]}.
\end{equation}

It remains to determine the coefficients $t_D^{(1,i)}$ and $r_{D}^{(1)}$.
We can do so with the help of the auxiliary quantum graph of Fig.
\ref{fig:fig5} (c).
We first recall that $t_D^{(1,i)}$ ($r_D^{(1)}$)
represents the sp contribution for the particle to go from lead $i$ (edge
$1$) to edge $1$ through the region $B$---$O$---$C$.
Inspecting Fig. \ref{fig:fig5} (c), we see that
$t_D^{(1,i)} = t_{O}^{(1,i)} + t_{O}^{(3,i)} P_3 + t_{O}^{(2,i)} P_2$
and
$r_D^{(1)} = r_{O}^{(1)} + t_{O}^{(3,1)} P_3 + t_{O}^{(2,1)} P_2$,
where for the $P$'s
\begin{equation}
  \left\{
    \begin{array}{l}
      P_3 = r_C^{(3)} \exp[2 i k \ell_3]
      \Big(
        r_{O}^{(3)} P_3 + t_{O}^{(2,3)} P_2 + t_{O}^{(1,3)}
      \Big)
      \\
      P_2 = r_B^{(2)} \exp[2 i k \ell_2]
      \Big(
        r_{O}^{(2)}  P_2 + t_{O}^{(3,2)} P_3 + t_{O}^{(1,2)}
      \Big).
      \label{eq:ps-tree}
    \end{array}
  \right.
\end{equation}
The solution of Eq. \eqref{eq:ps-tree} is given by Eq. \eqref{eq:ps-cross}
with the appropriate labels substitutions in \eqref{eq:ps-cross}:
$A \rightarrow C$, $1 \rightarrow 3$ and $f \rightarrow 1$.

\subsection{The Green's function solutions by eliminating, redefining
or regrouping scattering amplitudes}
\label{sec:cases}

\begin{figure}
  \centering
  \includegraphics*[width=0.5\textwidth]{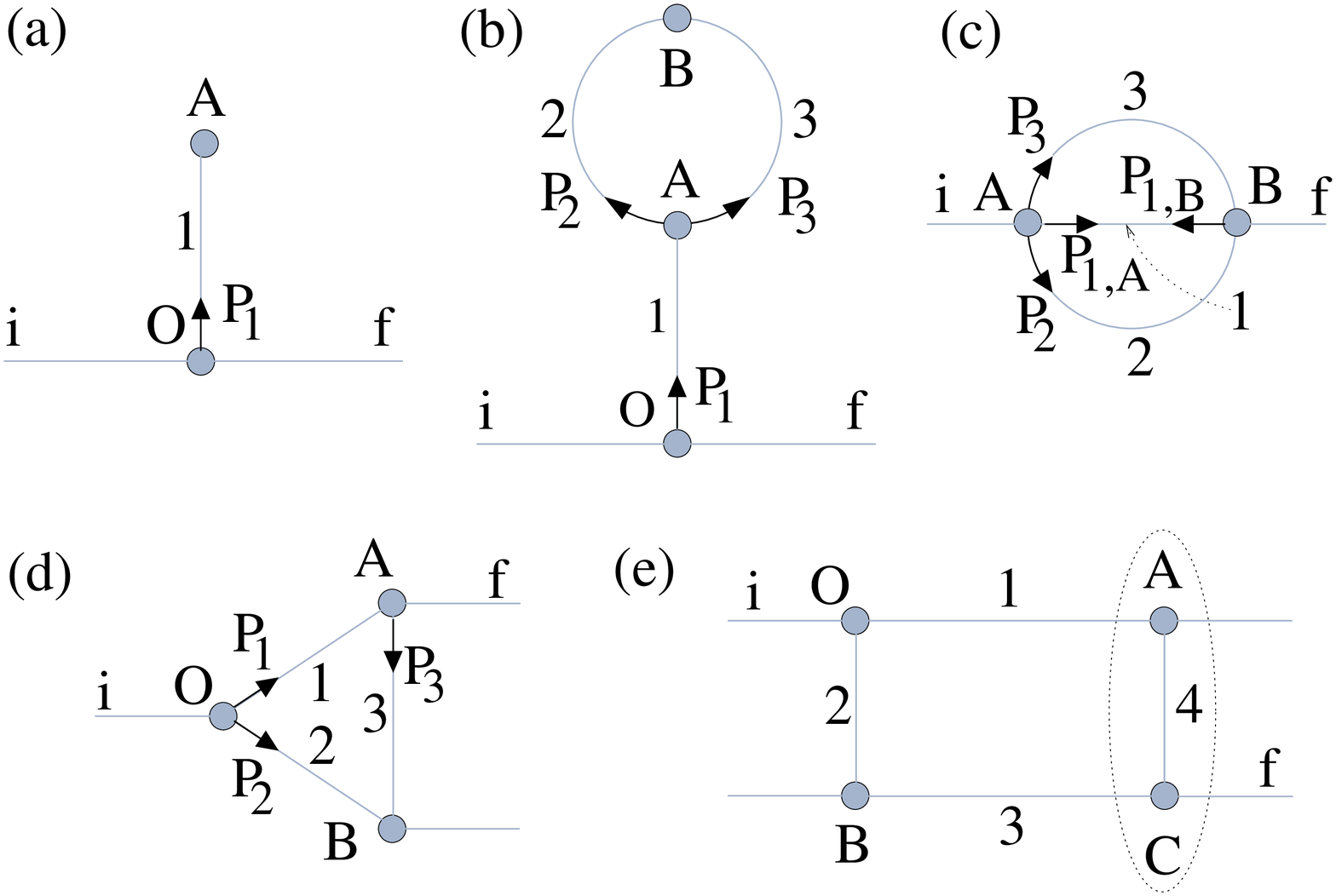}
  \caption{\label{fig:fig6} (Color online).
Several graphs whose $G$'s can be obtained from the solutions of other
topologies by eliminating, redefining or regrouping the vertices reflections
and transmissions quantum amplitudes.
(a) The cross shaped graph, Fig. 4, but with both the bottom edge and vertex
removed.
(b) The same as in (a), but with the simple vertex $A$ substituted by a
circle-like structure.
(c) A circle-like graph attached to two leads.
(d) Triangle (e) and rectangle graphs attached to semi-infinite leads.}
\end{figure}

A great advantage in writing the Green's function in terms of the general
scattering amplitudes of each vertex is that by setting appropriate values
for or regrouping these quantities, we can obtain $G$ for some graphs
based on the solutions for other topologies.

Indeed, for a vertex $j$ attached to two edges ($e_{j_1}$ and $e_{j_2}$),
to set $r_j^{(s)} = 0$ and $t_j^{(s,r)} = 1$ ($s,r = 1, 2$) is equivalent to
remove the vertex $j$ from the graph.
On the other hand, if for all $e_{j_r}$ we set $t_j^{(s,r)} = 0$ for the two
(one) vertices $j$ attached to the finite (semi-infinite) edge $e_{j_s}$, then
we eliminate $e_{j_s}$ from the structure.
For instance, consider the graph in Fig. \ref{fig:fig6} (a).
We obtain its exact $G_{f i}$, $G_{i i}$ and $G_{1 i}$ just by assuming
$t_{O}^{(2,i)} = t_{O}^{(2,1)} = 0$ for the solutions of the cross shaped
graph of Fig. \ref{fig:fig4}.

As for regrouping, the $G$'s for the graph in Fig. \ref{fig:fig6} (b) --
if $x_i$ and $x_f$ are not in the edges $2$ and $3$ -- follow from the
exact Green's functions for the graph of Fig. \ref{fig:fig6} (a) by just
supposing the whole region $A$---$B$---$A$ as a single vertex, say $C$,
and making the substitution $r_A^{(1)} \rightarrow r_C^{(1)}$.
From the Fig. \ref{fig:fig6} (b) we see that $r_C^{(1)}$ is given by
$r_C^{(1)} =r_A^{(1)} + t_A^{(2,1)} P_2 + t_A^{(3,1)} P_3$, with the $P$'s
obtained from
\begin{equation}
  \left\{
    \begin{array}{l}
      P_2 = r_B^{(2)} \exp[2 i k \ell_2] \Big(
        r_A^{(2)}  P_2 + t_A^{(3,2)}  P_3 + t_A^{(1,2)} \Big)
      \\ \ \ \ \ \ \
      + t_B^{(3,2)} \exp[i k (\ell_2 + \ell_3)] \Big(
        r_A^{(3)}  P_3 + t_A^{(2,3)}  P_2 + t_A^{(1,3)} \Big)
      \\
      P_3 = r_B^{(3)} \exp[2 i k \ell_3] \Big(
        r_A^{(3)}  P_3 + t_A^{(2,3)}  P_2 + t_A^{(1,3)} \Big)
      \\ \ \ \ \ \ \
      + t_B^{(2,3)} \exp[i k (\ell_2 + \ell_3)] \Big(
        r_A^{(2)}  P_2 + t_A^{(3,2)}  P_3 + t_A^{(1,2)}  \Big).
    \end{array}
  \right.
\end{equation}

Consider now the more involving example in Fig. \ref{fig:fig6} (c) and
$G_{1 1}$ for which both end points are in edge $1$, i.e.,
$0 < x_i, x_f < \ell_1$.
We define $r_C^{(1)}$ ($t_C^{(1,1)}$) as the resulting quantum amplitude for
the particle to hit the vertex $A$ from edge 1, to suffer all the multiple
scattering in edges $2$ and $3$ and finally to come back to edge 1 from
the vertex $A$ ($B$).
We likewise define $r_D^{(1)}$ and $t_D^{(1,1)}$ for the particle initially
hitting the vertex $B$.
So, we have that (dropping the superscripts $(1)$ and $(1,1)$ for
simplicity)
\begin{equation}
  G_{11}(x_f,x_i;k) = \frac{\mu}{i \hbar^2 k} \Big\{
  \exp[i k |x_f - x_i|] + \exp[i k (\ell_1 - x_i)]
  \Big( r_D P_{1,B} + t_D P_{1,A} \Big)
  + \exp[i k x_i] \Big( r_C P_{1,A} + t_C P_{1,B} \Big ) \Big\},
  \label{eq:gtadpole}
\end{equation}
where
\begin{equation}
  \left\{
    \begin{array}{l}
      P_{1,A} = \exp[i k x_f] + \exp[i k \ell_1] \Big(
        r_D  P_{1,B} + t_D  P_{1,A} \Big)
      \\
      P_{1,B} = \exp[i k (\ell_1 - x_f)] + \exp[i k \ell_1] \Big(
        r_C  P_{1,A} + t_C  P_{1,B} \Big).
    \end{array}
  \right.
\end{equation}
Solving the above system, the Green's function \eqref{eq:gtadpole} reads
\begin{align}
  G_{11}(x_f,x_i;k) = {}
  &
    \frac{\mu}{i \hbar^2 k} \, \frac{1}{g} \,
    \Big\{ g\exp[ik|x_f-x_i|] +
    r_C \exp[i k (x_f + x_i)]
    +r_D \exp[ i k (2\ell_1 - x_f + x_i)]\nonumber \\
  &
    + r_C \, r_D \exp[i k (2 \ell_1 + x_f - x_i)]
  + r_C \, r_D \exp[i k (2 \ell_1 - x_f + x_i)] \nonumber \\
  &
    + \Big(1 - t_C \exp[i k \ell_1] \Big) \exp[ i k (\ell_1 + x_f - x_i)]
  \nonumber \\
  & + \Big(1 - t_D \exp[i k \ell_1] \Big)
      \exp[-i k (\ell_1 - x_f + x_i)]
      \Big\},
\end{align}
with $g =
\Big(1-t_C \exp[ik\ell_1]\Big)
\Big(1-t_D\exp[ik\ell_1]\Big)-r_C \, r_D \exp[2ik\ell_1]$.

Above, the coefficient $r_C$ (see Fig. \ref{fig:fig6} (c)) is given by
$r_C = r_A^{(1)} + t_A^{(2,1)} P_2 + t_A^{(3,1)} P_3$, with
$P_2$ and $P_3$ obeying to
\begin{equation}
  \left\{
    \begin{array}{l}
      P_2 = r_B^{(2)} \exp[2 i k \ell_2] \left(
        r_A^{(2)} \, P_2 + t_A^{(3,2)} \, P_3 + t_A^{(1,2)} \right)
      \\ \ \ \ \ \ \ \,
      + t_B^{(3,2)} \exp[i k (\ell_2 + \ell_3)] \left(
        r_A^{(3)} \, P_3 + t_A^{(2,3)} \, P_2 + t_A^{(1,3)} \right)
      \\
      P_3 = r_B^{(3)} \exp[2 i k \ell_3] \left(
        r_A^{(3)} \, P_3 + t_A^{(2,3)} \, P_2 + t_A^{(1,3)} \right)
      \\ \ \ \ \ \ \ \,
      + t_B^{(2,3)} \exp[i k (\ell_2 + \ell_3)] \left(
        r_A^{(2)} \, P_2 + t_A^{(3,2)} \, P_3 +  t_A^{(1,2)} \right)
    \end{array}
  \right.
.
\label{eq:40}
\end{equation}
By its turn $t_C = t_A^{(2,1)} P_2 + t_A^{(3,1)} P_3$, where instead of Eq. 
(\ref{eq:40}) this time $P_2$ and $P_3$ satisfy to
\begin{equation}
  \left\{
    \begin{array}{l}
      P_2 = r_B^{(2)} \exp[2 i k \ell_2] \left(
        r_A^{(2)} \, P_2 + t_A^{(3,2)} \, P_3 \right)
      \\ \ \ \ \ \ \ \,
      + t_B^{(3,2)} \exp[i k (\ell_2 + \ell_3)] \left(
        r_A^{(3)} \, P_3 + t_A^{(2,3)} \, P_2 \right) +
      \exp[i k \ell_2] t_B^{(1,2)}
      \\
      P_3 = r_B^{(3)} \exp[2 i k \ell_3] \left(
        r_A^{(3)} \, P_3 + t_A^{(2,3)} \, P_2\right)
      \\ \ \ \ \ \ \ \,
      + t_B^{(2,3)} \exp[i k (\ell_2 + \ell_3)] \left(
        r_A^{(2)} \, P_2 + t_A^{(3,2)} \, P_3 \right) +
      \exp[i k \ell_3] t_B^{(1,3)}
    \end{array}
  \right.
.
\end{equation}
The amplitudes $r_D$ and $t_D$ are obtained from the expression for $r_C$ and
$t_C$ by just exchanging the indices $A \leftrightarrow B$.

Finally, if for both graphs of Fig. \ref{fig:fig6} (d) and (e), the $G$
initial and final points are, respectively, in the edges $i$ and $f$, 
the Green's function is simply
\begin{equation}
G_{fi}(x_f,x_i;k) = \frac{\mu}{i\hbar^2 k} T_{f i}
\exp[i k (x_f + x_i)].
\end{equation}

For the case of Fig. \ref{fig:fig6} (d), 
$T_{f i} = t_{O}^{(1,i)} \, P_1 + t_{O}^{(2,i)} \,P_2$, with $P_1$ and $P_2$
obtained from the following
\begin{equation}
  \left\{
    \begin{array}{l}
      P_1 = r_A^{(1)} \exp[2 i k \ell_1] \Big(
        r_{O}^{(1)}  P_1 + t_{O}^{(2,1)}  P_2 \Big)
      + \exp[i k \ell_1] \Big(
        t_A^{(3,1)}  P_3 + t_A^{(f,1)} \Big)
      \\
      P_2 = r_B^{(2)} \exp[2 i k \ell_2] \Big(
        r_{O}^{(2)}  P_2 + t_{O}^{(1,2)}  P_1 \Big)
      \\ \ \ \ \ \ \
      + t_B^{(3,2)} \exp[i k (\ell_2 + \ell_3)] \Big(
        r_A^{(3)}  P_3 + t_A^{(f,3)} \Big)
      \\ \ \ \ \ \ \
      + t_B^{(3,2)} t_A^{(1,3)} \exp[i k (\ell_1 + \ell_2 + \ell_3)]
      \Big( r_{O}^{(1)}  P_1 + t_{O}^{(2,1)}  P_2 \Big)
      \\
      P_3 = r_B^{(3)} \exp[2 i k \ell_3] \Big(
        r_A^{(3)}  P_3 + t_A^{(f,3)} \Big)
      \\ \ \ \ \ \ \
      + t_B^{(2,3)} \exp[i k (\ell_2 + \ell_3)] \Big(
        r_{O}^{(2)}  P_2 + t_{O}^{(1,2)}  P_1 \Big)
      \\ \ \ \ \ \ \
      + r_B^{(3)} t_A^{(1,3)} \exp[i k (\ell_1 + 2 \ell_3)]
      \Big( r_{O}^{(1)}  P_1 + t_{O}^{(2,1)}  P_2 \Big),
    \end{array}
  \right.
  \label{system}
\end{equation}
with $P_3$ an auxiliary family of infinite trajectories, introduced just
to help in the recursive definitions of $P_1$ and $P_2$
(see Fig. \ref{fig:fig6} (d)).
The solution of the above system put into the expression for $T_{f i}$ yields
the final exact Green's function.

For $G_{f i}$ for the graph of Fig. \ref{fig:fig6} (e) we can use the above
same set of equations if we treat the region comprising vertices $A$ and $C$
of Fig. \ref{fig:fig6} (e) as a single effective vertex, corresponding
to $A$ in Fig. \ref{fig:fig6} (d).
Thus, by using the previous analysis, we find that we need only
to make the following substitutions in the Green's function expression
for the graph of Fig. \ref{fig:fig6} (d) so to get that for Fig. 
\ref{fig:fig6} (e):
\begin{align}
  r_A^{(1)}  \rightarrow {}  & \, r_A^{(1)} +
  t_A^{(4,1)} \, r_{C}^{(4)} \, t_A^{(1,4)} \exp[2 i k \ell_4]/g,
  \nonumber \\
  t_A^{(f,1)} \rightarrow {} & t_A^{(4,1)} \, t_C^{(f,4)}
  \exp[i k \ell_4]/g, \nonumber \\
  t_A^{(3,1)} \rightarrow {} & t_A^{(4,1)} \, t_C^{(3,4)}
  \exp[i k \ell_4]/g, \nonumber \\
  r_A^{(3)} \rightarrow {} & r_C^{(3)} +
  t_C^{(4,3)} \, r_{A}^{(4)} \, t_C^{(3,4)} \exp[2 i k \ell_4]/g,
  \nonumber \\
  t_A^{(f,3)}  \rightarrow {} & t_C^{(f,3)} +
  t_C^{(4,3)} \, r_A^{(4)} t_C^{(f,4)}
  \exp[2 i k \ell_4]/g, \nonumber \\
  t_A^{(1,3)} \rightarrow {} &
  t_C^{(4,3)} \, t_A^{(1,4)}
  \exp[i k \ell_4]/g, \nonumber
\end{align}
where $g = 1 - r_A^{(4)} \, r_C^{(4)} \exp[2 i k \ell_4]$.

%%% Local Variables:
%%% mode: latex
%%% TeX-master: "green-qg-pr"
%%% ispell-local-dictionary: "american"
%%% End:
\section{Eigenstates and scattering states in quantum graphs}
\label{sec:cesog}

From the previous Sec. we have seen that different techniques enable one
to obtain $G$ in a relatively straightforward way.
Moreover, we also have mentioned that the calculation of the wave function 
in certain contexts might be lengthy.
Therefore, a natural question is how easily one can extract from $G$ the system 
eigenvalues, eigenstates and scattering states, thus allowing to bypass 
the more traditional approach of directly solving the Schr\"odinger equation.
Next we give some examples along this line.
For definiteness, we concentrate on the graph of Fig. \ref{fig:fig6} (a).

\subsection{Eigenstates}

The explicit expression for the Green's function with $x_i$
in lead $i$ and $x_f$ in lead $f$ is
(Fig. \ref{fig:fig6} (a))
\begin{align}
  G_{f i}(x_f,x_i;k) = {}
  &
    \frac{\mu}{i \hbar^2 k} \, T_{f i} \, \exp[ik(x_f+x_i)],
\nonumber \\
  T_{f i} = {}
  &
    t_{O}^{(f,i)} + \frac{t_O^{(1,i)} \, r_A^{(1)} \, t_{O}^{(f,1)} \exp[2 i k \ell_1]}
    {1 - r_{O}^{(1)} \, r_{A}^{(1)} \exp[2 i k\ell_1]}.
    \label{eq:gtbif}
\end{align}
For both $x_i$ and $x_f$ ($0 < x_i, x_f < \ell_1$, $x_f > x_i$) in the
edge $1$, we get
\begin{align}
  G_{1 1}(x_f,x_i;k) = {}
  & \frac{\mu}{i \hbar^2 k}
    \frac{1}{\Big(1 - r_{O}^{(1)} \, r_{A}^{(1)} \exp[2 i k\ell_1] \Big)}
    \nonumber \\
  &  \times \Big(
\exp[-i k x_i] + r_{O}^{(1)} \exp[i k x_i] \Big)
%\nonumber \\
%  &  \times
\, \Big(
\exp[i k x_f] + r_{A}^{(1)} \exp[2 i k \ell_1]\exp[- i k x_f]
\Big).
  \label{eq:gtbl}
\end{align}

For open graphs, like that in Fig. \ref{fig:fig6} (a), depending on the
characteristics of the vertices, the system may support bound
states\footnote{A trivial textbook example is the usual $\delta$-function
potential in the line.
If its strength $\gamma$ is negative, it has exactly one bound state.}.
In these cases, the eigenstates are calculated from the residues of
$G(x_f,x_i;k)$ at the poles $k = k_n$ \cite{Book.2006.Economou}, which
give the problem eigenenergies through $E_n = \hbar^2k_n^2/(2 \mu)$.

By inspecting the above Green's functions, we see that they can diverge
(consequently presenting poles \cite{PRL.1995.74.3503}) only if
$g(k = k_n) = 0$, with
\begin{equation}
g(k) = 1 - r_{O}^{(1)}(k) \, r_{A}^{(1)}(k) \exp[2 i k \ell_1].
\label{eq:ev-green}
\end{equation}
As a concrete example, consider the vertex $O$ being a generalized $\delta$
interaction (here attached to $N = 3$ edges, Fig. \ref{fig:fig6} (a))
of strength $\gamma$ \cite{PRL.1995.74.3503}.
Then, for simplicity setting $\hbar = \mu = 1$, the reflection coefficients 
for the vertex $O$ are given by (see Appendix \ref{app:boundary-conditions})
\begin{equation}
r_{O}^{(1)}(k) = r_{O}^{(i)}(k) = r_{O}^{(f)}(k) = r_{O}(k) =
\frac{2 \gamma - (N-2) i k}{Nik-2\gamma} =
\frac{2 \gamma - i k}{3 i k - 2 \gamma},
\label{eq:r-delta-gen}
\end{equation}
and the transmission coefficients by
\begin{equation}
t_{O}^{(1,i)}(k) = t_{O}^{(f,1)}(k) = t_{O}^{(f,i)}(k) =
t_{O}(k) = \frac{2 i k}{N i k - 2 \gamma} =
\frac{2 i k}{3 i k - 2 \gamma}.
\label{eq:t-delta-gen}
\end{equation}
For the vertex $A$, as discussed in the Appendix 
\ref{app:boundary-conditions}, we take the boundary condition 
$- \psi'(A) = \lambda \psi(A)$, which is equivalent to the following 
reflection coefficient
\begin{equation}
r_{A}(k) = \frac{i k + \lambda}{i k - \lambda}.
\label{eq:rA}
\end{equation}

It is a well-known fact that any pole of the scattering amplitudes in the
upper half of complex $k$-plane along the imaginary axis represents a
bound energy \cite{Book.1989.Chadan}.
For example, for the usual (1D) Dirac $\delta$-function with intensity
$\gamma < 0$ (attractive $\delta$), the transmission coefficient is
$t_{\delta} = i k /(i k - \gamma)$.
In this case, the unique negative energy of the system reads
$E_1 = k_{1}^2/2 = - \gamma^2/2$, where $k_1 = i |\gamma|$ is the only pole
of $t_{\delta}(k)$ \cite{JPA.1997.30.3937,PRA.1988.37.973}.

So, for our graph the eigenvalues are obtained from the following 
transcendental equation (with Re$[k_n] = 0$ and Im$[k_n] > 0$)
\begin{equation}
g(k_n) =  1 - \left(\frac{2 \gamma - i k_n}{3 i k_n - 2 \gamma}\right)
  \left(\frac{i k_n + \lambda}{i k_n - \lambda}\right)
  \exp[i 2 k_n \ell_1] = 0.
%\label{eq:ev-green}
\end{equation}
Further, using the formula
($g'(k_n) \equiv \left. dg(k)/dk \right|_{k=k_n}$)
\begin{equation}
  \lim_{E \rightarrow E_{n}} \frac{(E-E_n)}{g(k)} =
  \frac{1}{2} \lim_{k \rightarrow k_{n}} \frac{(k^2 - k_n^2)}{g(k)} =
  \frac{k_n}{g'(k_n)},
\end{equation}
the residues of Eq. \eqref{eq:gtbif} are obtained from
\begin{align}
\psi_n^{(f)}(x_f) \, {\psi_{n}^{(i)}}^{*}(x_i) = {}
  &
    \frac{1}{2} \lim_{k \rightarrow k_{n}}(k^2-k_n^2) \, G_{f i}(x_f,x_i;k)
    \nonumber \\
  = {}
  &
    \Big\{{\mathcal N}_G(k_n) \, t_O(k_n) \, \exp[i k_n x_f] \Big\} \,
    \Big\{
    {\mathcal N}_G(k_n) \, t_O(k_n) \, \exp[i k_n x_i] \Big\},
\label{eq:wave-if}
\end{align}
and of Eq. \eqref{eq:gtbl} from
\begin{align}
\psi_n^{(1)}(x_f) \, {\psi_{n}^{(1)}}^{*}(x_i) = {}
  &
    \frac{1}{2} \lim_{k \rightarrow k_{n}} (k^2-k_n^2) \, G_{1 1}(x_f,x_i;k)
    \nonumber \\
  = {}
  & \Big\{{\mathcal N}_G(k_n) \,
    \Big(\exp[- i k_n x_f] + r_O^{(1)}(k_n) \exp[i k_n x_f] \Big)\Big\}
    \nonumber \\
  &
    \times
    \Big\{{\mathcal N}_G(k_n) \,
    \Big(\exp[- i k_n x_i] + r_O^{(1)}(k_n) \exp[i k_n x_i]\Big)\Big\}.
\label{eq:wave-11}
\end{align}
Observe that in the above Eqs., because after the substitution 
$k_n = i \kappa_n$ all the terms become real-valued functions, the complex 
conjugation, in this particular case, makes no practical difference.
Finally
\begin{equation}
  {\mathcal N}_G(k_n) = \frac{1}{\sqrt{i g'(k_n) \, r_O^{(1)}(k_n)}}.
\label{eq:norm-green}
\end{equation}

Note that for the poles $k_n = i \kappa_n$, with $\kappa_n > 0$, the wave 
functions in both leads have the general form $\psi_n(x) = {\mathcal N} 
\exp[- \kappa_n x]$ (recall that $x \geq 0$).
Hence, they decay away from the origin (vertex $O$) exponentially,
as it should be.
The ${\mathcal N}$'s also lead to the correct normalization for the
eigenstates.
Important to mention that the same results follow from the direct solution 
of the Schr\"odinger equation with the appropriate boundary conditions 
(which is done in the Appendix \ref{app:boundary-conditions}).

\begin{figure}[t!]
  \centering
  \includegraphics*[width=0.45\textwidth]{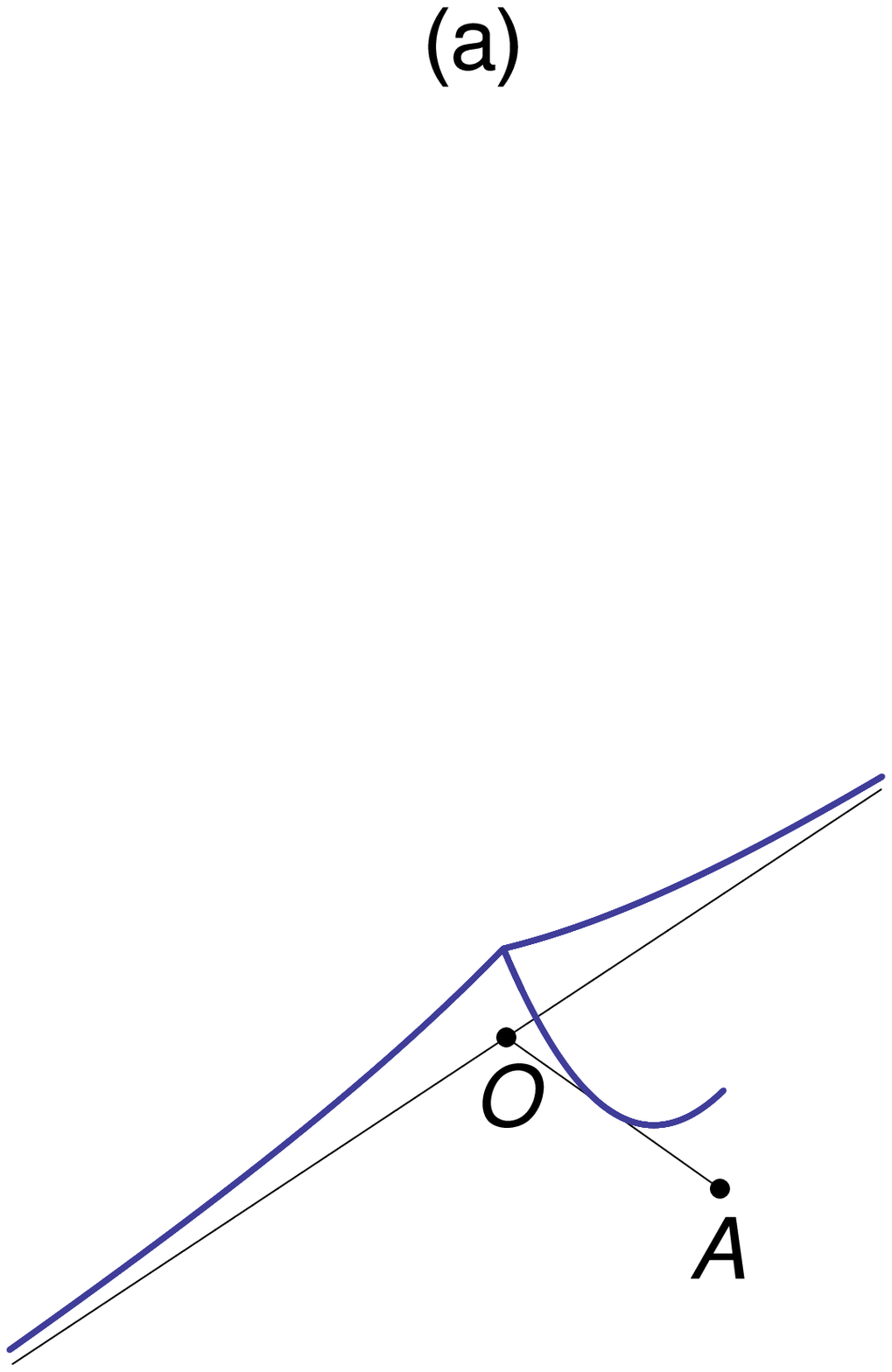}
  \includegraphics*[width=0.45\textwidth]{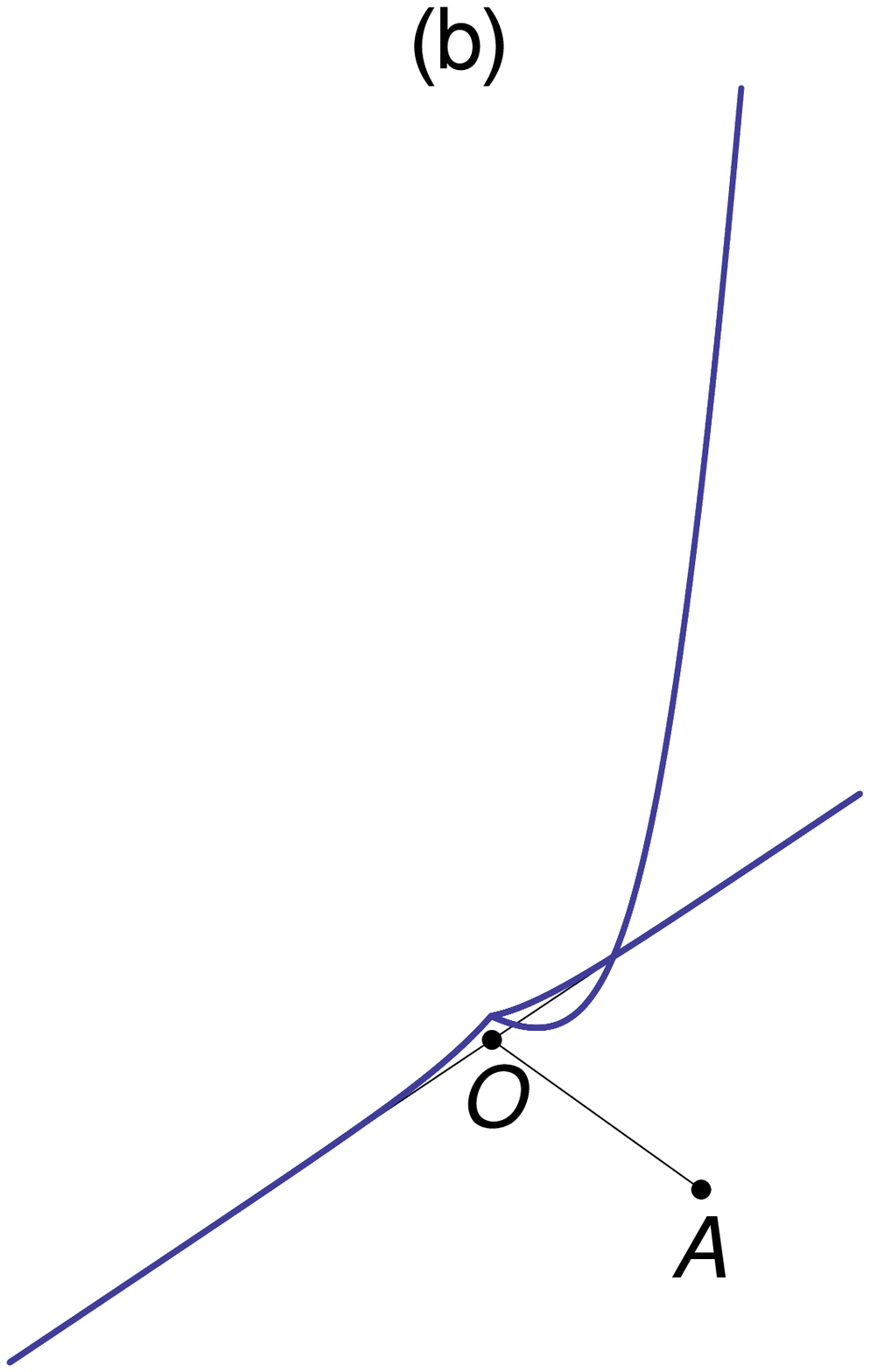}
  \caption{\label{fig:fig7} (Color online).
The bound eigenstates probability distribution along the quantum graph of
Fig. \ref{fig:fig6} (a), here with $\ell_1 = 1$.
The vertex $O$ is a $\delta$ interaction of strength $\gamma=-3/2$.
The boundary condition at the vertex $A$ is given by
$-\psi'(A) = \lambda \psi(A)$, with $\lambda=-2$.
(a) $|\psi_1(x)|^2$ for which $\kappa_1=0.463618$ and
(b) $|\psi_2(x)|^2$ for which $\kappa_2=2.022448$.
}
\end{figure}

As a numerical example, consider $\gamma = -3/2$, $\lambda=-2$,
and $\ell_1 = 1$.
Then, the system has two bound eigenstates, $n = 1, 2$.
In Fig. \ref{fig:fig7} we show the corresponding $|\psi_n(x)|^2$.
The first (second) eigenstate, with $\kappa_1 = 0.463618$ 
($\kappa_2 =2.022448$), is mainly due to the attractive $\delta$ potential
(to the boundary condition at the vertex $A$\footnote{Positive values for 
$\lambda$ cannot give rise to eigenstates ``associated'' to the vertex $A$.}).
This can verified in Fig. \ref{fig:fig7}:
$|\psi_1|^2$ ($|\psi_2|^2$) is much more concentrated around the vertex
$O$ ($A$).

\subsection{Scattering}
\label{sec:scattering}

Consider again the Green function $G_{f i}$, Eq. \eqref{eq:gtbif},
for the open graph of Fig. \ref{fig:fig6} (a).
As already discussed, the quantity $|T_{f i}|^2$ (in $G_{f i}$) can be 
interpreted as the total probability for a particle of wave number $k$ 
incident from the lead $i$ to be transmitted to the lead $f$.
Similarly, supposing $x_i$ and $x_f$ in lead $i$, we have
\begin{align}
G_{i i}(x_f,x_i;k) = {}
  & \frac{\mu}{i \hbar^2 k}
    \Big\{ \exp[i k |x_f - x_i|] + R_{i} \exp[i k (x_f + x_i)] \Big\},
    \nonumber \\
  R_{i} = {}
  &
    r_{O}^{(i)} + \frac{t_O^{(1,i)} \, r_A^{(1)} \, t_{O}^{(i,1)} \exp[2 i k \ell_1]}
    {1 - r_{O}^{(1)} \, r_{A}^{(1)} \exp[2 i k\ell_1]}.
\end{align}
Then,  $|R_{i}|^2$ represents the total probability for a particle of
wave number $k$ incident from the lead $i$ to be reflected to the lead $i$.
By choosing different quantum amplitudes for the vertices,
we naturally get different scattering patterns from $R_{i}$ and $T_{f i}$.

\begin{figure}[t!]
  \centering
  \includegraphics*[width=0.5\textwidth]{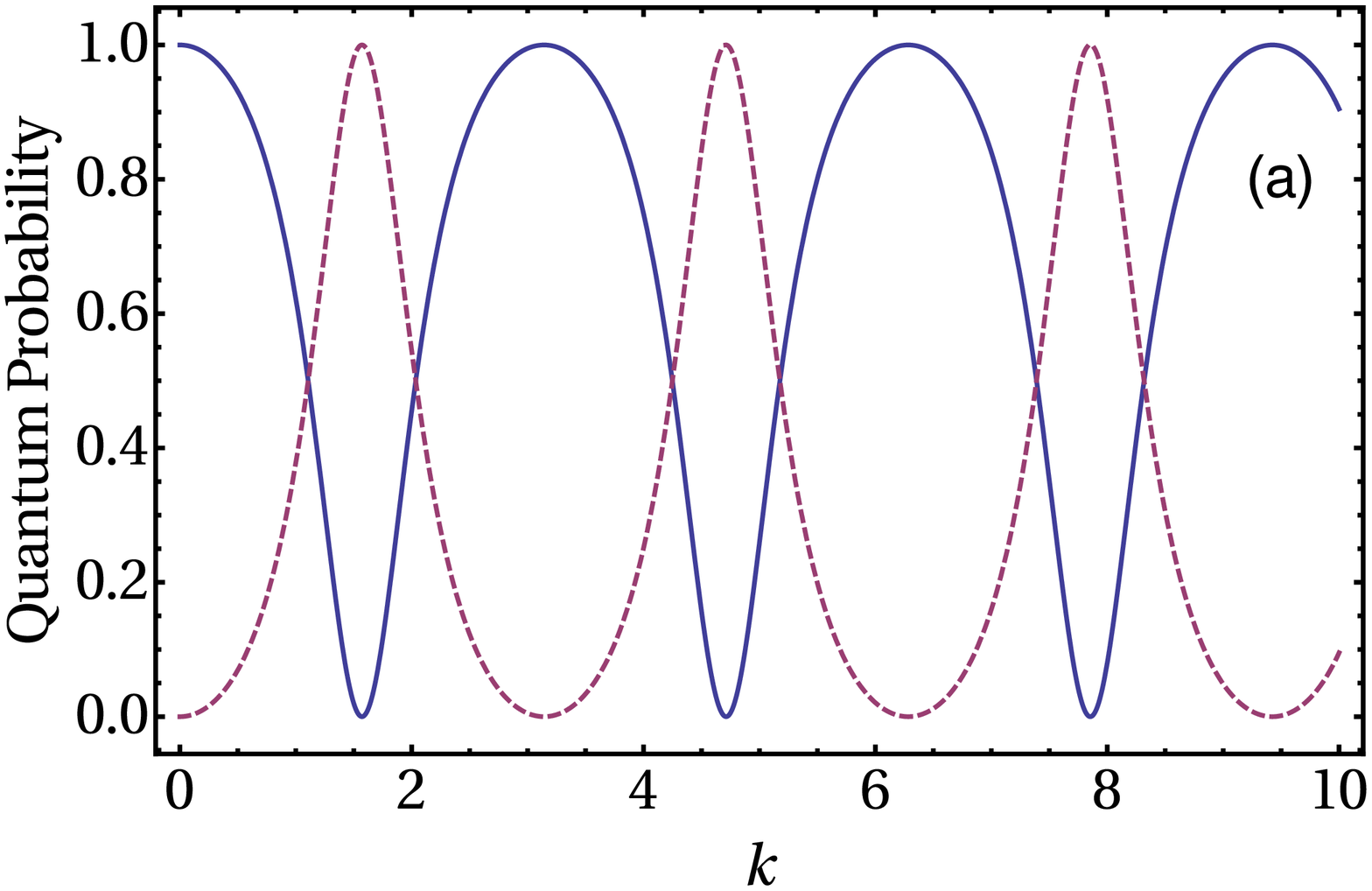}%
  \includegraphics*[width=0.5\textwidth]{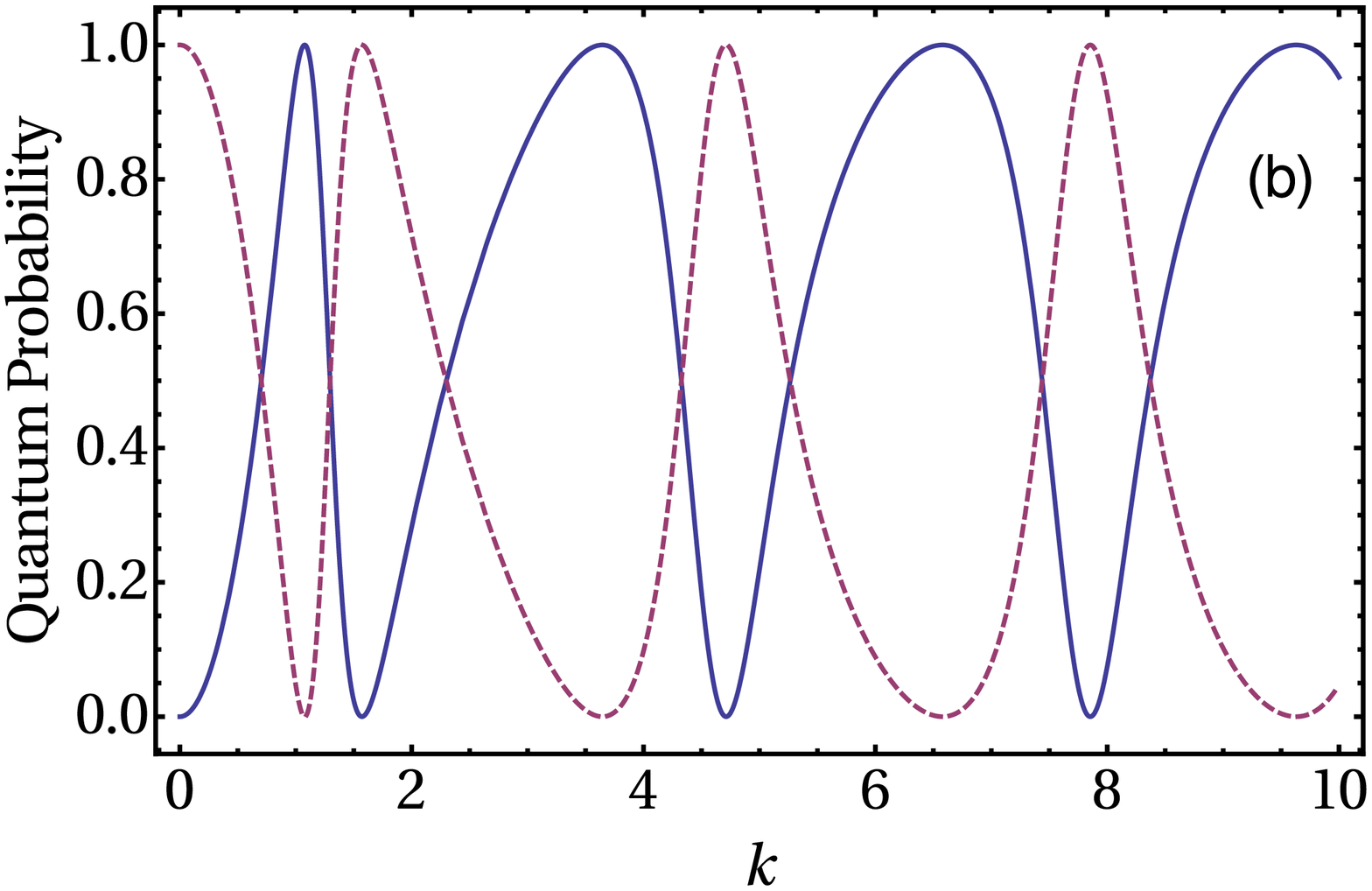}\\
\includegraphics*[width=0.5\textwidth]{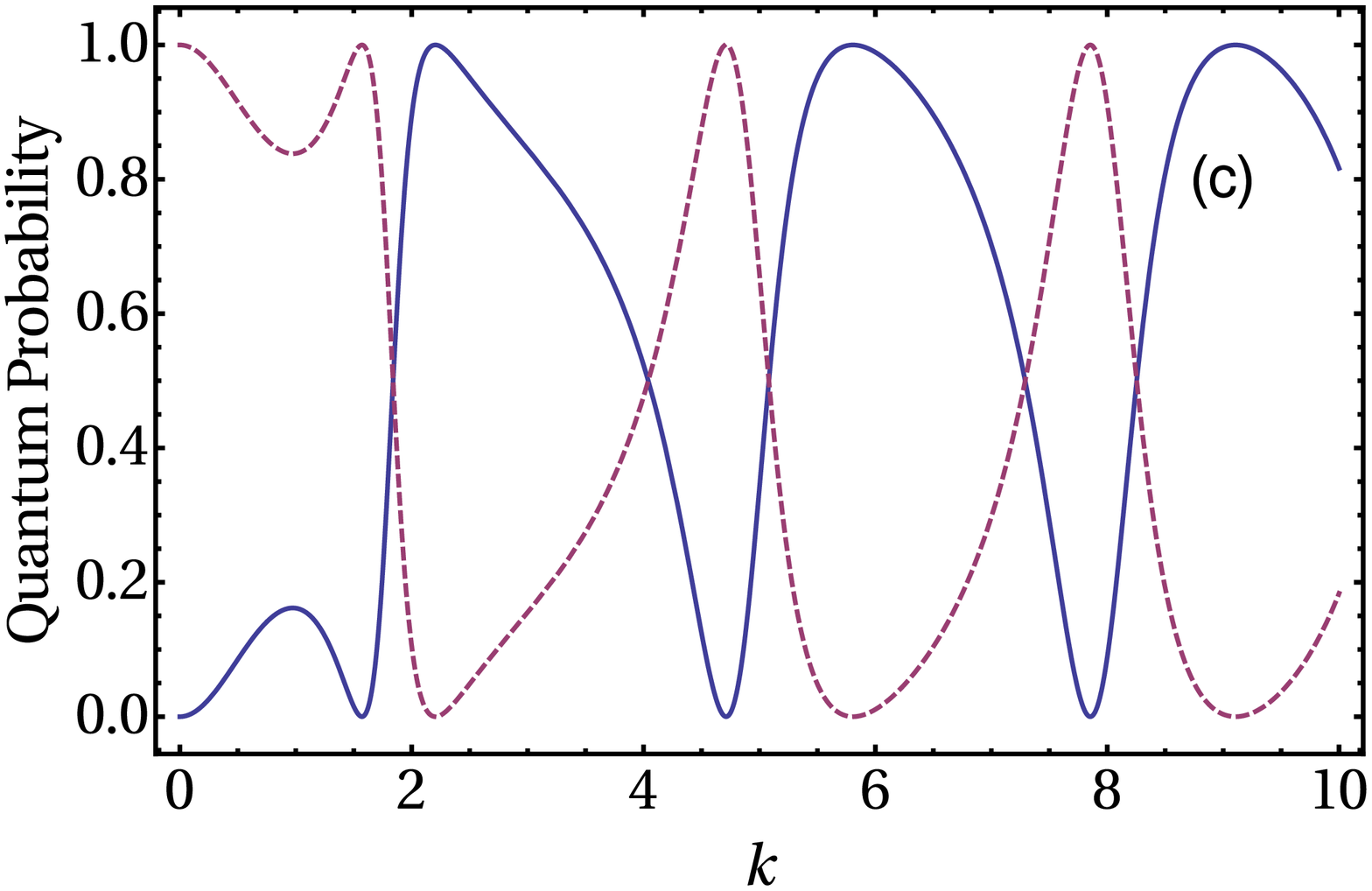}\\
  \caption{\label{fig:fig8} (Color online).
    The transmission $|T_{f i}|^2$ (solid line) and reflection
    $|R_{i}|^2$ (dashed) probabilities as function of $k$ for the
    quantum graph of Fig. \ref{fig:fig6} (a).
    In all cases $\ell_1 = 1$ and $\lambda = 0$ (Neumann-Kirchhoff
    boundary conditions at $A$).
    The values of $\gamma$ at $O$ are:
    (a) 0, (b) 1, and (c) -3/2.
      }
\end{figure}

To illustrate possible different scattering behavior for this graph, we 
assume the Neumann-Kirchhoff boundary conditions (Appendix 
\ref{app:boundary-conditions}) at the vertex $A$, so we set $\lambda = 0$
in Eq. (\ref{eq:rA}).
For $O$, we consider three values for the parameter $\gamma$: 
(a) $\gamma=0$ (so, also Neumann-Kirchhoff); and the 
generalized $\delta$ of strengths (b) $\gamma = 1$ and (c) 
$\gamma = -3/2$.
The resulting $|R_{i}|^2$ and $|T_{f i}|^2$ as function of $k$ are
shown in Fig. \ref{fig:fig8}, where distinctions in the scattering
probabilities are clearly observed.
In all cases $\ell_1 = 1$.

%%% Local Variables:
%%% mode: latex
%%% TeX-master: "green-qg-pr"
%%% ispell-local-dictionary: "american"
%%% End:
\section{Representative quantum graphs}
\label{sec:rqg}

So far we have discussed the general ideas of how to use the energy domain 
Green's function method to study quantum graphs through the explicit 
calculation of arbitrary cases.
But in the literature one can find certain topologies which are particularly 
convenient and flexible to model many distinct quantum phenomena.
For instance, the examples already addressed in Sec. \ref{sec:ogfqggp},
Fig. \ref{fig:fig6}, are indeed proper structures to construct logic
gates for quantum information processing
\cite{PRL.2009.102.180501,Science.2013.339.791}.
In special, the graph in Fig. \ref{fig:fig6} (b) can act as a phase 
shifter, whereas that in Fig. \ref{fig:fig6} (e) could functioning as a 
basis-changing gate.

Other very important examples include:
\begin{itemize}
\item The widely analyzed (with the most distinct purposes
\cite{Book.2013.Portugal,PRA.2009.79.012325,SLoQC.2008.1.1,
PRA.2006.73.032341,Inproceedings.2002.Moore},
like to investigate scattering features of 3D graphs
\cite{PRA.2005.71.012306}) hypercube;
\item The binary tree \cite{ToC.2008.4.169,ToC.2009.5.119,PRA.1998.58.915},
e.g., useful to highlight differences between classical and quantum walks
\cite{QIP.2002.1.35} as well as to test
the speed up gain -- which is actually exponential -- in searching
algorithms based on quantum dynamics \cite{Inproceedings.2003.Childs}.
We should observe that the graph of Fig. \ref{fig:fig5} (a) is in
fact an extension of a binary tree, being a fragment of a large-scale
ternary tree network \cite{RPP.2013.76.096001};
\item Triangular Sierpi\'{n}ski-like structures \cite{NPB.2010.828.515},
a nice illustration of graphs which in the limit of infinite vertices
would be fractal.
It has been considered in connection with molecular assembling
\cite{NatureChem.2015.7.389} and with the mathematics of logical games
like the Hanoi tower \cite{TMI.1995.17.52,IJMMNO.3.251.2012}.
\end{itemize}

Given the relevance of the above mentioned three graph systems, in the
present section we show in details how to calculate the exact Green's
function for each one of these problems.

\subsection{Cube}
\label{sec:closedgraphs}

\begin{figure}
  \centering
  \includegraphics*[width=0.50\textwidth]{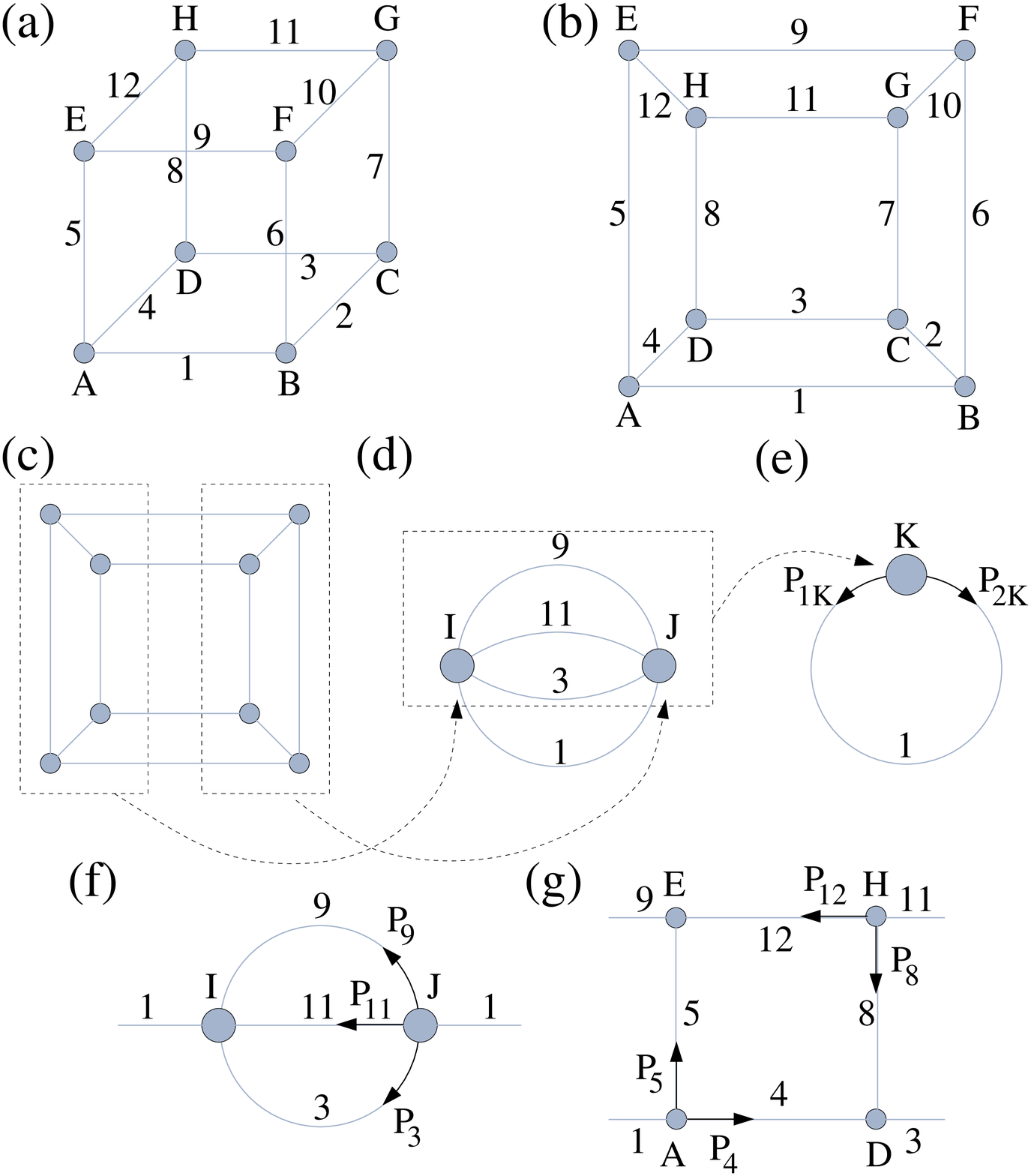}
  \caption{\label{fig:fig9}  (Color online).
    A cube quantum graph.
    (a) The letters represent the vertices indices and the integers
    the edges indices.
    (b) A cube graph planar representation.
    (c)-(e) Regrouping procedures (see the main text).
    (f) Auxiliary graph to determine the total $R$'s and $T$'s.
    (g) The inner structure of vertex $I$.
    The $P_l$'s indicate the sp families.}
\end{figure}

The Green's function for closed quantum graphs can be obtained by the
regrouping technique discussed in the previous sections.
Thus, we will use this procedure to get the Green's function for the cube 
quantum graph of Fig. \ref{fig:fig9} (a) (where all edges have length $\ell$).
In Fig. \ref{fig:fig9} (b) we show a planar representation of the cube 
graph.
For concreteness, let us suppose both the initial and final positions in 
the edge $1$ (see Fig. \ref{fig:fig9} (a)).
The first step to simplify the calculations is to view the two regions 
marked by dashed lines in Fig. \ref{fig:fig9} (c) as two vertices 
$I$ and $J$, Fig. \ref{fig:fig9} (d).
The second is a further regrouping, in which we represent $I$ and $J$ 
as a single vertex $K$, Fig. \ref{fig:fig9} (e).
Therefore, we end up reducing the original cube to a simple circular graph.

Now, consider Fig. \ref{fig:fig9} (e), with $x_f > x_i$ ($x \in (0, \ell)$
increases anti-clockwise from vertex $K$).
We then define for $K$ the total reflection and transmission amplitudes
$R^{(\pm)}$ and $T^{(\pm)}$ (where the superscript $+$ ($-$) indicates
that the scattering process takes place at $\ell$ ($0$)).
In this way, all the information about the internal structure of the cube 
graph are contained in these $K$ vertex coefficients.
Thus, for the circular graph of Fig. \ref{fig:fig9} (e), the Green's function 
can be written as
\begin{align}
  G_{11}(x_f,x_i;k)= {}
  &
    \frac{\mu}{i \hbar^2 k}
    \left\{ 
    \exp[i k (x_f-x_i)]
    + \exp[i k x_i] \, 
    \left(R^{(-)} \, P_{1K} + T^{(-)} \, P_{2K} \right) \right. 
\nonumber \\
  &
    + \left. 
\exp[ik (\ell - x_i)] \, \left(R^{(+)} \, P_{2K} + T^{(+)} \, 
P_{1K}\right) \right\},
  \label{eq:gcqg-cube}
\end{align}
with $P_{1K}$ and $P_{2K}$ given by
\begin{equation}
  \left\{
    \begin{array}{l}
      P_{1K} = \exp[i k x_f] + \exp[ i k \ell]
\left(R^{(+)} \, P_{2K} + T^{(+)} \, P_{1K} \right), \\
      P_{2K}=\exp[i k (\ell-x_f)] + \exp[ i k \ell]
\left(R^{(-)} \, P_{1K} + T^{(-)} \, P_{2K} \right).
    \end{array} 
  \right.
\end{equation}
Solving the above system, the Green's function \eqref{eq:gcqg-cube} reads
\begin{align}
  G_{11}(x_f,x_i;k)= {}
  &\frac{\mu}{i \hbar^2 k} \frac{1}{g}
    \Big\{ \Big(1 - T^{(-)} \exp[ i k \ell] \Big) \exp[ik(x_f-x_i)]
    \nonumber\\
  &
    + R^{(-)} \exp[ik(x_f+x_i)] + R^{(+)} \exp[ik(2\ell-x_f-x_i)]
\nonumber\\
  &+
\Big(T^{(-)} + \Big(R^{(+)} \, R^{(-)} - T^{(+)} \, T^{(-)}\Big)
\exp[ i k \ell]\Big) \exp[ik(\ell-x_f+x_i)]\Big\},
  \label{gfcqg}
\end{align}
with
\begin{equation}
  g = (1-T^{(+)} \exp[ i k \ell]) \, (1-T^{(-)} \exp[ i k \ell])
 - R^{(+)} \, R^{(-)} \exp[2 i k \ell].
\label{eq:roots-cube}
\end{equation}

Next, we must determine the coefficients $R$'s and $T$'s.
We do so with help of the auxiliary quantum graph in Fig. \ref{fig:fig9}
(f).
We recall that $T^{(\pm)}$ ($R^{(\pm)}$) represents the paths contribution
for the particle going from edge 1 to edge 1 by means of a transmission through
(reflection from) the vertex $K$.
Inspecting Fig. \ref{fig:fig9} (e) and (f), we see that the transmission
from $I$ ($J$) to $J$ ($I$) yields $T^{(-)}$ ($T^{(+)}$).
Similarly, the reflection from $I$ ($J$) leads to $R^{(-)}$ ($R^{(+)}$).
We start with $T^{(-)}$, then
\begin{align}
  T^{(-)} = {} &
  t_{I}^{(3,1)} \, \exp[ i k \ell] \,
  \Big(
  r_{J}^{(3)} \, P_3 + t_{J}^{(9,3)} \, P_9 + t_{J}^{(11,3)} \, P_{11} + t_{J}^{(1,3)}
  \Big)
  \nonumber \\
  & + t_{I}^{(9,1)} \, \exp[ i k \ell] \,
  \Big(
  r_{J}^{(9)} \, P_9 + t_{J}^{(3,9)} \, P_3 + t_{J}^{(11,9)} \, P_{11} + t_{J}^{(1,9)}
  \Big)
  \nonumber \\
  & + t_{I}^{(11,1)} \, \exp[ i k \ell] \,
  \Big(
  r_{J}^{(11)} \, P_{11} + t_{J}^{(3,11)} \, P_3 + t_{J}^{(9,11)} \, P_{9} +
  t_{J}^{(1,11)}
  \Big),
    \label{tad1}
\end{align}
where the $P$'s are
\begin{equation}
  \left\{
    \begin{array}{l}
      P_3 = r_{I}^{(3)} \, \exp[2 i k \ell] \, \left(
            r_{J}^{(3)} \, P_3 + t_{J}^{(9,3)} \, P_9
            + t_{J}^{(11,3)} \, P_{11} + t_{J}^{(1,3)}
            \right)
      \\ \ \ \ \ \,\,
      + \,\, t_{I}^{(9,3)} \, \exp[2 i k \ell] \, \left(
             r_{J}^{(9)} \, P_9 + t_{J}^{(3,9)} \, P_3
        + \,\, t_{J}^{(11,9)} \, P_{11} + t_{J}^{(1,9)} \right)
      \\ \ \ \ \ \,\,
      +\,\, t_{I}^{(11,3)} \, \exp[2 i k \ell] \, \left(
            r_{J}^{(11)} \, P_{11} + t_{J}^{(3,11)} \, P_3
        +\,\, t_{J}^{(9,11)} \, P_{9} + t_{J}^{(1,11)}\right)
      \\
      P_9 = r_{I}^{(9)} \, \exp[2 i k \ell] \, \left(
            r_{J}^{(9)} \, P_9 + t_{J}^{(3,9)} \, P_3 +
            t_{J}^{(11,9)} \, P_{11} + t_{J}^{(1,9)} \right)
      \\ \ \ \ \ \,\,
      +\,\, t_{I}^{(3,9)} \, \exp[2 i k \ell] \, \left(
            r_{J}^{(3)} \, P_3 + t_{J}^{(9,3)} \, P_9 +
        t_{J}^{(11,3)} \, P_{11} + t_{J}^{(1,3)}\right)
      \\ \ \ \ \ \,\,
      +\,\, t_{I}^{(11,9)} \, \exp[2 i k \ell] \, \left(
            r_{J}^{(11)} \, P_{11} + t_{J}^{(3,11)} \, P_3 +
        t_{J}^{(9,11)} \, P_{9} + t_{J}^{(1,11)} \right)
      \\
      P_{11} = r_{I}^{(11)} \, \exp[2 i k \ell] \, \left(
              r_{J}^{(11)} \, P_{11} + t_{J}^{(3,11)} \, P_3+
              t_{J}^{(9,11)} \, P_{9} + t_{J}^{(1,11)}\right)
      \\ \ \ \ \ \,\,
      +\,\, t_{I}^{(3,11)} \, \exp[2 i k \ell] \, \left(
            r_{J}^{(3)} \, P_3 + t_{J}^{(9,3)} \, P_9 +
            t_{J}^{(11,3)} \, P_{11} + t_{J}^{(1,3)}\right)
      \\ \ \ \ \ \,\,
      +\,\, t_{I}^{(9,11)} \, \exp[2 i k \ell] \, \left(
            r_{J}^{(9)} \, P_{9} + t_{J}^{(3,9)} \, P_3 +
            t_{J}^{(11,9)} \, P_{11} + t_{J}^{(1,9)}\right).
    \end{array}
  \right.
  \label{tad}
\end{equation}

For $R^{(+)}$ we have
\begin{equation}
  R^{(+)} =
  r_{J}^{(1)} + t_{J}^{(3,1)} \, P_3 + t_{J}^{(9,1)} \, P_9 + t_{J}^{(11,1)} \, P_{11},
\end{equation}
where the $P$'s are those in Eq. (\ref{tad}).
We obtain $T^{(+)}$ and $R^{(-)}$ from $T^{(-)}$ and $R^{(+)}$ by the simple
substitution $I \leftrightarrow J$.

Finally, we shall obtain $r_{I \, (J)}$ and $t_{I \, (J)}$ in terms of the 
original vertices coefficients.
As one might expect, because the cube symmetry the quantum amplitudes for $I$
and $J$ can be derived from each other by a direct indices relabeling
\footnote{We obtain the coefficients for $J$ by considering the corresponding
  formulas for $I$ and performing the indices changes:
  $A \rightarrow B$,  $D \rightarrow C$,  $E \rightarrow F$,  $H \rightarrow G$,
  $4 \rightarrow 2$,  $5 \rightarrow 6$,  $8 \rightarrow 7$,
  $12 \rightarrow 10$.}.
So, we just discuss in details the vertex $I$.
Moreover, such type of procedure is  also possible for the distinct
$r_I^{(s)}$'s and $t_I^{(s,r)}$'s in Eq. \eqref{tad}:
we can calculate, say,  $r_{I}^{(1)}$, $t_{I}^{(1,3)}$, $t_{I}^{(1,11)}$, and
then to infer the expressions for the others $r_I$'s and $t_I$'s by proper
exchanges of vertices and edges labels.

From Fig. \ref{fig:fig9} (g), depicting the inner structure of $I$,
we can write
\begin{align}
  r_{I}^{(1)} = {} & r_A^{(1)}+t_A^{(4,1)} P_4 + t_A^{(5,1)} P_{5},
  \nonumber \\
  t_{I}^{(1,11)}  = {} & t_H^{(8,11)} P_8 + t_H^{(12,11)} P_{12},
  \nonumber \\
  t_{I}^{(1,3)}  = {} & t_{D}^{(4,3)} \, \exp[ i k \ell] \,
\left(r_A^{(4)} \, P_4 + t_A^{(5,4)} \, P_5 + t_A^{(1,4)} \right) + 
t_D^{(8,3)} \, \exp[i k \ell] \,
\left(r_H^{(8)} \, P_8 + t_H^{(12,8)} \, P_{12} \right),
\end{align}
where
\begin{equation}
  \left\{
    \begin{array}{l}
      P_4 = r_D^{(4)} \, \exp[2 i k \ell] \, \left(r_A^{(4)} \, P_4
        +t_A^{(5,4)} \, P_5+t_A^{(1,4)}\right)
      \\ \ \ \ \ \,\,
      +t_D^{(8,4)}\, \exp[2 i k \ell] \, \left(r_H^{(8)} \, P_{8}+
        t_H^{(12,8)} \, P_{12}\right)\\
      P_5 = r_E^{(5)} \, \exp[2 i k \ell] \, \left(r_A^{(5)} \, P_5 +
        t_A^{(4,5)} \, P_4 + t_A^{(1,5)}\right)
      \\ \ \ \ \ \,\,
      +t_E^{(12,5)} \, \exp[2 i k \ell] \, \left(r_H^{(12)} \, P_{12}+
        t_H^{(8,12)} \, P_{8}\right)\\
      P_8=t_D^{(4,8)} \, \exp[2 i k \ell] \, \left(r_A^{(4)} \, P_4 +
        t_A^{(5,4)} \, P_5 + t_A^{(1,4)}\right)
      \\ \ \ \ \ \,\,
      +r_D^{(8)} \, \exp[2 i k \ell] \, \left(r_H^{(8)} \, P_{8}+
        t_H^{(12,8)} \, P_{12}\right)\\
      P_{12}=t_E^{(5,12)} \, \exp[2 i k \ell] \, \left(r_A^{(5)} \, P_5 +
        t_A^{(4,5)} \, P_4 + t_A^{(1,5)}\right)
      \\ \ \ \ \ \,\,
      +r_E^{(12)} \, \exp[2 i k \ell] \, \left(r_H^{(12)} \, P_{12}+
        t_H^{(8,12)} \, P_{8}\right).\\
    \end{array}
  \right.
  \label{eq:rcube}
\end{equation}

For $t_I^{(1,9)}$ we take the final expression for $t_I^{(1,3)}$ and perform
the interchanges $D \leftrightarrow E$, $4 \leftrightarrow 5$ and
$8 \leftrightarrow 12$.
Note this is exactly the effect of a specular reflection across the
diagonal $A$---$H$ of Fig. \ref{fig:fig9} (g).
Actually, we can obtain all other scattering amplitudes by using this
artifact of specular reflections of indices about a proper symmetry axis
of the square in Fig. \ref{fig:fig9} (g).
For instance, for $t_I^{(3,11)}$, $r_I^{(9)}$ and $t_I^{(11,1)}$, the
indices exchanges applied, respectively, to $t_I^{(1,9)}$, $r_I^{(1)}$ and
$t_I^{(1,11)}$, would be those resulting from reflections by an axis perpendicular
to edges 4 and 12 ($A \leftrightarrow D$, $E \leftrightarrow H$ and
$5 \leftrightarrow 8$), perpendicular to edges 5 and 8 ($A \leftrightarrow E$,
$D \leftrightarrow H$ and $4 \leftrightarrow 12$), and in the diagonal
$E$---$D$ ($A \leftrightarrow H$, $4 \leftrightarrow 8$ and
$5 \leftrightarrow 12$).

\subsubsection{Closed cube eigenenergies}

Now, let us examine the closed cube graph eigenstates supposing all the
vertices having the same properties.
Hence, for the cube eight vertices we assume the previously discussed
generalized $\delta$ interaction.
Since the coordination number for this topology is $N=3$, for any vertex
we set (see Eqs. (\ref{eq:r-delta-gen}) and (\ref{eq:t-delta-gen}))
$r = (2 \gamma - i k)/(3 i k - 2 \gamma)$ and $t = 2 i k/(3 i k - 2 \gamma)$.
The eigenenergies come from the poles of Green's function, i.e., the roots of
Eq. (\ref{eq:roots-cube}):
$g = (1 - T(k) \exp[i k \ell])^2 - {R(k)}^2 \exp[2 i k \ell] = 0$
(observe that in this very symmetric case, $T^{(+)} = T^{(-)} = T$ and
$R^{(+)} = R^{(-)} = R$, with $R$ and $T$ obtained from the calculations
described in the previous Sec.).
In the Table \ref{tab:tab1} we show the resulting first ten eigenvalues for
$\gamma = 0$ and $\gamma = 1$ (with $\mu = \hbar = 1$).

\begin{table}[htb]
  \centering
  \begin{tabular}{@{}ccc@{}}
    \toprule
    \multirow{2}{*}{} & \multicolumn{2}{c}{$\gamma$} \\ \cmidrule(l){2-3}
    State             &  0           &  1    \\     \midrule
    1                 & 1.230959     & 1.094322 \\
    2                 & 1.919633     & 1.642395 \\
    3                 & 3.141593     & 2.190764 \\
    4                 & 4.372552     & 3.141593 \\
    5                 & 5.052226     & 3.516328 \\
    6                 & 6.283185     & 5.177393 \\
    7                 & 7.514145     & 6.283185 \\
    8                 & 8.193819     & 7.602957 \\
    9                 & 9.424778     & 8.273085 \\
    10                & 10.65574     & 9.424778 \\
    \bottomrule
  \end{tabular}
  \caption{
    The first ten numerically calculated $k_n$ values (from $g=0$, see
    Eq. (\ref{eq:roots-cube})) for the cube quantum graph.
    All the vertices are assumed generalized $\delta$ interactions of
    strength $\gamma=0$ (so, Neumann-Kirchhoff) and $\gamma=1$.}
  \label{tab:tab1}
\end{table}

In order to check the eigenvalues found through the Green's function
approach, one can directly solve the Schr\"{o}dinger equation.
Along the edge $s \, (= 1, \ldots, 12)$, the component $\psi_s(x_s)$ of the
total wave function $\Psi$ is the solution of
(where for simplicity we drop the subscript notation for $x$)
\begin{equation}
  -\frac{d^{2}}{dx^{2}}\psi_{s}(x) = k^{2}\psi_{s}(x),
\end{equation}
with $k=\sqrt{2\mu E}/\hbar$ and the origin for the edges taken in the
vertices $A$, $C$, $F$ and $H$.
Thus, the $\psi$'s have the form
\begin{equation}
  \psi_{s}(x) = {\mathcal A}_{s} \exp[i k x] + {\mathcal B}_s \exp[-i k x].
\end{equation}
The coefficients ${\mathcal A}_{s}$ and ${\mathcal B}_{s}$ are determined
by the boundary conditions, corresponding to a delta potential on the
vertices (see the discussion in the Appendix 
\ref{app:boundary-conditions-1}).
Therefore
\begin{align}
  \psi_1(0) = \psi_4(0) = \psi_5(0)= {} & \psi(A)
  &  
  \psi_1'(0) + \psi_4'(0) + \psi_5'(0) = {} & 2 \gamma \, \psi(A)
                                              \nonumber\\
  \psi_2(0) = \psi_3(0) = \psi_7(0) = {} & \psi(C), 
  &                                                    
  \psi_2'(0) + \psi_3'(0) + \psi_7'(0) = {} & 2 \gamma \, \psi(C),
  \nonumber \\
  \psi_6(0) = \psi_9(0) = \psi_{10}(0) ={} & \psi(F), 
  &
  \psi_6'(0) + \psi_9'(0) + \psi_{10}'(0) = {} & 2 \gamma \, \psi(F),
  \nonumber \\
  \psi_8(0) = \psi_{11}(0) = \psi_{12}(0) = {} &  \psi(H), 
  &
  \psi_8'(0) + \psi_{11}'(0) + \psi_{12}'(0) = {} & 2 \gamma \, \psi(H),
  \nonumber \\
  \psi_1(0) = \psi_2(0) = \psi_6(0) = {} & \psi(B), 
  &
  \psi_1'(0) + \psi_2'(0) + \psi_6'(0) = {} & - 2 \gamma \, \psi(B),
  \nonumber \\
  \psi_3(0) = \psi_4(0) = \psi_8(0) = {} & \psi(D), 
  &
  \psi_3'(0) + \psi_4'(0) + \psi_8'(0) = {} & - 2 \gamma \, \psi(D),
  \nonumber \\
  \psi_5(0) = \psi_9(0) = \psi_{12}(0) = {} & \psi(E), 
  &
  \psi_5'(0) + \psi_9'(0) + \psi_{12}'(0) = {} & - 2 \gamma \, \psi(E),
  \nonumber \\
  \psi_7(0) = \psi_{10}(0) = \psi_{11}(0) = {} & \psi(G), 
  &
  \psi_7'(0) + \psi_{10}'(0) + \psi_{11}'(0) = {} & - 2 \gamma \, \psi(G).
  \label{eq:system-cube}
\end{align}
From the above system of equations -- plus the normalization condition
$\sum_{s=1}^{s=12} \int_{0}^{\ell} dx \, |\psi_s(x)|^2 = 1$ -- one gets the
eigenfunctions and eigenvalues.
By solving Eq. (\ref{eq:system-cube}) -- e.g, numerically -- one finds that
the eigenvalues from the Green's functions are exactly those from the
Schr\"odinger equation, as it should be.

\subsubsection{Scattering by attaching leads to the quantum cube graph}
\label{sec:cube_scatt}

\begin{figure}
   \centering
   \includegraphics*[width=0.2\textwidth]{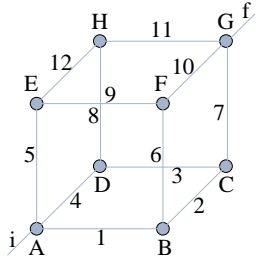}
   \caption{\label{fig:fig10}  (Color online).
   The original quantum closed cube graph is attached to two leads (at the
   vertices $A$ and $G$), thus becoming an open graph structure.}
\end{figure}

One also can study transmission through (as well as reflection from)
the original closed cube by attaching leads to it.
In Fig. \ref{fig:fig10} we display a possible configuration for the system,
where leads are added to the vertices $A$ and $G$ of our previous
very symmetric graph.
For the now modified vertices $A$ and $G$, we also assume a $\delta$
interaction of strength $\gamma$, only recalling that in this case these
two vertices have a coordination number $N=4$ (instead of $N=3$).
Just as an illustration, for $x_i$ in lead $i$ and $x_f$ in lead $f$
(see Fig. \ref{fig:fig10}), the Green's function reads
\begin{equation}
G_{fi}(x_f,x_i;k) = \frac{\mu}{i \hbar^2 k}  \, T_{f i} \, \exp[i k (x_f + x_i)].
\end{equation}

\begin{figure}[h]
   \centering
   \includegraphics*[width=0.5\textwidth]{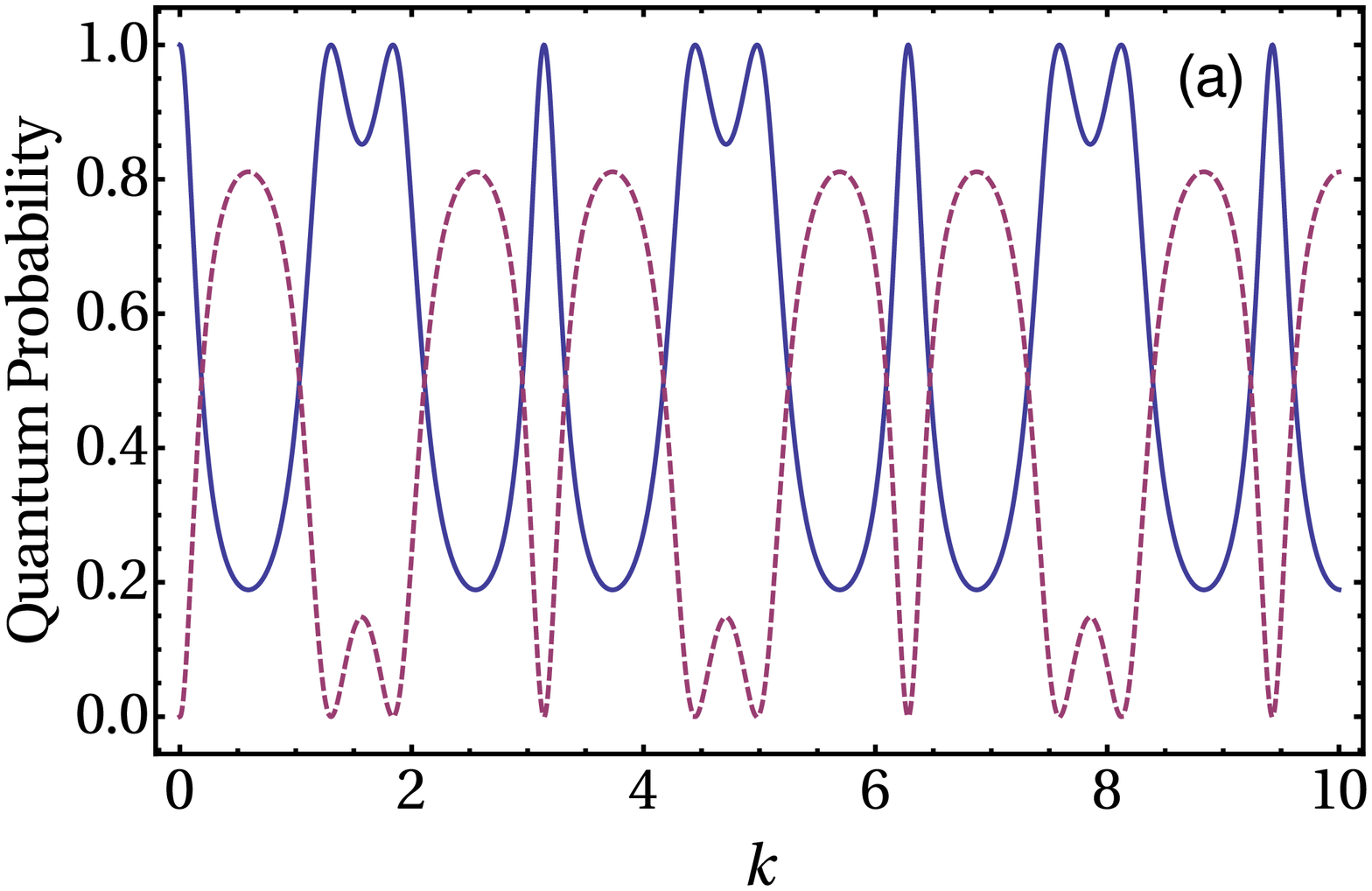}%
   \includegraphics*[width=0.5\textwidth]{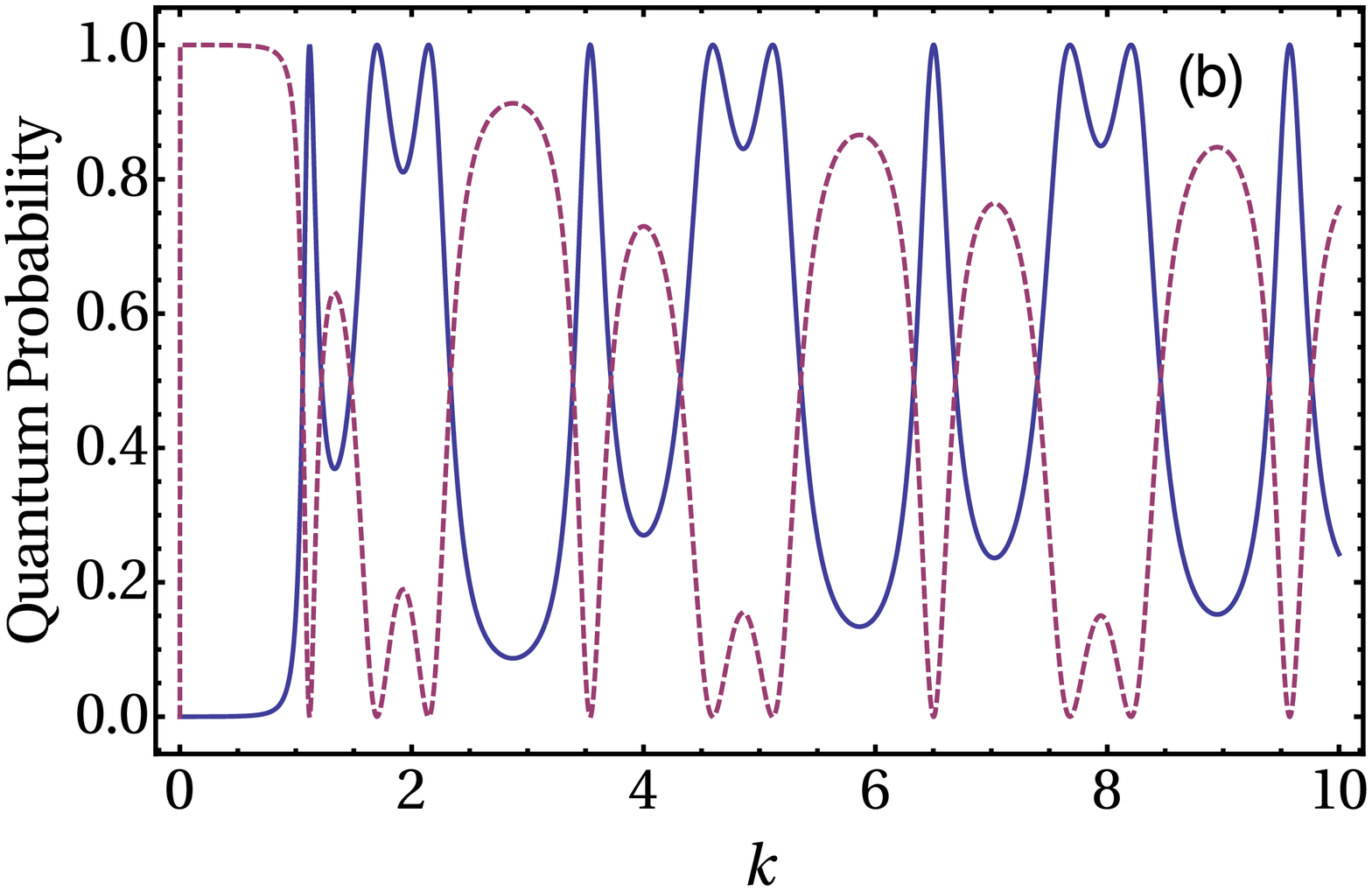}\\
   \caption{\label{fig:fig11}  (Color online).
      The transmission $|T_{f i}|^2$ (solid line) and reflection
      $|R_{i}|^2$ (dashed) probabilities for the open cube graph of Fig.
      \ref{fig:fig10}.
      All the vertices are generalized $\delta$ interactions of
      strength (a) $\gamma=0$ and (b) $\gamma=1$.
      Here $\mu = \hbar = 1$.
   }
\end{figure}

Calculating $T_{f i}$ (and also $R_i$) using the discussed techniques,
we show in Fig. \ref{fig:fig11} the transmission and reflection probabilities
as function of $k$ for $\gamma = 0$ and $\gamma = 1$.
Since for the former the individual edges transmission and reflections
coefficients are not function of $k$, we do not see $|T_{f i}|^2$ tending
to 1 for $k$ increasing (as slowly seen for $\gamma = 1$).

\subsection{Binary tree}
\label{sec:binarytree}

\begin{figure}
  \centering
  \includegraphics*[width=0.5\textwidth]{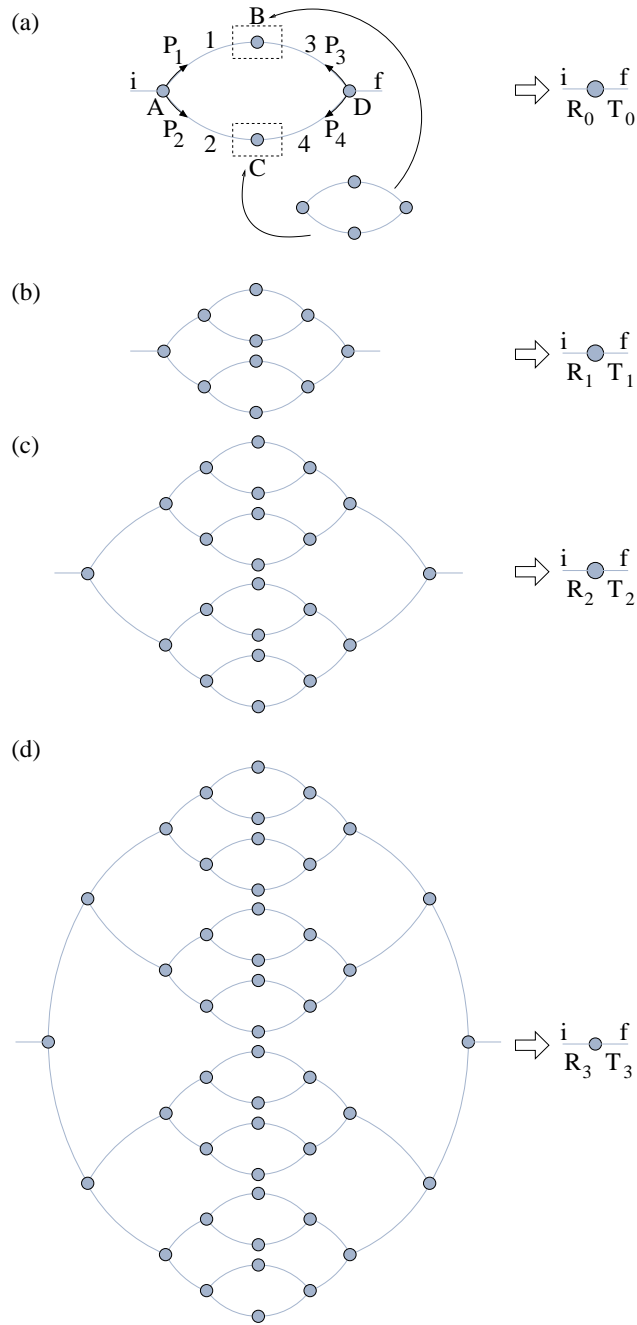}
  \caption{\label{fig:fig12} (Color online).
    Binary tree quantum graphs (attached to leads $i$ and $f$) with
    different number of recursive compositions $l$.
    The way a single composition (by insertion) is performed is illustrated
    in (a).
    By using the regrouping procedure to calculate the Green's function, 
    one can reduce the original structure to a simple graph comprising an 
    unique effective vertex linked to two leads (depicted in the right 
    panels).
    At each level $l$, the rescaled system has the same global transmission
    $T_l$ and reflection $R_l$ amplitudes of the corresponding original graph.
    Here it is shown, (a) the initial basic topology ($l=0$), and (b) $l=1$, 
    (c) $l=2$, and (d) $l=3$, insertions.}
\end{figure}

As previously emphasized, the general way the Green's function can be
written in terms of arbitrary quantum coefficients -- encompassing 
`blocks' of vertices and edges -- allows one to use a recursive procedure
to obtain the system full solution.
This is a particularly useful protocol for graphs displaying a 
hierarchical structure, as the case of the binary tree depicted in Fig. 
\ref{fig:fig12} (which illustrates three `levels' ($l=1,2,3$) of the
graph construction by insertions).
In the following we assume all the edges having the same length $\ell$ 
(so, Fig. \ref{fig:fig12} is not shown in scale).

\begin{figure}
  \begin{center}
    \includegraphics*[width=0.5\textwidth]{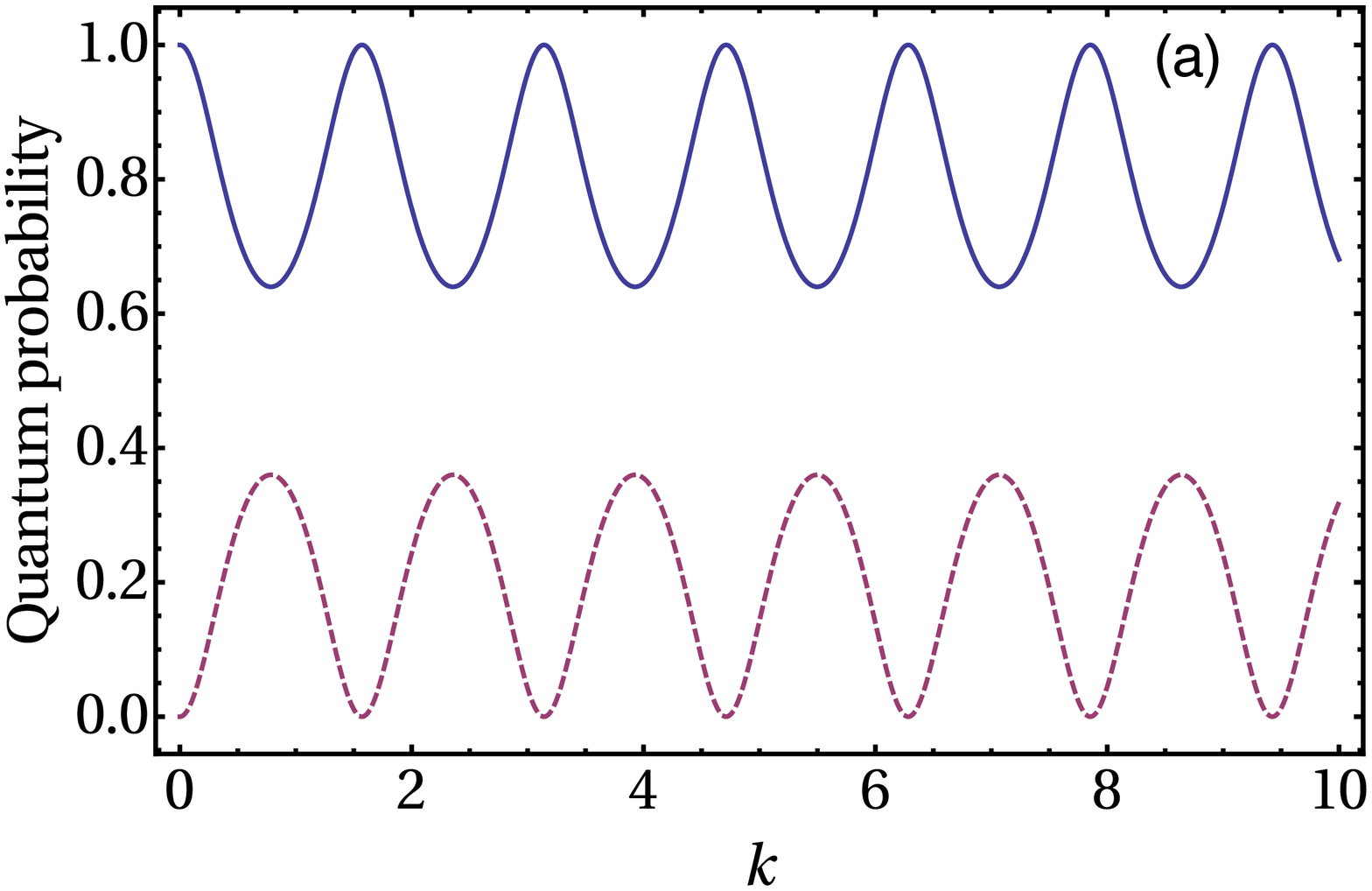}%
    \includegraphics*[width=0.5\textwidth]{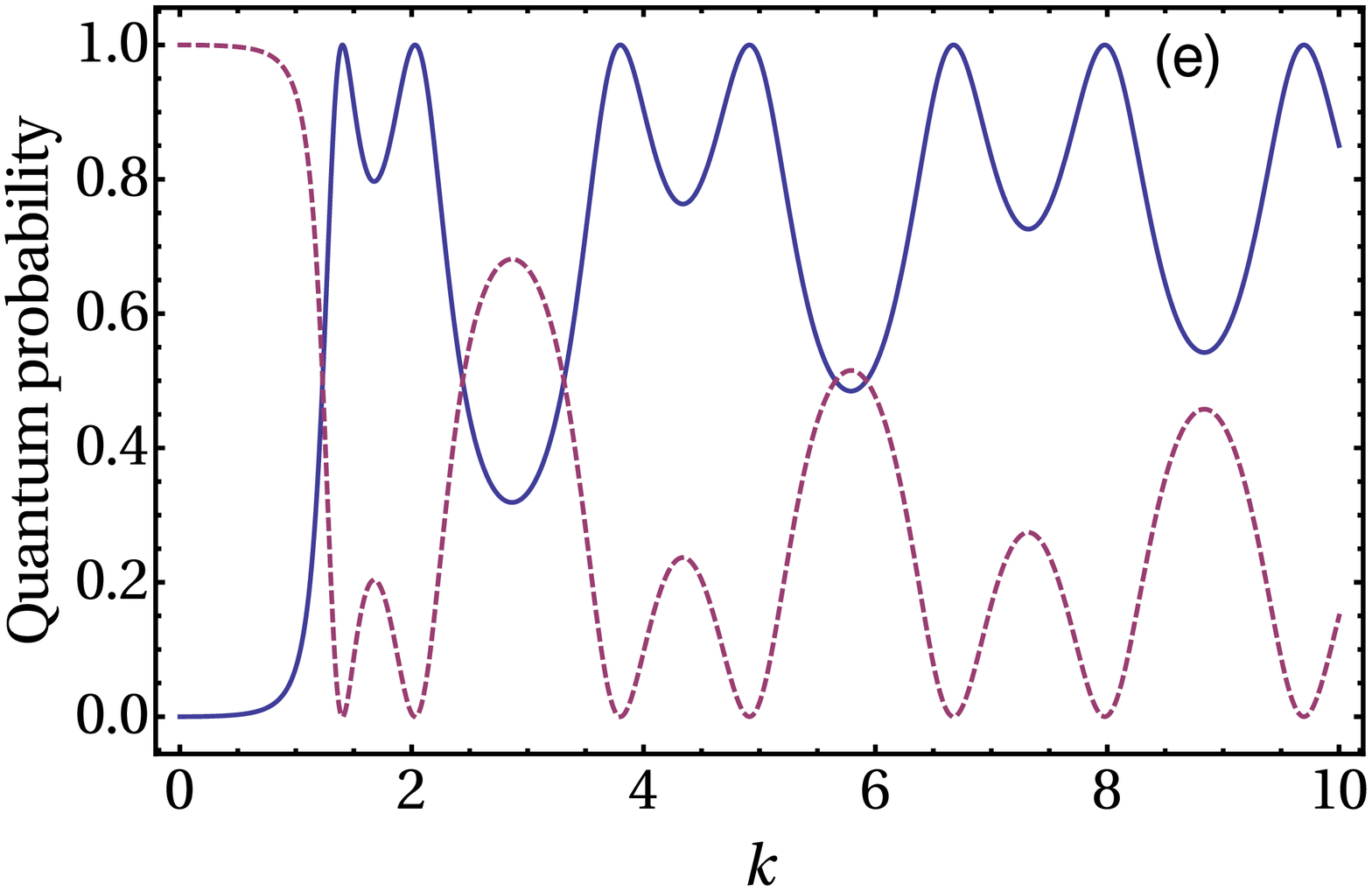}\\
    \includegraphics*[width=0.5\textwidth]{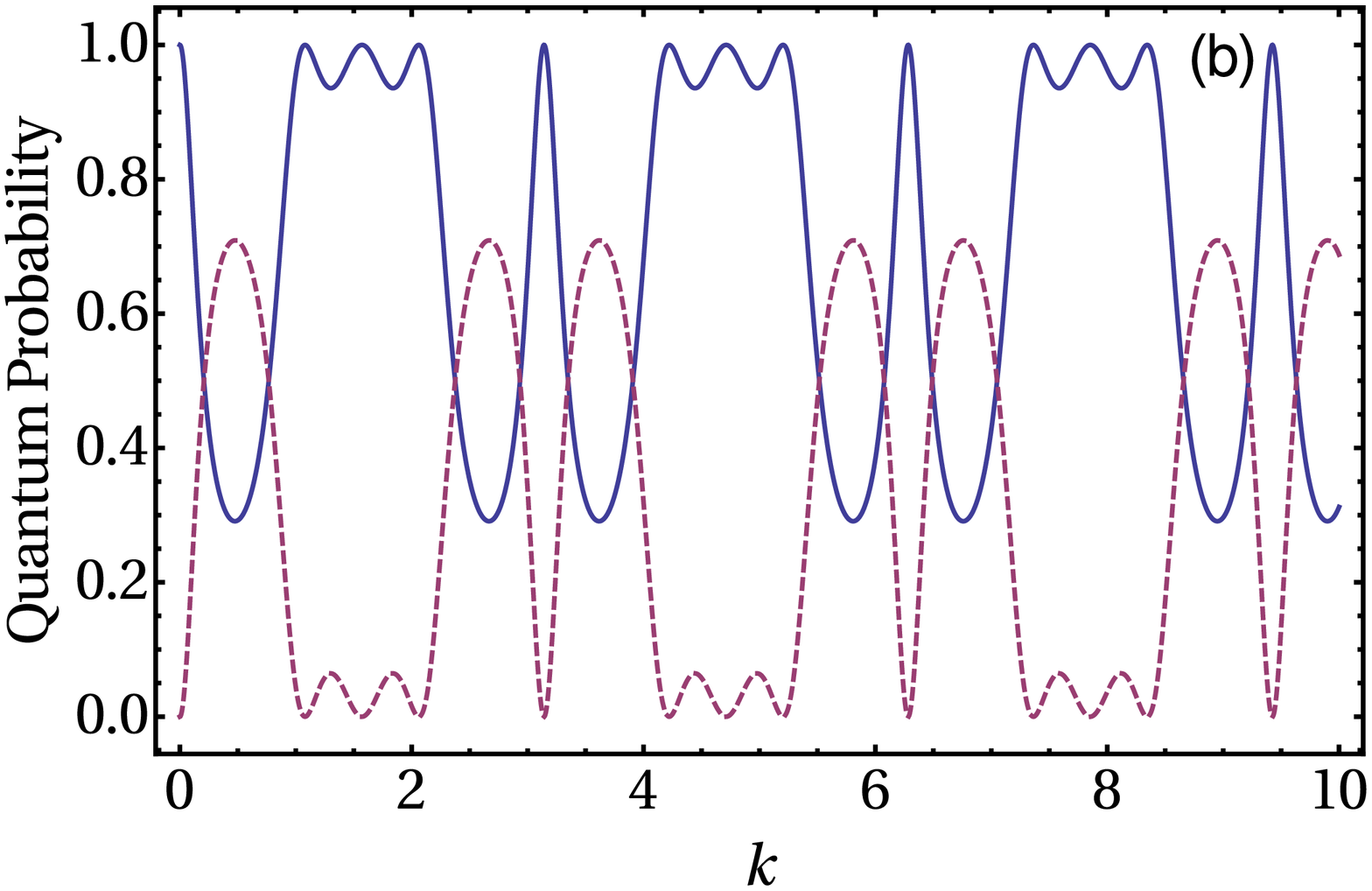}%
    \includegraphics*[width=0.5\textwidth]{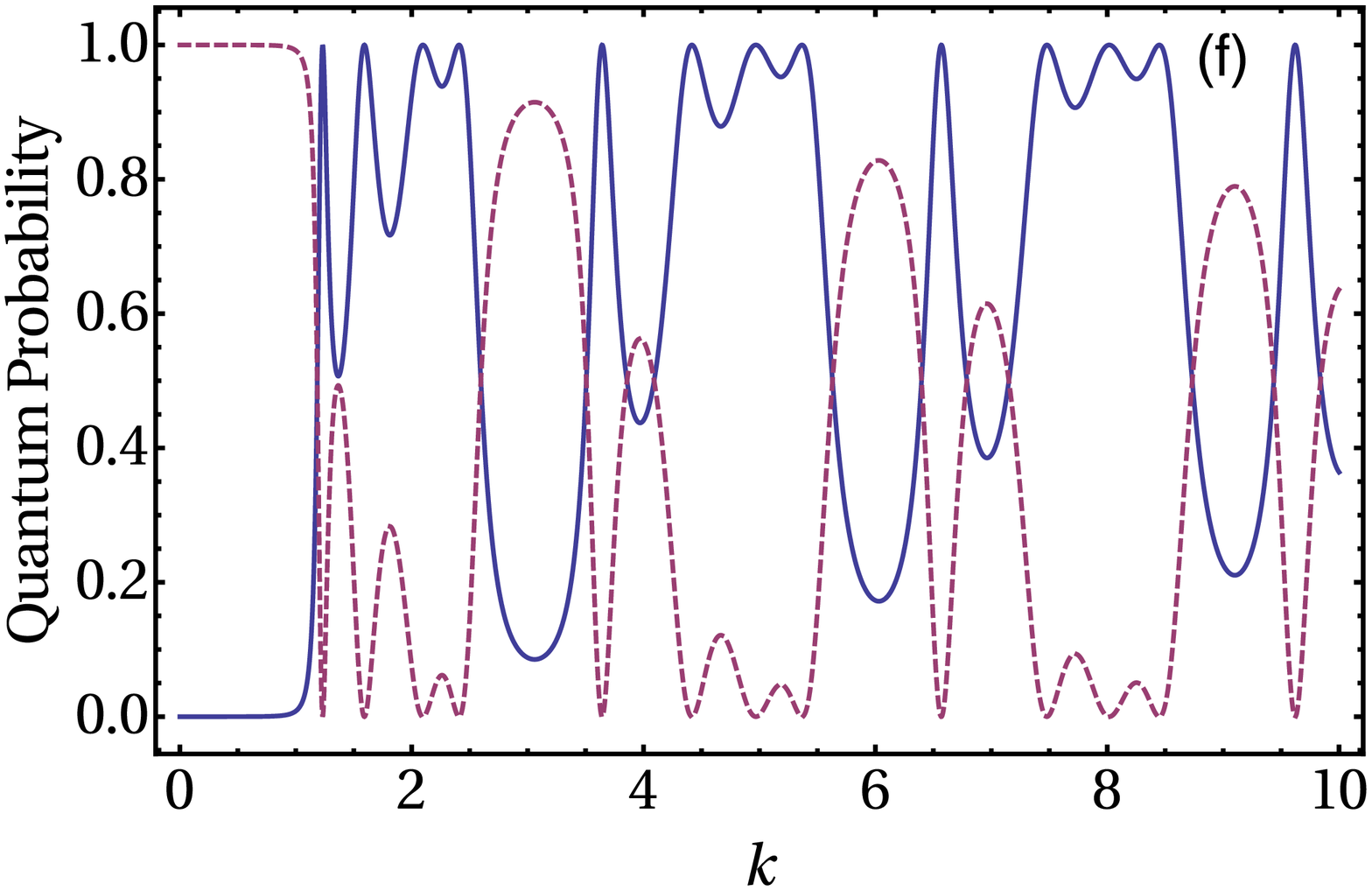}\\
    \includegraphics*[width=0.5\textwidth]{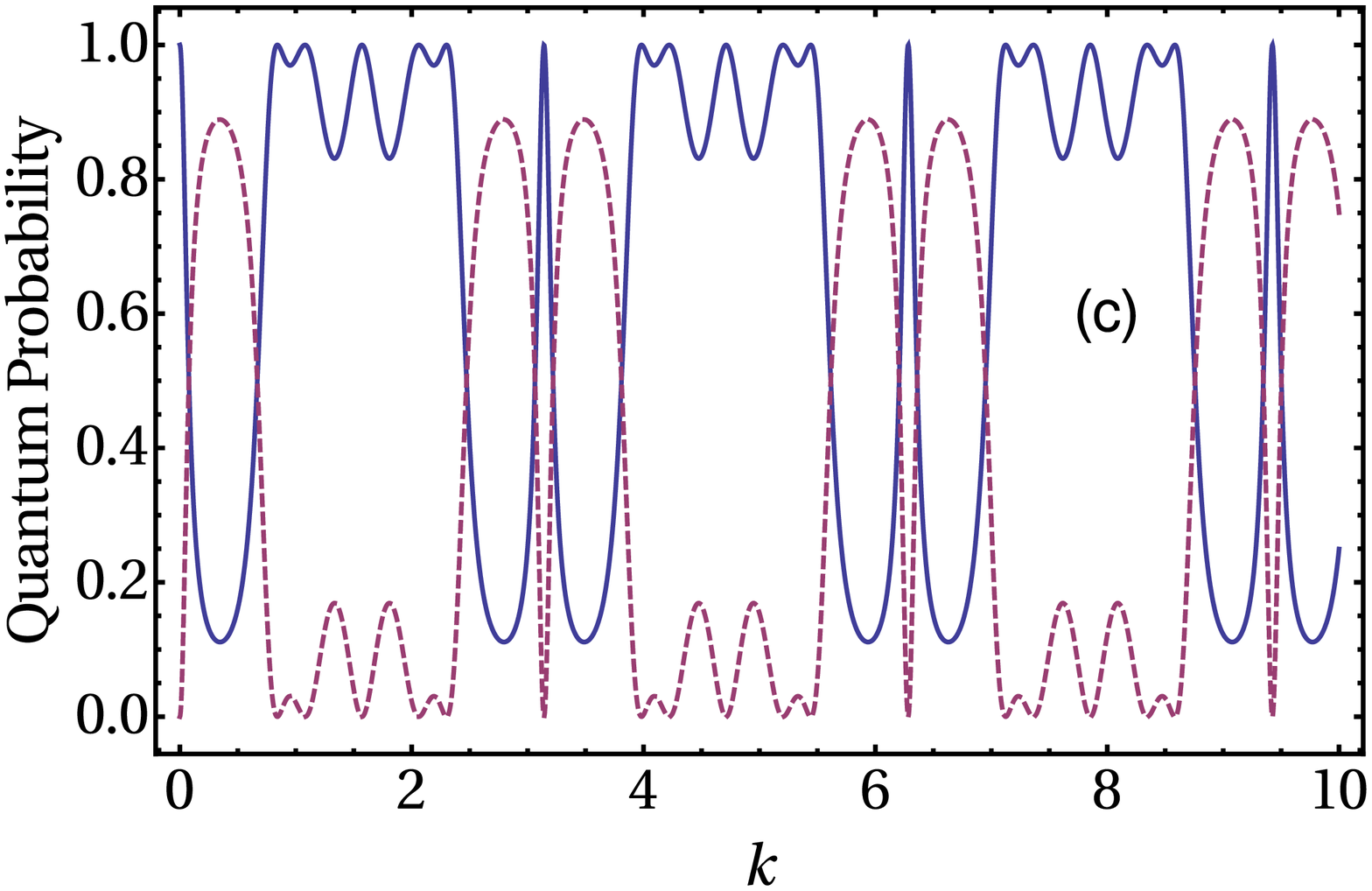}%
    \includegraphics*[width=0.5\textwidth]{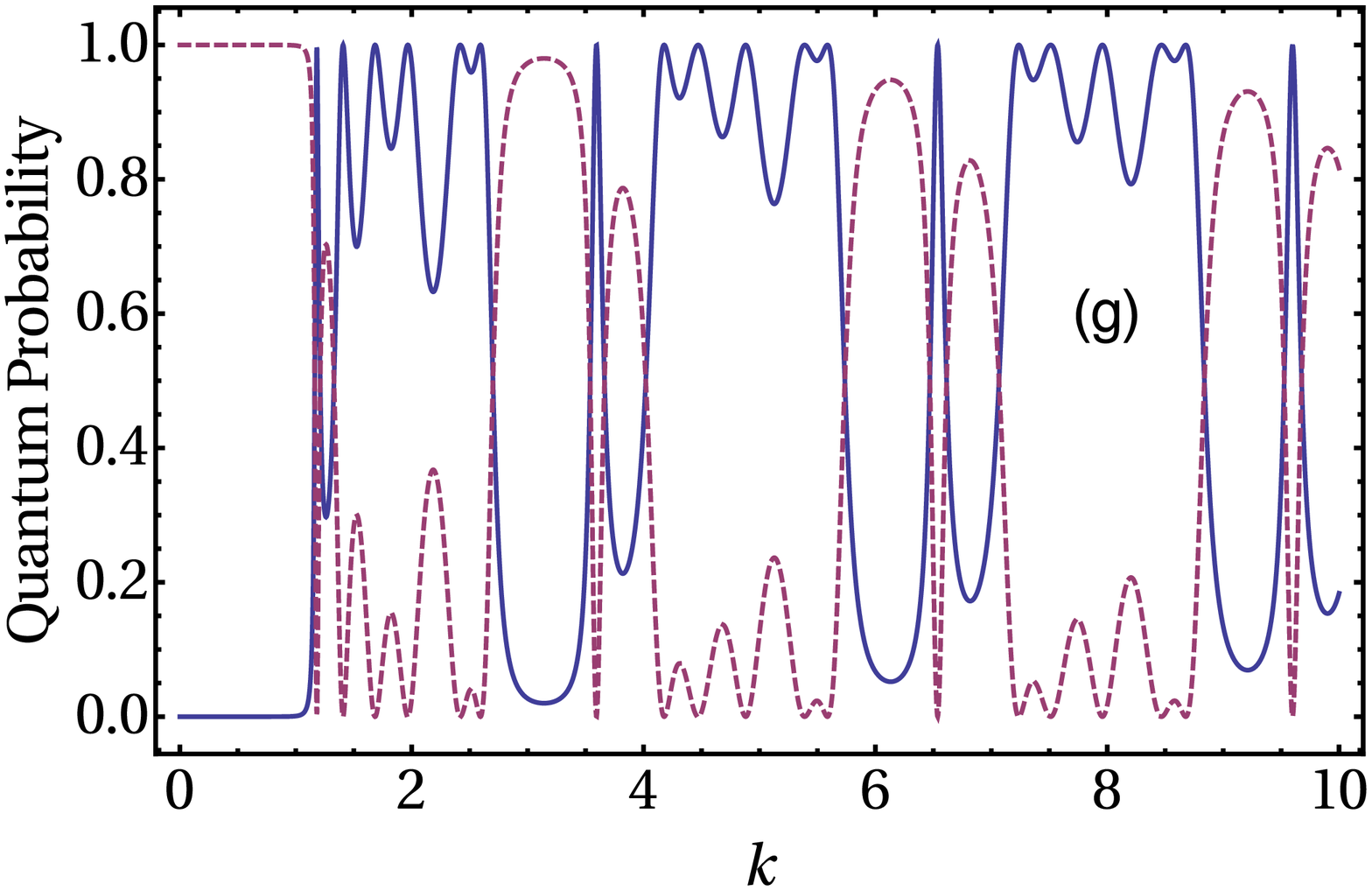}\\
    \includegraphics*[width=0.5\textwidth]{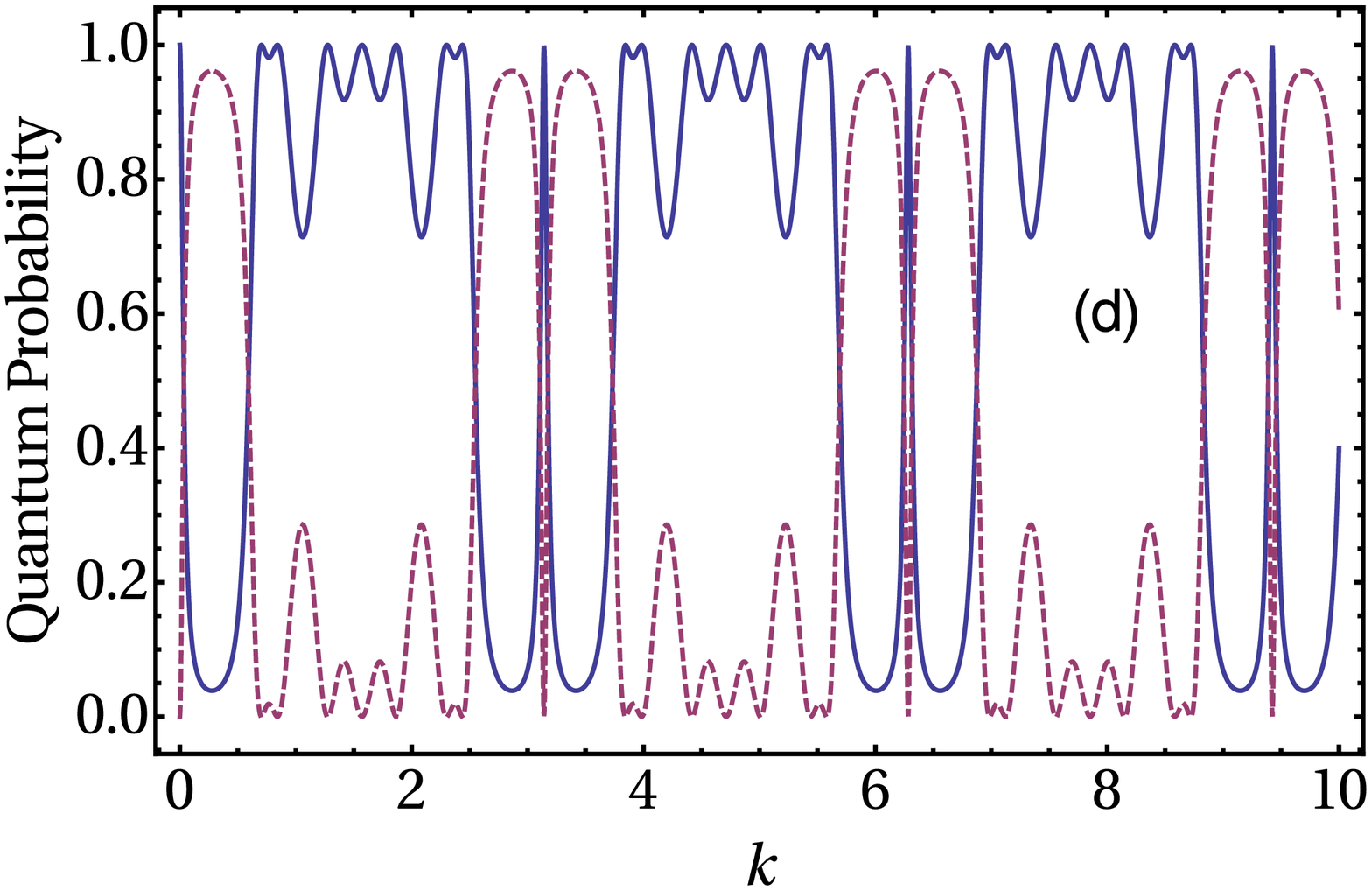}%
    \includegraphics*[width=0.5\textwidth]{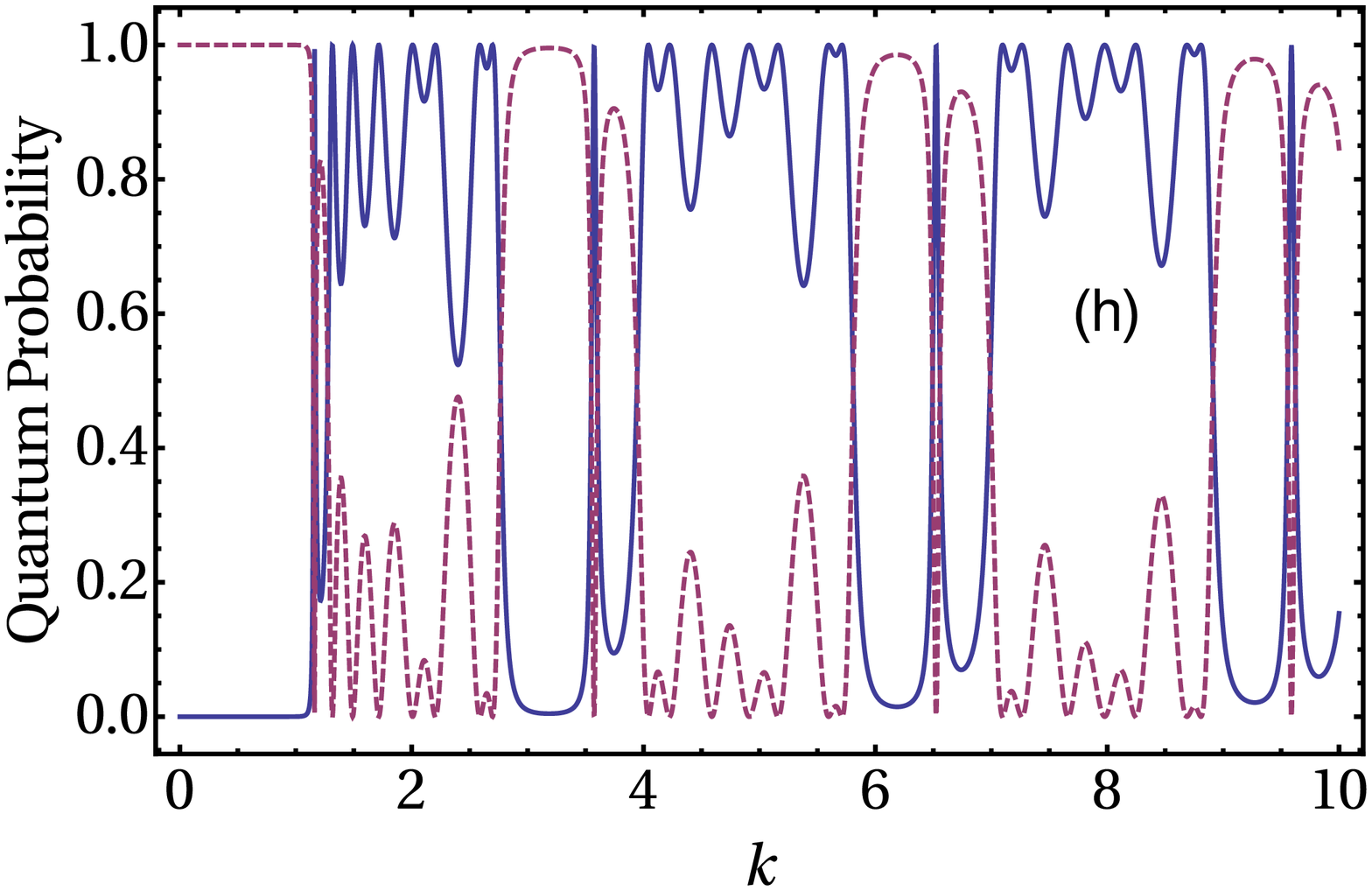}\\
  \end{center}
  \caption{ \label{fig:fig13} (Color online).
    The transmission $|T_{l}|^2$ (solid line) and reflection
    $|R_{l}|^2$ (dashed) probabilities for the binary trees of
    Fig. \ref{fig:fig12}.
    The vertices are generalized $\delta$ interactions of strength
    (a)-(d) $\gamma=0$ and (e)-(h) $\gamma=1$.
    All the edges have length $\ell = 1$.
    Here also $\mu = \hbar = 1$.
    The quantum probabilities for the graphs of Fig. \ref{fig:fig12} (a),
    (b), (c) and (d) are shown, respectively, in (a) and (e), (b) and (f),
    (c) and (g) and (d) and (h).
  }
\end{figure}

Using the Green's function method, let us derive the transmission and
reflection quantum amplitudes for the basic structure (so level $l=0$)
of Fig. \ref{fig:fig12} (a).
In fact, such calculation is similar to that of $r_I^{(1)}$ and $t_I^{(11,1)}$
for the graph of Fig. \ref{fig:fig9} (g).
By grouping the four vertices $A$, $B$, $C$ e $D$ in a single vertex
$l=0$ (right panel of Fig. \ref{fig:fig12} (a)), the global reflection
coefficient $R_{0}$, from $i$ to $i$, is given by
(see the left panel of Fig. \ref{fig:fig12} (a))
\begin{equation}
  R_{0} = r_A^{(i)} + t_A^{(1,i)} \, P_1 + t_A^{(2,i)} \, P_2
\end{equation}
where
\begin{equation}
  \left\{
    \begin{array}{l}
      P_1 = r_B^{(1)} \, \exp[2 i k \ell] \,
      \left(
      r_A^{(1)} \, P_1 + t_A^{(2,1)} \, P_2 + t_A^{(i,1)}
      \right)
      \\ \ \ \ \ \,\,
      +\,\, t_B^{(3,1)} \, \exp[2 i k \ell] \, \left(
      r_D^{(3)} \, P_3 + t_D^{(4,3)} \, P_4
      \right) \\
      P_2 = r_C^{(2)} \, \exp[2 i k \ell] \, \left(
      r_A^{(2)} \, P_2 + t_A^{(1,2)} \, P_1 + t_A^{(i,2)} \right)
      \\ \ \ \ \ \,\,
      +\,\, t_C^{(4,2)} \, \exp[2 i k \ell] \, \left(
      r_D^{(4)} \, P_4 + t_D^{(3,4)} \, P_3\right) \\
      P_3 = t_B^{(1,3)} \, \exp[2 i k \ell] \, \left(
      r_A^{(1)} \, P_1 + t_A^{(2,1)} \, P_2 + t_A^{(i,1)} \right)
      \\ \ \ \ \ \,\,
      +\,\, r_B^{(3)} \, \exp[2 i k \ell] \, \left(
      r_D^{(3)} \, P_3 +  t_D^{(4,3)} \, P_4\right) \\
      P_4 = t_C^{(2,4)} \, \exp[2 i k \ell] \, \left(
      r_A^{(2)} \, P_2 + t_A^{(1,2)} \, P_1 + t_A^{(i,2)} \right)
      \\ \ \ \ \ \,\,
      +\,\, r_C^{(4)} \, \exp[2 i k \ell] \, \left(
      r_D^{(4)} \, P_4 + t_D^{(3,4)} \, P_3\right).
    \end{array}
  \right.
  \label{eq:ralpha}
\end{equation}

The transmission coefficient $T_{0}$ (from $i$ to $f$) follows from
\begin{equation}
  T_{0}=t_A^{(1,i)} \, P_1 + t_A^{(2,i)} \, P_2,
\end{equation}
where the $P$'s are given by Eq. \eqref{eq:ralpha}, but for which we 
exchange all the indices (including those of the $P$'s) as:
$1 \leftrightarrow 3$, $2 \leftrightarrow 4$, $A \leftrightarrow D$ and
$i \leftrightarrow f$.
Solving the system \eqref{eq:ralpha} we get $R_{0}$ and $T_{0}$.
We observe that the reflection (for $f \rightarrow f$) and transmission 
(for $f \rightarrow i$) are acquired, respectively, from the expressions 
$R_0$ and $T_0$ by just applying the above same exchange of indices.

Then, we can substitute the vertices $B$ and $C$ by our basic graph 
structure, as schematically represented in Fig. \ref{fig:fig12} (a).
This leads to the graph of Fig. \ref{fig:fig12} (b) (level $l=1$) of 
quantum amplitudes $R_1$ ($i \rightarrow i$) and $T_1$ ($i \rightarrow f$).
These latter coefficients are exactly those for $R_0$ and $T_0$, 
but where in the place of $r_B$, $r_C$, $t_B$ and $t_C$ we use the
corresponding $R_{0}$ and $T_{0}$.
Such process can be repeated any number of times, with $R_l$ and $T_l$ 
always directly obtained from $R_{l-1}$ and $T_{l-1}$.

As a numerical example, consider the edges with the same length $\ell=1$ 
and Dirac $\delta$ interactions of intensity $\gamma$ (for $\gamma = 0$
and $\gamma = 1$) as the boundary conditions (see Appendix 
\ref{app:boundary-conditions-1}) at all the vertices.
For the vertices with $N=2$ edges (say $B$ and $C$) we have 
$r = \gamma/(ik-\gamma)$ and $t = ik/(ik-2\gamma)$ and for those with 
$N=3$ (say $A$ and $D$) $r = (2\gamma-ik)/(3ik-2\gamma)$ and 
$t = 2ik/(3ik-2\gamma)$.
In Fig. \ref{fig:fig13} we show the reflection $|R_l|^2$ 
($i \rightarrow i$) and transmission $|T_l|^2$ ($i \rightarrow f$)
probabilities for the basic structure (Fig. \ref{fig:fig12} (a)) and 
for the three levels of insertions for the binary tree (Fig. 
{\ref{fig:fig12} (b)--(d)).
As it should be expected, for higher $l$'s the patterns of reflection and 
transmission, as function of $k$, become much more complex.
Also, we do not observe any systematic increasing of $|T_l|^2$ as $k$ 
increases because the rich interference behavior -- due to the wave 
propagation along the distinct edges -- takes place for any value of $k$.

\subsection{Sierpi\'{n}ski-like graphs}
\label{sec:egf}

One of the many reasons for the interest in self-similar lattices is 
their utility to model systems which are self-assembled from an original 
backbone (the motif of the replication), the case of certain complex 
molecules \cite{NatureChem.2015.7.389}.
Sierpi\'{n}ski graphs are very nice examples of structures which can be 
recursively generated from a basic building block.
They originate from the Sierpi\'{n}ski gasket, a well-known fractal 
object introduced by Sierpi\'{n}ski in 1915 \cite{TMI.1995.17.52}.

Sierpi\'{n}ski graphs have been studied in relation to small-world networks
\cite{JPA.2006.39.11739}.
Also, Sierpi\'{n}ski gaskets have been analyzed in
\cite{Inproceedings.2004.Bondarenko,Inproceedings.2005.Bondarenko},
where Neumann-Kirchhoff boundary conditions were considered.
However, the most general case of arbitrary reflection and transmission
amplitudes for the vertices are still not well explored in the literature.

Here we shall address procedures similar to those of the previous section,
allowing one to derive the scattering Green's function for the Sierpi\'{n}ski
graph.
We present a schematic method to regroup the multiple stages of the graph
(up to stage $n$), leading to the total $R$ and $T$ amplitudes for the 
whole composition in terms of the basic vertices $A$, $B$, $C$
(Fig. \ref{fig:fig14}) scattering coefficients.
But the construction next is not a simple repetition of the binary tree
graph calculation.
One must take into account that part of the edges change their lengths from
one Sierpi\'{n}ski stage to another.
This means that in fact $R_n$ and $T_n$ are not trivial functions of the
actual edges lengths at each stage $n$.

\begin{figure}
  \begin{center}
    \includegraphics*[width=0.6\textwidth]{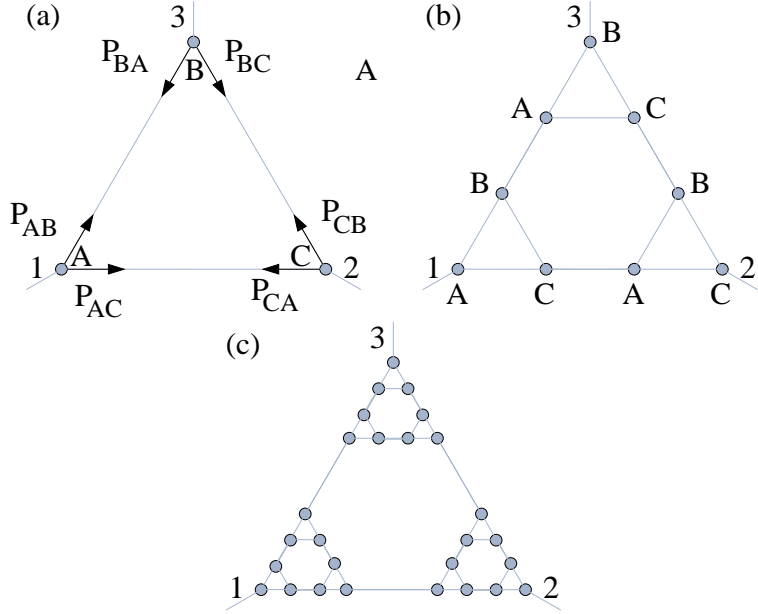}
  \end{center}
  \caption{\label{fig:fig14} (Color online).
    Finite Sierpi\'{n}ski graphs with different number $n$ of recursive
    stages:
    (a) $n=1$ (the initial $\Delta_{ABC}$ structure, main text),
    (b) $n=2$, and
    (c) $n=3$.
    The $P$'s in (a) represent proper infinite families of scattering
    paths useful to calculate the Green's function.
    The generation of new vertices from (a) to (b) illustrates the
    elementary transformation to $\Delta_{ABC}$, the basic step leading
    to the successive graph stages.}
\end{figure}

In Fig. \ref{fig:fig14} we show three different stages ($n = 1, 2, 3$)
of a Sierpi\'{n}ski graph.
The basic step to go from $n$ to $n+1$ involves a transformation in all
the fundamental equilateral triangles $\Delta_{ABC}$ of the graph $n$.
For instance, starting from $n=1$ (the basic configuration of Fig. 
\ref{fig:fig14} (a), with all the three edges of length $\ell_1 = \ell$),
$n=2$ is created by adding two extra vertices to each side of $\Delta_{ABC}$, 
as illustrated in Fig. \ref{fig:fig14} (b).
To obtain $n=3$, the procedure is repeated for the three $\Delta_{ABC}$ in
Fig. \ref{fig:fig14} (b), leading thus to the 9 triangles 
$\Delta_{ABC}$ of  Fig. \ref{fig:fig14} (c), and so on and so forth.
Note that at the stage $n \ (= 1, 2, \ldots)$, all the sides of the triangles 
$\Delta_{ABC}$ have a same length $\ell_n = \ell/3^{(n-1)}$.

Since at any stage $n$ the graph always has exactly three semi-infinite 
leads, the scattering matrix is of order $3$ and given by 
(see Appendix \ref{app:flux})
\begin{equation}
  S_n =
  \left(
    \begin{array}{ccc}
      R_{n}^{(1)} & T^{(1, 2)}_{n} & T^{(1, 3)}_{n}  \\
      T^{(2, 1)}_{n} & R^{(2)}_{n} & T^{(2, 3)}_{n}  \\
      T^{(3, 1)}_{n} & T^{(3, 2)}_{n} & R^{(3)}_{n}
    \end{array} \right).
\end{equation}
Above, $R_{n}^{(a)}$ and $T_{n}^{(b,a)}$ are the resulting reflection (from
lead $a \ (=1, 2, 3)$) and transmission (from lead $a$ to lead $b$,
with $a \neq b$ and $a,b = 1, 2, 3$) amplitudes for the group of $3^{n}$
vertices constituting the Sierpin\'nski graph at stage $n$
(see Fig. \ref{fig:fig14}).

The Green's function for the transmission case of the Sierpi\'{n}ski
graph of stage $n$ is given by (for $x_f$ in lead $b$ and $x_i$ in lead $a$)
\begin{equation}
  G_{b a}(x_f,x_i;k) = \frac{\mu}{i \hbar^2 k}
  T_{n}^{(b, a)} \exp[i k (x_f + x_i)].
\end{equation}
For the reflection case the Green's function reads (for $x_i$ and
$x_f$ in lead $a$)
\begin{equation}
  G_{a a}(x_f,x_i;k) = \frac{\mu}{i \hbar^2 k}
  \Big(
  \exp[ik|x_f-x_i|]+R_{n}^{(a)} \exp[i k (x_f + x_i)]
  \Big).
\end{equation}

For simplicity, we next assume that all the elementary vertices 
$V = A, B, C$ (Fig. \ref{fig:fig14} (a)) have the same scattering 
properties along any edge, thus $r_V^{(a)} = r$ and $t_V^{(b,a)} = t$.
Hence, for all $n$ it holds that $R_n^{(a)} = R_n$ and $T_n^{(b,a)} = T_n$.
Because so, the specific leads we choose to calculate $R_n$ and $T_n$
will not alter the final expression.
In this way, for $n=1$, Fig. \ref{fig:fig14} (a), we consider the
reflection from lead $2$ and the transmission from lead $1$ to lead $2$,
or
(recalling that $\ell_1$ is just $\ell$)
\begin{equation}
T_{1}(\ell_1) = t \, (P_{AB} + P_{AC}), \qquad 
R_{1}(\ell_1) = r + t \, (P_{CA} + P_{CB}),
\end{equation}
where (see Fig. \ref{fig:fig14} (a))
\begin{equation}
  \left\{
    \begin{array}{l}
      P_{AB} = \exp[i k \ell_1] \, (r \, P_{BA} + t \, P_{BC})    \\
      P_{AC} = \exp[i k \ell_1] \, (r \, P_{CA} + t \, P_{CB} + t)\\
      P_{BC} = \exp[i k \ell_1] \, (r \, P_{CB} + t \, P_{CA} + t)\\
      P_{BA} = \exp[i k \ell_1] \, (r \, P_{AB} + t \, P_{AC})    \\
      P_{CA} = \exp[i k \ell_1] \, (r \, P_{AC} + t \, P_{AB})    \\
      P_{CB} = \exp[i k \ell_1] \, (r \, P_{BC} + t \, P_{BA})
    \end{array}
  \right. .
  \label{eq:sistema_sierpinski}
\end{equation}

Solving the system of equations in \eqref{eq:sistema_sierpinski}, we get
the transmission and reflection coefficients of the Sierpi\'{n}ski graph
stage $n=1$, Fig. \ref{fig:fig14} (a), as
\begin{equation}
R_{1}(\ell_1) = r + \frac{
2 t^2 \big(r + (t^2 - r^2) \exp{[ik\ell_1]} \big)
\exp{[2 i k\ell_1]}}
{\big(1 - (r + t) \exp{[i k \ell_1]} \big)
\big(1 + t \exp{[ik\ell_1]} + (t^2 - r^2) \exp{[2 i k \ell_1]}\big)}
\label{eq:R1_sierpinski}
\end{equation}
and
\begin{equation}
T_{1}(\ell_1) = \frac{t^2 \big(1 + (t - r) \exp{[ik\ell_1]} \big)
\exp{[ik\ell_1]}}
{\big(1 - (r + t) \exp{[i k \ell_1]}\big)
\big(1 + t \exp{[ik\ell_1]} + (t^2 - r^2)\exp{[2 i k \ell_1]}\big)}.
\label{eq:T1_sierpinski}
\end{equation}

Finally, given the system hierarchical character, the scattering
coefficients for the stage $n+1$ can be recursively obtained from those
of stage $n$.
Indeed, from the geometry of the graph formation process, depicted in
Fig. \ref{fig:fig14}, and from Eqs. (\ref{eq:R1_sierpinski}) and
(\ref{eq:T1_sierpinski}), one concludes after some straightforward
reasoning that
\begin{equation}
R_{n+1}(\ell_{n+1}) = R_{n}(\ell_n/3) +
\frac{2 [T_{n}(\ell_n/3)]^2 \big(R_n(\ell_n/3) + 
([T_n(\ell_n/3)]^2 - [R_n(\ell_n/3)]^2)
\exp{[ik\ell/3]}\big)\exp{[2ik\ell/3]}}
{D_n(\ell_n/3)}
\label{eq:Rn_sierpinski}
\end{equation}
and
\begin{equation}
T_{n+1} = \frac{ [T_n(\ell_n/3)]^2 \big(1 + (T_n(\ell_n/3) - 
R_n(\ell_n/3)) \exp{[ik\ell/3]}\big)\exp{[ik\ell/3]}}{D_n(\ell_n/3)},
\label{eq:Tn_sierpinski}
\end{equation}
for
\begin{equation}
D_{n}(L) \equiv \big( 
1 - (R_{n}(L) + T_{n}(L)) \exp[i k L] \big) \, 
\big(1 + T_n(L) \exp[i k L] + ([T_n(L)]^2 - [R_n(L)]^2) 
\exp[2 i k L] \big).
\end{equation}
Observe that the above equations correctly account for the reduction by
a factor three in the fundamental triangles $\Delta_{ABC}$ edges length
of the successive stages of the Sierpi\'{n}ski graph.

Setting $\ell = \ell_ 1 = 1$ and the same delta point interaction of strength
$\gamma$ at all the elementary vertices $A$, $B$, and $C$, we show in Fig.
\ref{fig:fig15} ($\gamma=0$) and Fig. \ref{fig:fig16} ($\gamma=1$), the
behavior of the reflection and transmission coefficients as function of $k$
for the Sierpi\'{n}ski graph stage $n$, up to $n=5$.
We notice that as $n$ increases, the system becomes more and more
selective to which $k$'s (or equivalently, energies) can be transmitted
through the structure.
This effect is stronger for $\gamma = 1$ (Fig. \ref{fig:fig16}) since
then the elementary $r$'s and $t$'s are also $k$-dependent.
So, in this respect the Sierpi\'{n}ski graph at the different stages
contrasts with the binary tree at different levels, Fig. \ref{fig:fig13},
for which there is not a such filter-like phenomenon.

\begin{figure}
  \begin{center}
    \includegraphics*[width=0.5\textwidth]{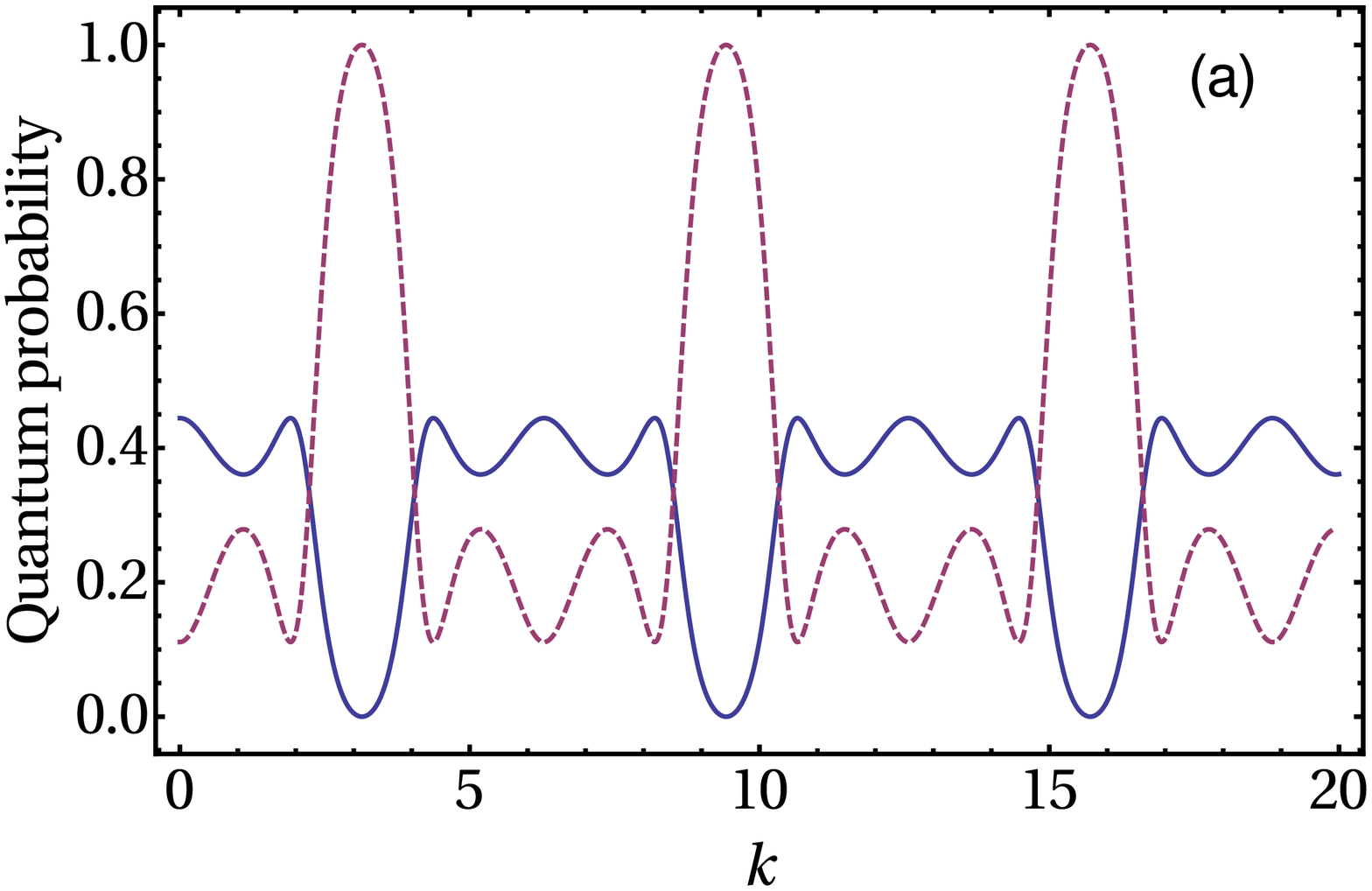}%
    \includegraphics*[width=0.5\textwidth]{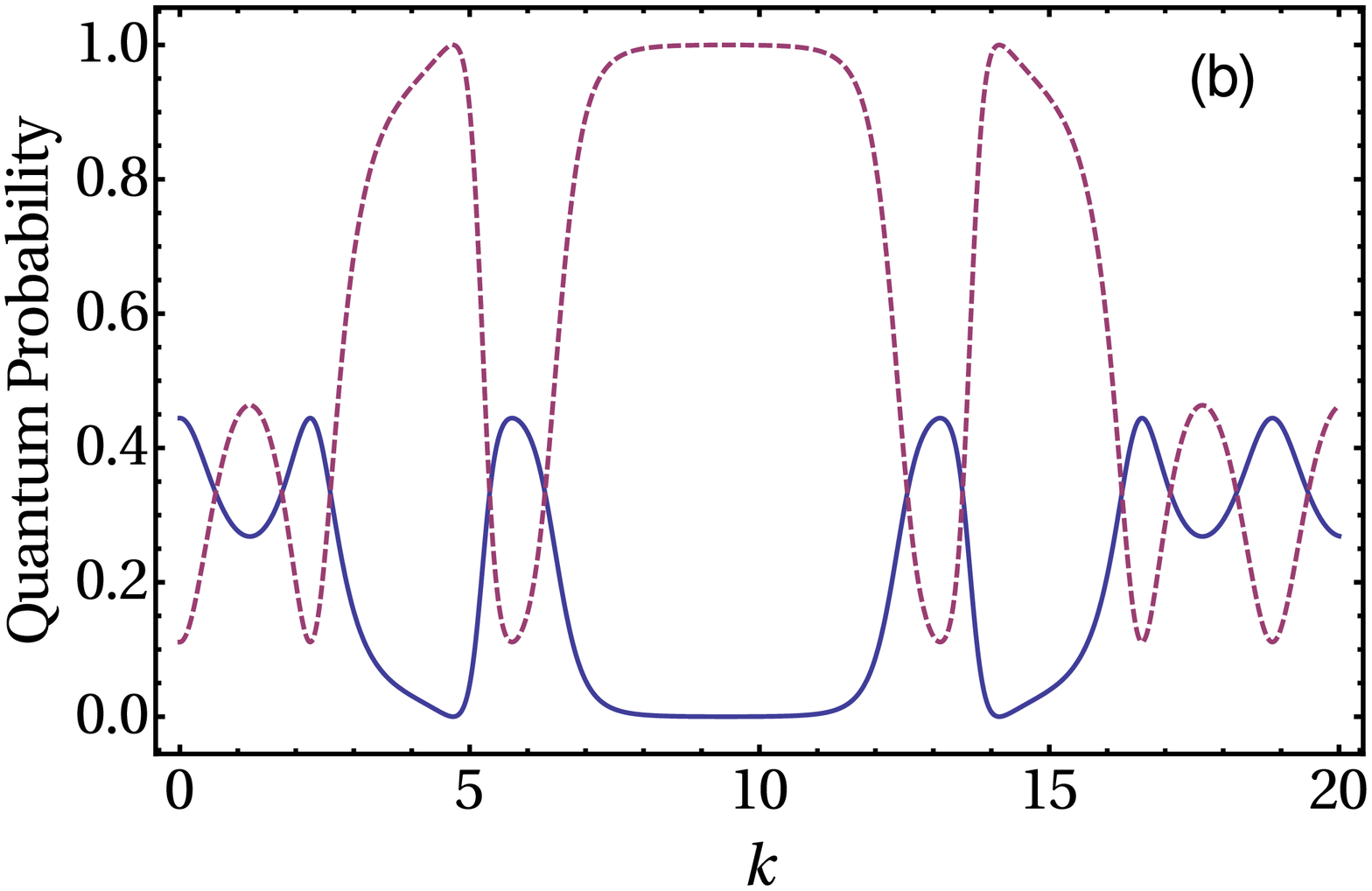}\\
    \includegraphics*[width=0.5\textwidth]{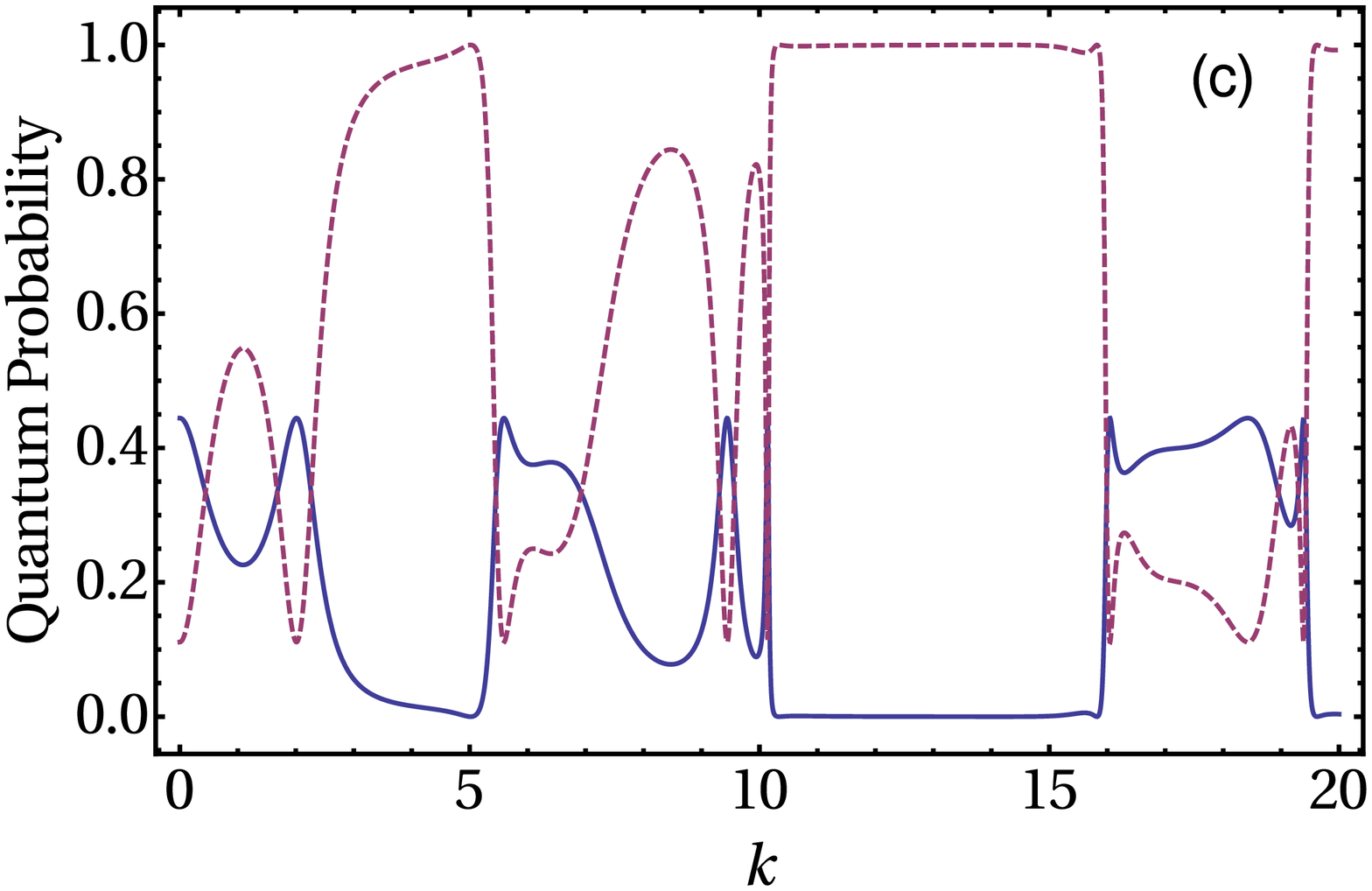}%
    \includegraphics*[width=0.5\textwidth]{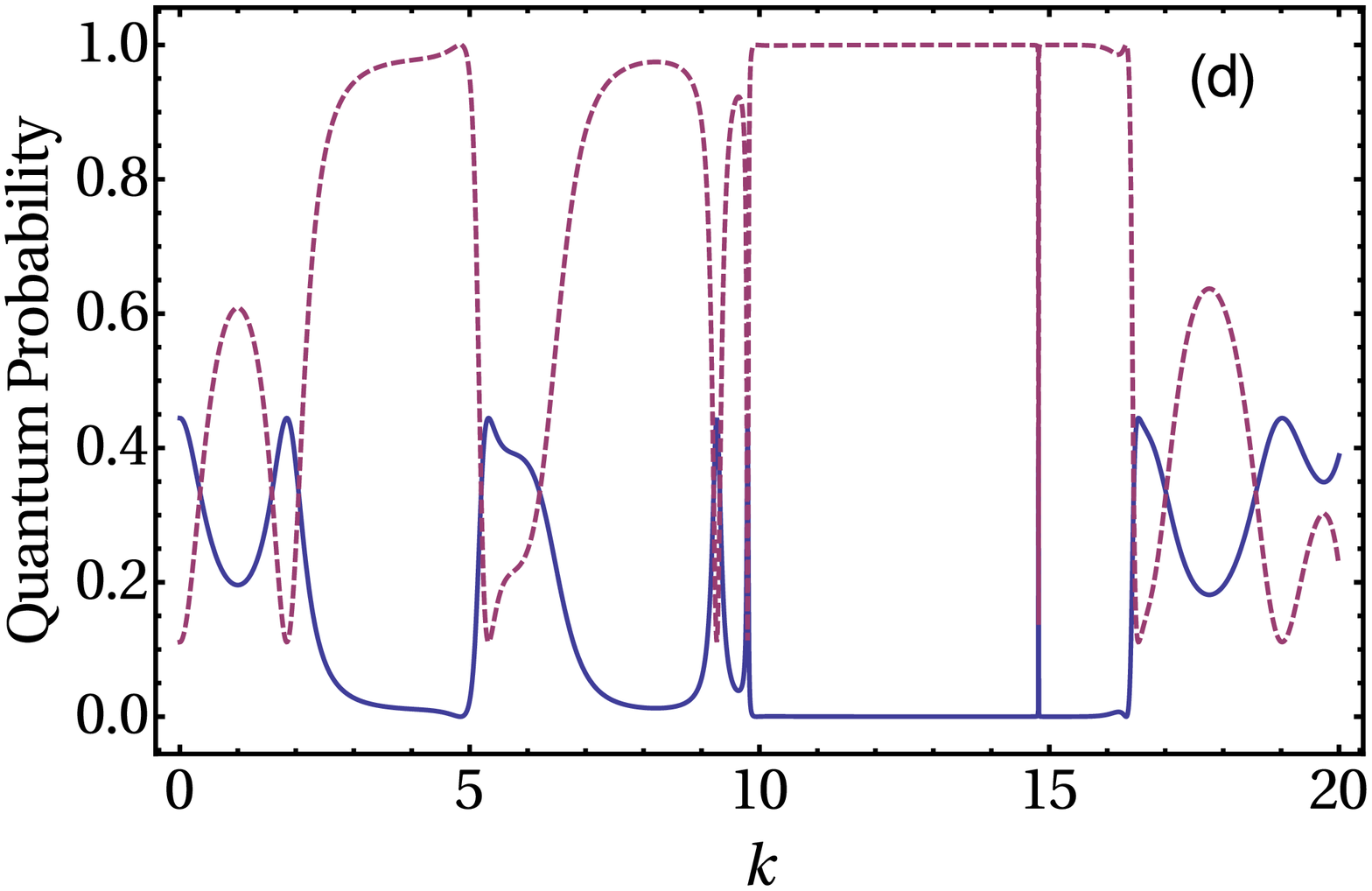}\\
    \includegraphics*[width=0.5\textwidth]{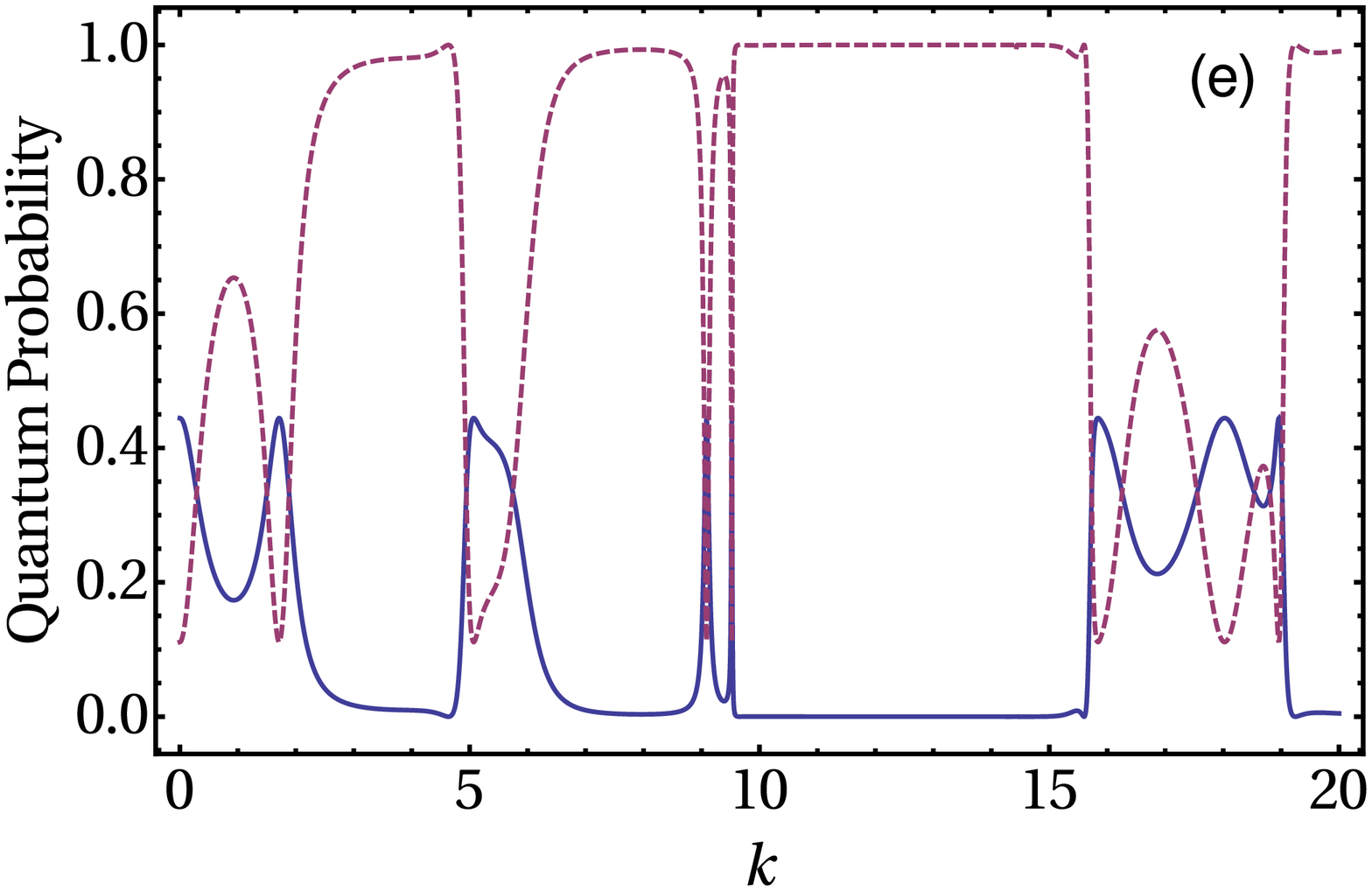}
  \end{center}
  \caption{\label{fig:fig15} (Color online).
    The transmission $|T_n|^2$ (solid line) and reflection $|R_n|^2$
    (dashed) probabilities for the stage $n$ of the Sierpi\'{n}ski
    graph, Fig. \ref{fig:fig14}.
    Here $\ell = \ell_1 = 1$ and at any elementary vertex $A$, $B$,
    and $C$, we assume a same generalized $\delta$ interaction of 
    strength $\gamma=0$.
    The cases $n=1$, $n=2$, $n=3$, $n=4$, and $n=5$, are displayed,
    respectively, in (a), (b), (c), (d) and (e).
  }
\end{figure}

\begin{figure}
  \begin{center}
    \includegraphics*[width=0.5\textwidth]{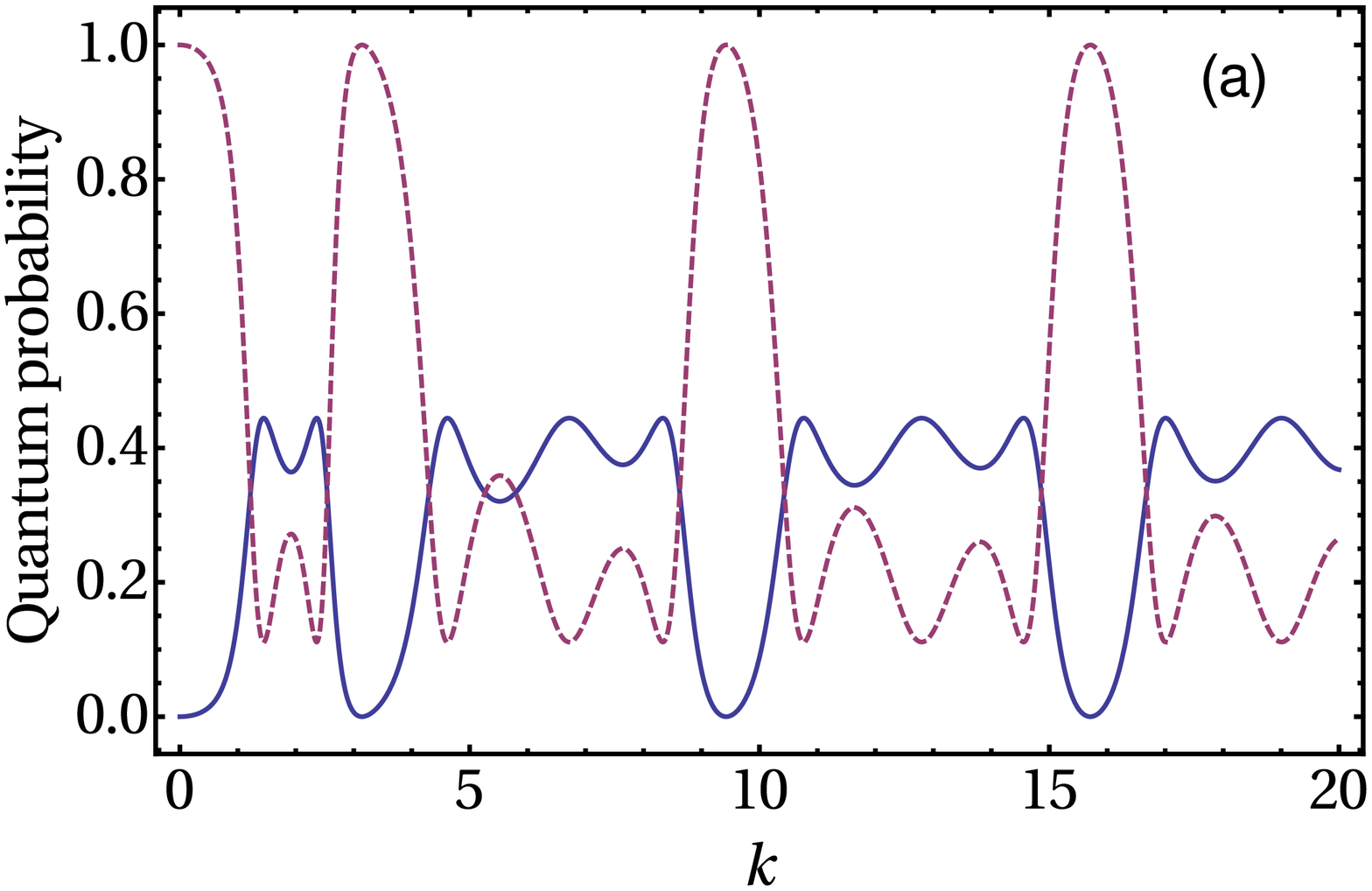}%
    \includegraphics*[width=0.5\textwidth]{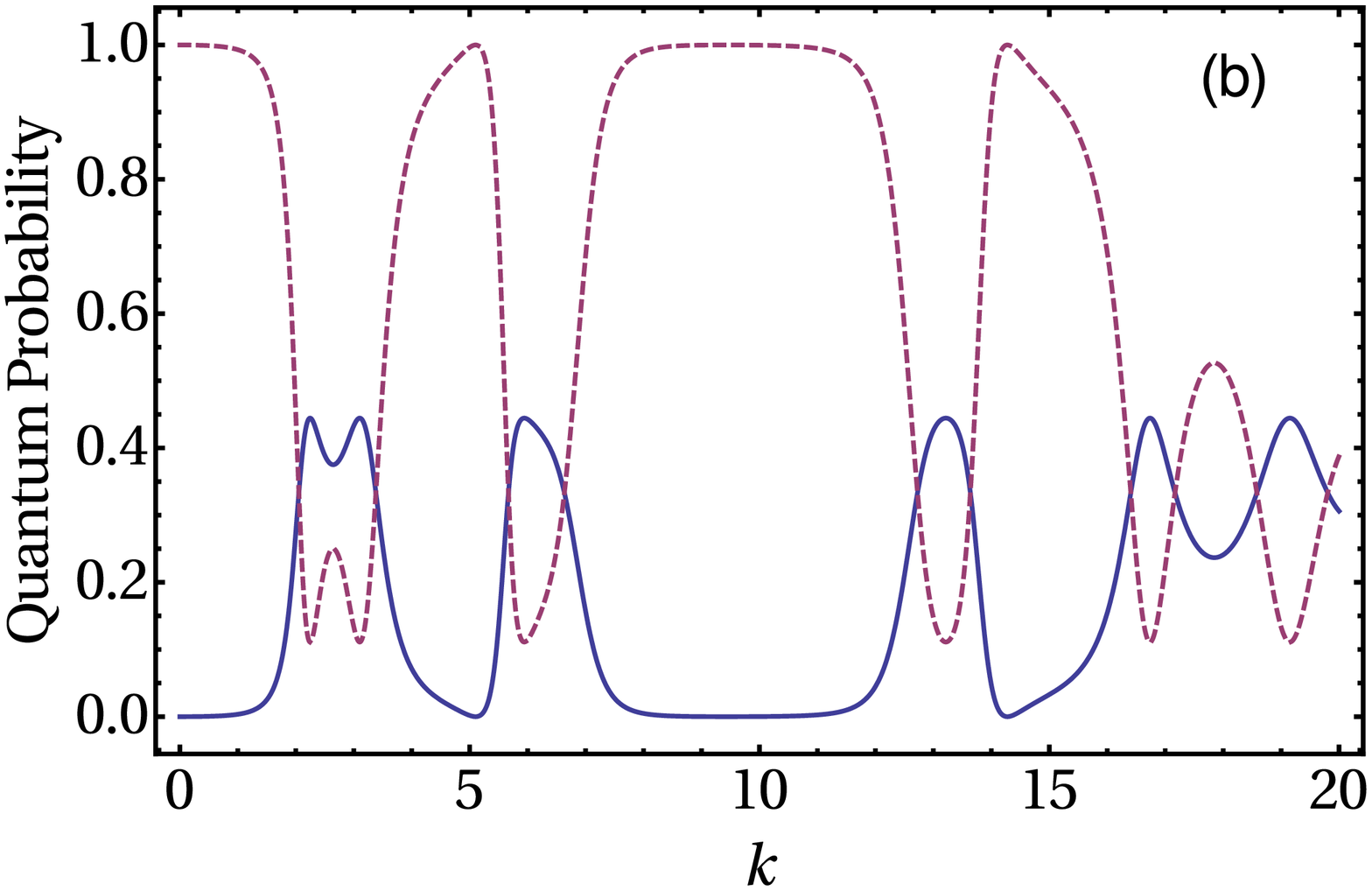}\\
    \includegraphics*[width=0.5\textwidth]{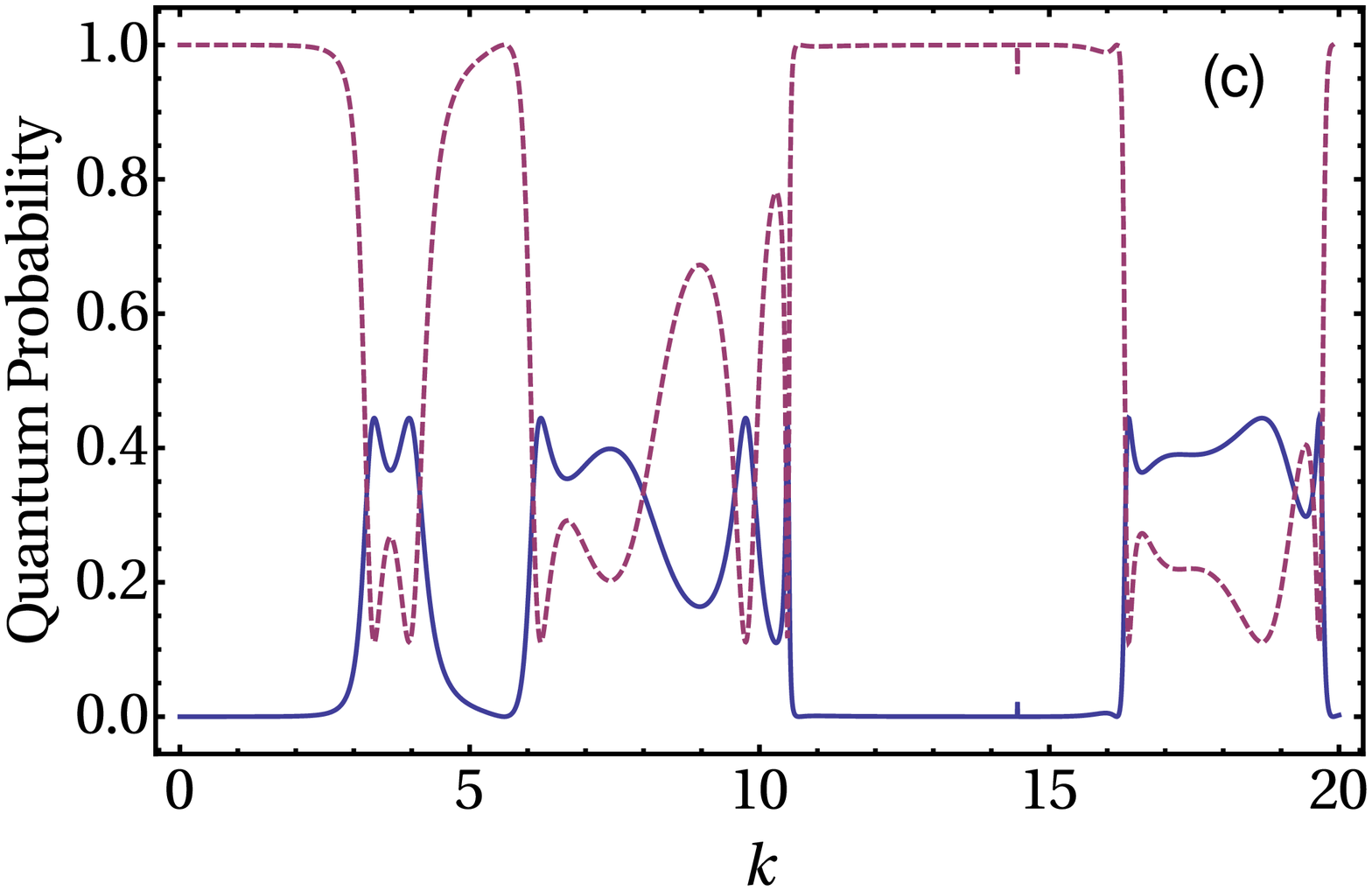}%
    \includegraphics*[width=0.5\textwidth]{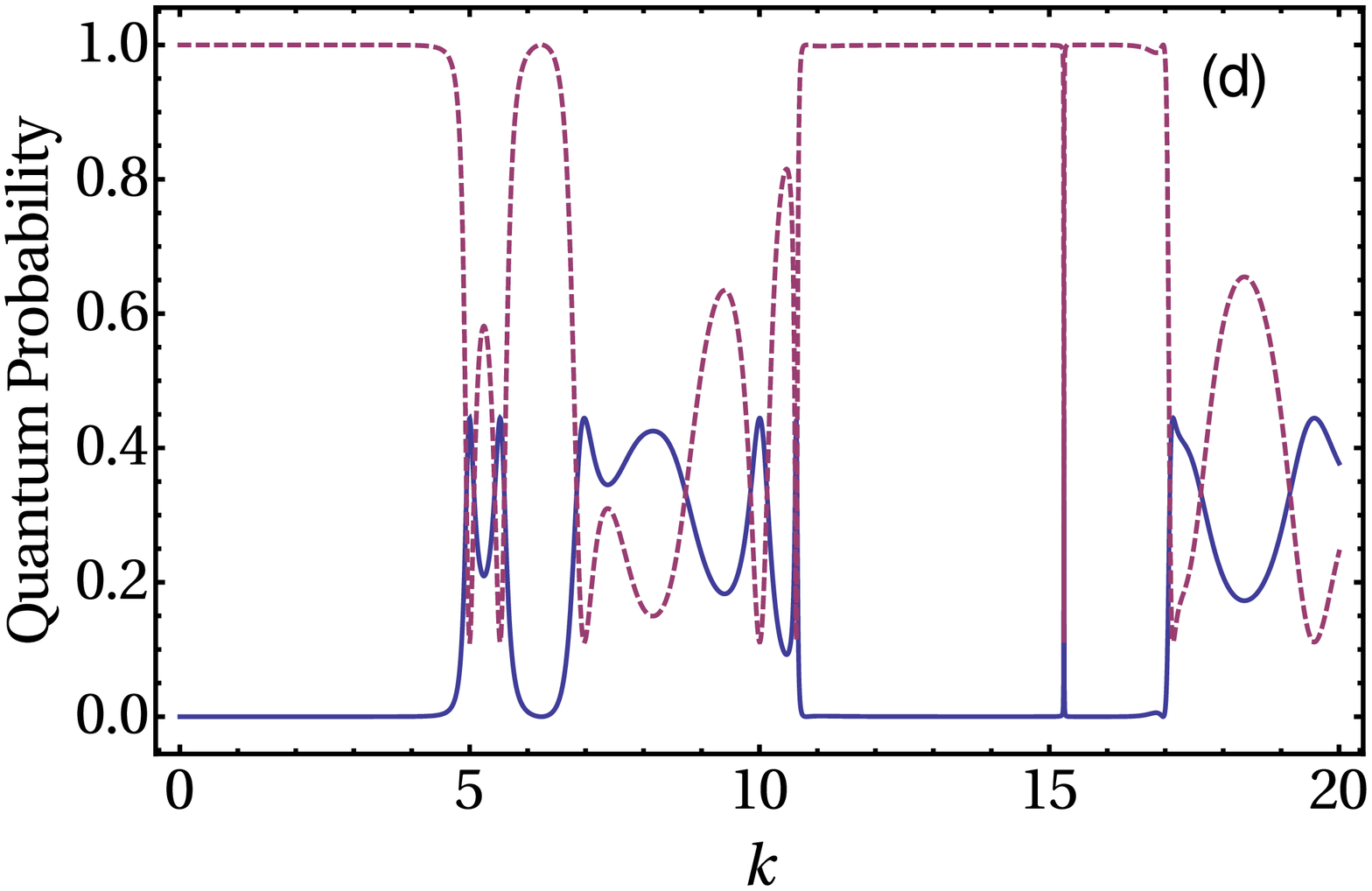}\\
    \includegraphics*[width=0.5\textwidth]{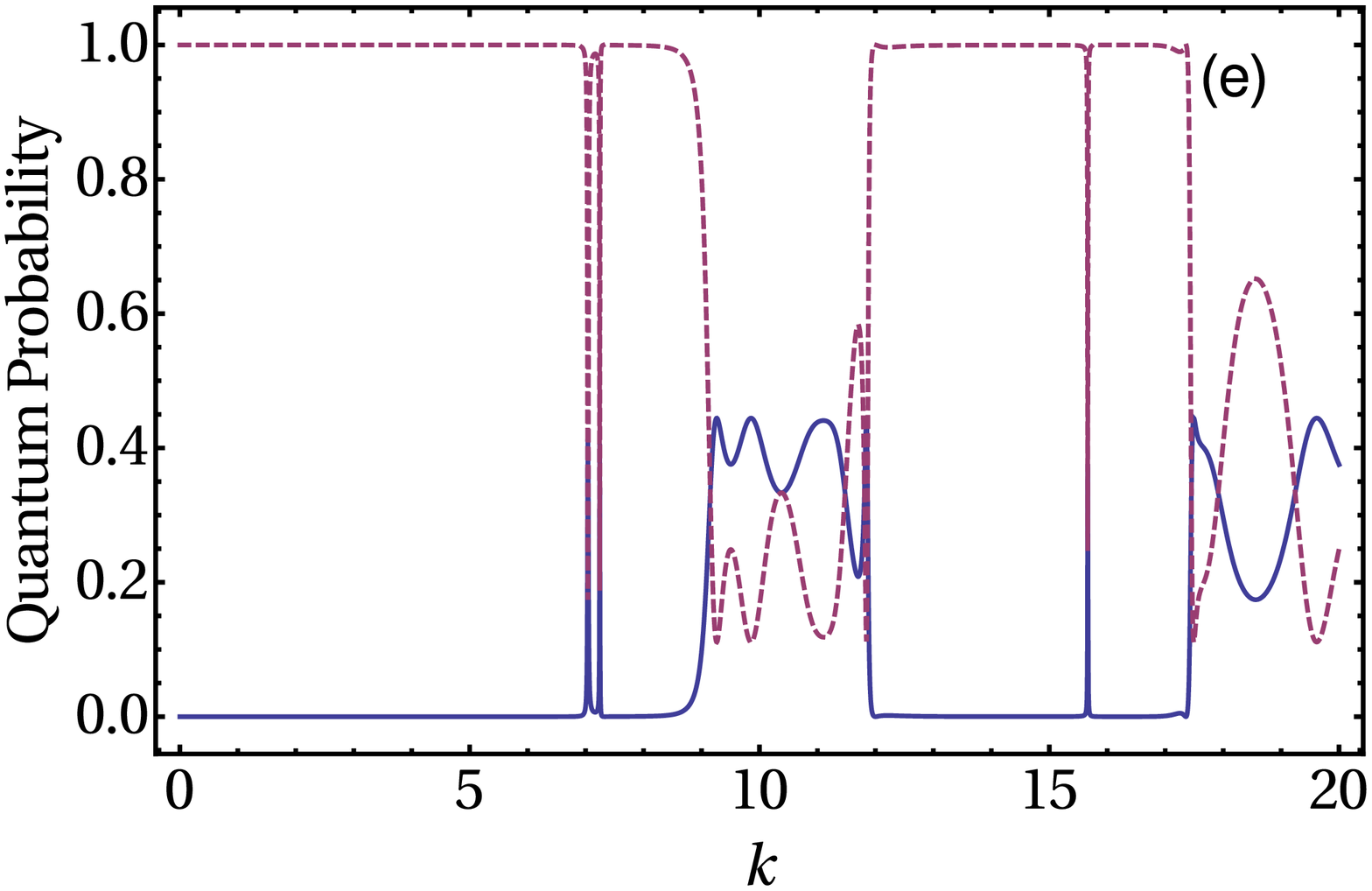}
  \end{center}
  \caption{\label{fig:fig16} (Color online).
    The same as in Fig. \ref{fig:fig15}, but for $\gamma=1$.
  }
\end{figure}

%%% Local Variables:
%%% mode: latex
%%% TeX-master: "green-qg-pr"
%%% ispell-local-dictionary: "american"
%%% End:
\section{Quasi-bound states in quantum graphs}
\label{sec:qbsqg}

\subsection{Basic aspects}

\begin{figure}
  \centering
  \includegraphics*[width=0.5\textwidth]{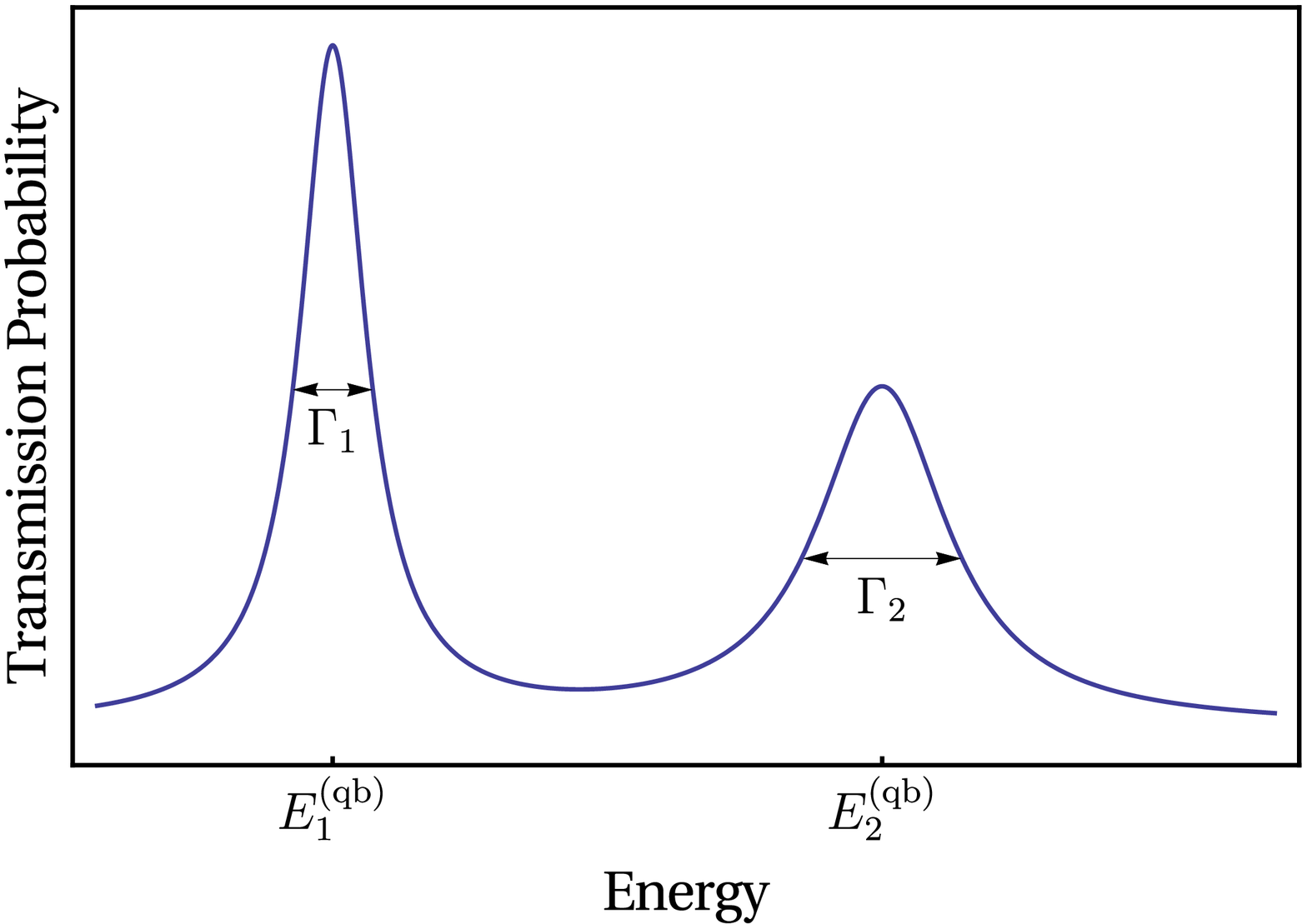}
  \caption{\label{fig:fig17} (Color online).
    Typical profile of the transmission probability as function of
    energy for a system displaying two quasi-bound states at
    energies $E_{1}^{(\rm qb)}$ and $E_{2}^{(\rm qb)}$.
    The quasi-bound states widths, here $\Gamma_1$ and $\Gamma_2$,
    are usually defined as the half height widths of the corresponding
    peaks.}
\end{figure}

As a last application for the Green's function approach reviewed so far,
we finally consider a context not usually addressed for the present quantum
systems (but see \cite{PLA.2014.378.1461}): quasi-bound states.
For a general treatment of such problems using $G$ -- however not discussing
quantum graphs -- we cite \cite{AA.2014.4.027109}.

In quantum mechanics, a quasi-bound state is a type of resonance,
associated to the geometry and (local) features of the system potential $V$.
Suppose a quantum particle of total energy $E = \hbar^2 k^2/(2 \mu)$, whose
value is assumed in a certain range $\Sigma_E$.
Also suppose a region ${\mathcal V}$ of the space in which $V$ is attractive
or has the generic shape of a well.
It might be that the potential cannot confine infinitely
the particle to ${\mathcal V}$.
In other words, for energies $E \in \Sigma_E$ the potential $V$ does not
support true bound states localized in ${\mathcal V}$.
However, for specific $E = E^{(\rm{qb})} \in \Sigma_E$, $V$ may be able to
trap the particle in ${\mathcal V}$ during a very long time
$\tau$ \cite{Book.1981.Landau}.
Such $\tau$ is called the lifetime of the quasi-bound state of energy
$E^{(\rm{qb})}$.

The concept of quasi-bound states is ubiquitous, and has been used
to explain a large number of phenomena.
For instance, tunneling ionization rates \cite{PRL.1998.81.2663},
diffraction in time \cite{PRA.2011.83.043608}, decay of cold atoms in
quasi-one-dimensional traps \cite{EPL.2006.74.965}, and certain
condensed-matter experiments \cite{PRL.2005.95.066801}, just to mention
few examples.

We begin our analysis with the simple linear quantum graph of Fig.
\ref{fig:fig3} (a), Sec. \ref{subsec:cuafggq}.
It is formed by two vertices, $A$ and $B$, joined together by an edge of
length $\ell_1$.
Each vertex is also attached to a semi-infinite lead.
Now, we take for the vertices delta interactions of a same strength $\gamma$.
If $\gamma \rightarrow + \infty$, then $r_A = r_B = -1$ and $t_A = t_B = 0$
(see Sec. \ref{sec:cesog}), which is equivalent to Dirichlet boundary
conditions at $A$ and $B$.
Then, the graph system becomes equivalent to an infinite square well.
In fact, for $k = k_n = n \pi/\ell_1$ with $n=1,2,3,\dots$ (so, for well
determined energies $E_n = n^2 \pi^2 \hbar^2/(2 \mu \ell_1^2)$), an
acceptable standing solution is
$\psi_1(x) = {\mathcal C} \, \sin[k_n x]$ along the edge and vanishing
$\psi$'s at the leads.
This is a proper stationary wave function of infinite lifetime\footnote{Note
that due to the Heisenberg uncertainty principle,
$\Delta E \, \Delta t \sim \hbar$, if the energy is exactly determined,
then $\Delta E = 0$ and the state lifetime is infinite once
$\Delta t \rightarrow + \infty$ \cite{Book.2011.Sakurai}.},
hence a genuine bound state (in the sense that these $\psi$'s are
(not-scattering) eigenstates of the problem Hamiltonian).

Further, if for this same graph we set arbitrary boundary conditions
resulting in non-zero transmission amplitude at least for one of the two
vertices, the quantum particle initially localized in the edge cannot
remain there, eventually it will escape due to tunneling.
But as explained above, embedded in the continuous spectrum of $k$ there
may exist a discrete set of values $k_n^{(\rm{qb})}$ corresponding to the
quasi-energies $E_n^{(\rm{qb})} = \hbar^2 {k_n^{(\rm{qb})}}^2/(2 \mu)$ of widths
$\Gamma_n = \hbar/\tau_n$ \cite{Book.1998.Merzbacher}.
A direct way to determine these $k_n^{(\rm{qb})}$'s is through a
scattering approach.
Defining transmission $T(k)$ and reflection $R(k)$ amplitudes for the
relevant ${\mathcal V}$ region (for contexts where only $R$ is defined,
see below), it is a well known fact \cite{Book.2011.Sakurai} that $|T(k)|^2$
exhibits a pronounced peak for $k$ around $k_n^{(\rm{qb})}$.
Moreover, the $\Gamma$'s are given by the half height width of the
corresponding peaks.
Such behavior is schematically illustrated in Fig. \ref{fig:fig17} (and
also concretely observed in some examples in the previous Secs.).

Finally, to frame the problem in terms of the Green's function formalism,
we address $G$ for the graph of Fig. \ref{fig:fig3} (a) with both $x_f$ and
$x_i$ in lead $i$.
Also, to illustrate the situation one can define only a reflection
coefficient for the region ${\mathcal V}$ (see next), we assume for vertex
$B$ boundary conditions leading to a zero transmission amplitude, i.e.,
the reflection probability from vertex $B$ is exactly 1.
In this way we can generally write $r_B = \exp[i \phi_B]$, for $\phi_B(k)$
a wavenumber dependent phase \cite{JOB.7.S77.2005}.
For $A$, we consider arbitrary boundary condition corresponding to generic
$r_A$ and $t_A$.
Note then that the global transmission amplitude $i \rightarrow f$
(crossing $A$--$B$) must be zero because $t_B = 0$.
Hence, any manifestation of a quasi-bound state should be
identified in the phase of $R_{i i}(k) = \exp[i \phi_R(k)]$.

Following the convention that $0 \leq x < + \infty$ in lead $i$ (with the
origin at $A$), we have
\begin{equation}
G_{i i}(x_f, x_i;k) = \frac{\mu}{i \hbar^2 k} \left(
  \exp[i k |x_f - x_i|] + R_{i i}(k) \, \exp{[i k (x_f + x_i)]}\right),
\label{greenpotl}
\end{equation}
where $R_{i i}$ is easily derived from the previous sums over paths
construction, or (already setting $r_B = \exp[i \phi_B]$)
\begin{equation}
  R_{i i}(k) = r_{A}^{(i)}(k) +
  \frac{t_{A}^{(1,i)}(k) \, t_{A}^{(i,1)}(k) \,
    \exp{[i \, (2 k\ell_{1} + \phi_B(k))]}}
       {1 - r_ {A}^{(1)}(k) \, \exp{[i \, (2 k\ell_{1} + \phi_B(k))]}}.
\label{eq:R-qb}
\end{equation}
Using the relations in Eq. (\ref{eq:rt-relations}) for the vertex $A$
quantum amplitudes $r_A$ and $t_A$, it is a little tedious but
straightforward to prove that
$R_{i i} \, {R_{i i}}^{*} = 1$.
So, as previously mentioned we can write $R_{i i}(k) = \exp[i \phi_R(k)]$,
with $\phi_R(k)$ coming from Eq. (\ref{eq:R-qb}).

The natural question now is how to characterize a quasi-bound state from
the function $\phi_R(k)$.
This is a textbook analysis \cite{Book.1998.Merzbacher}, but answered next
by means of a very simple heuristic argument.
The system wave function, with $x$ in lead $i$, is (for
${\mathcal N}$ a proper normalization constant)
\begin{equation}
  \psi(x) = {\mathcal N} \,
  \Big\{\exp{[-i k x]} + R_{i i}(k) \, \exp{[+ i k x]} \, \Big\} =
 {\mathcal N} \,
  \Big\{\exp{[-i k x]} + \exp{[+ i \, (k x + \phi_R(k))]} \, \Big\}.
\label{wavpotl}
\end{equation}
It represents the scattering process of plane wave incoming from lead
$i$, being scattered at the graph region $A$--$B$, and then being reflected
back to lead $i$.
Observe that if for a $k = k^{(\rm{qb})}$
\begin{equation}
  \phi_R(k^{(\rm{qb})}) = (2 m + 1) \, \pi,
  \label{eq:phase-cond}
\end{equation}
with $m=0,1,\ldots$, Eq. (\ref{wavpotl}) yields
$\psi(x) \propto \sin[k^{(\rm{qb})} \, x]$.
Although here not a real bound state, this is exactly the sine-type of solution
for the edge region -- thus similar to a stationary standing wave -- in
the already discussed case the graph is equivalent to an infinite square
well.
Therefore, the quasi-bound wavenumber must be those $k = k^{(\rm{qb})}$
verifying  Eq. (\ref{eq:phase-cond}).
The quasi-bound state width is related to a $\Delta k$ around
$k^{(\rm{qb})}$ for which $\phi_R$ mod $2 \pi$ is close enough to 
$\pi$.

At this point, it should be clear the benefits of the Green's function
method to treat quantum graphs quasi-bound states.
On the one hand, the behavior of transmission and reflection probabilities
is a direct route to determine the quasi-bound energies.
On the other hand, the Green's function is a very appropriate tool to
calculate such quantities, especially for involving topologies.
Furthermore, $G$ can be used to obtain transition amplitudes {\em to} and
{\em from} specific parts of a graph, allowing a precise selection
of the region of interest ${\mathcal V}$.
In the following we will discuss recurrence protocols to calculate
global $R$ and $T$ for different quantum graphs, also illustrating how
to identify the quasi-bound states from such expressions.
We should mention that most of the procedures explained in details below
have been developed with distinct purposes in different previous works 
\cite{JPA.1998.31.2975,PRA.2002.66.062712,JPA.2003.36.545,PRA.2011.84.042343,PLA.2014.378.1461} and are somehow related to the general idea of the 
transfer matrix method \cite{PR.513.191.2012}.

\subsection{Recurrence formulas for the reflection and transmission
coefficients}
\label{sec:7-2}

Next we discuss the derivation of recurrence formulas for the quantum 
graphs global transmission and reflection amplitudes by means of the 
present sum over scattering paths technique.
For convenience, in the following we address only linear graphs (for the
more general case, see Sec. \ref{sec:7-4}).
% and use slight different notation and coordinate convention than 
% those in the previous Sections.

So, consider the linear open quantum graph in Fig. \ref{fig:fig18},
composed by a left semi-infinite lead $i$ and vertices named 
$l \ (=1, 2, \ldots, N)$.
Along the lead, the spatial coordinate $x$ ranges from $+\infty$ to 0 
(with the origin at the vertex 1).
For the edge $e_l$ (between vertices $l$ and $l+1$), $x$ goes from
0 (at vertex $l$) to $\ell_l$ (at vertex $l+1$). 

From the simplification procedures of Sec. \ref{sec:simplification}, we
can get the Green's function for the case where $x_i$ is in the
lead $i$ and $x_f$ is in the edge $e_l$ (see Fig. \ref{fig:fig18}) as
\begin{align}
  G_{l i}(x_f,x_i;k) = {} & \frac{\mu}{i\hbar^2k}
  \frac{T_{(1,l)}^{(+)}}
       {\left(
        1 - R_{(1,l)}^{(-)} \, R_{(l+1,N)}^{(+)} \, \exp{[2 i k \, \ell_l]}
       \right)}
       \Big(
       \exp[i k \, (x_{f} + x_{i})]
       + R_{(l+1,N)}^{(+)} \,
           \exp[ik \, (2 \ell_l - x_{f} + x_{i})]
       \Big).
  \label{eq:green_qe}
\end{align}
In the above, for $l_b \geq l_a$, the subscript ($l_a,l_b$) indicating the 
full block of vertices and edges from $l_{a}$ to $l_{b}$, and the superscript
$(+/-)$ meaning incoming from the left/right, then $T_{(l_a,l_b)}^{(\pm)}$
($R_{(l_a,l_b)}^{(\pm)}$) represents the global transmission (reflection)
coefficient across (from) such $l_a$---$l_b$ graph block.
Note that $T_{l,l}^{(\pm)} = t_l^{(\pm)}$ and $R_{l,l}^{(\pm)} = r_l^{(\pm)}$,
for $t_l$ and $r_l$ the quantum amplitudes of the individual vertex $l$.

\begin{figure}
  \centering
  \includegraphics*[width=0.5\textwidth]{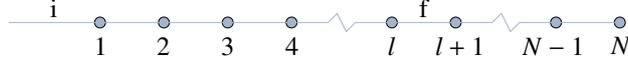}
  \caption{\label{fig:fig18} (Color online).
    A linear graph composed by a semi-infinite lead $i$ (at the left) 
    attached to a series of $N$ simply connected vertices.
    This structure allows quasi-bound states.}
\end{figure}

\begin{figure}
  \centering
  \includegraphics*[width=0.5\textwidth]{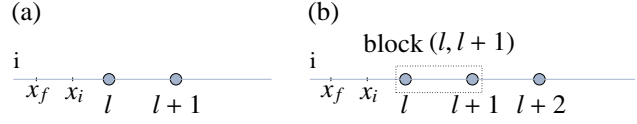}
  \caption{\label{fig:fig19} (Color online).
    Linear graphs with (a) two and (b) three simply connected
    vertices attached to left and right leads.
    In (b) it is exemplified the construction process of a block structure.}
\end{figure}

These $T_{(l_a,l_b)}^{(\pm)}$ and $R_{(l_a,l_b)}^{(\pm)}$ are recursively obtained 
in terms of the reflection and transmission coefficients of each individual 
vertex.
To see how, consider the graph composed of two vertices, $l$ and $l+1$,
an edge $e_l$, and two, left and right, leads.
We also assume both $x_i, x_f$  in the left lead, Fig. \ref{fig:fig19} (a).
Performing the sum over all scattering paths, the Green's function for
the graph in Fig. \ref{fig:fig19} (a) reads
\begin{align}
  G_{i i}(x_f,x_i;k) = {}
  &
    \frac{\mu}{i\hbar^2k}
    \Biggl(
    \exp[ik |x_f - x_i|] +
    r^{(+)}_{l} \, \exp[i k (x_f + x_i)] \nonumber \\
& + \frac{t^{(+)}_{l} \, r^{(+)}_{l+1} \, t^{(-)}_{l} \,
\exp{[2 i k \ell_l]}}
    {1 - r^{(-)}_{l} \, r^{(+)}_{l+1} \, \exp[2 i k \ell_l]}
    \exp{[i k (x_f + x_i)]}
    \Biggr).
\end{align}

From the above expression it is easy to identify a global reflection
coefficient from the left of block $(l,l+1)$, Fig. \ref{fig:fig19} (b), or
\begin{equation}
R^{(+)}_{(l,l+1)} = r^{(+)}_{l} +
\frac{t^{(+)}_{l} \, r^{(+)}_{l+1} \, t^{(-)}_{l} \, \exp[2 i k \ell_l]}
{1 - r^{(-)}_{l} \, r^{(+)}_{l+1} \, \exp{[2 i k \ell_l]}}.
\label{eq:r_doisv}
\end{equation}
Similarly, calculating $G$ for $x_i, x_f$ in the right lead, we also can 
identify a global reflection coefficient from the right of this same 
block, given by
\begin{equation}
R^{(-)}_{(l,l+1)} = r^{(-)}_{l+1} +
\frac{t^{(-)}_{l+1} \, r^{(-)}_{l} \, t^{(+)}_{l+1} \, \exp[2 i k \ell_l]}
{1 - r^{(-)}_{l} \, r^{(+)}_{l+1} \, \exp{[2 i k \ell_l]}}.
\label{eq:r_doisve}
\end{equation}

Now, considering the case in which $x_i$ ($x_f$) is in the left (right) 
lead, then
\begin{equation}
  G_{f i}(x_f,x_i;k) =
  \frac{\mu}{i\hbar^2k}
  \frac{t^{(+)}_{l} \, t^{(+)}_{l+1} \, \exp[i k \ell_l]}
{\left(1 - r^{(-)}_{l} \, r^{(+)}_{l+1} \, \exp{[2 i k \ell_l]}\right)}
\exp[ik (x_f + x_i)],
\end{equation}
naturally yielding 
\begin{equation}
  T^{(+)}_{(l,l+1)} = \frac{t^{(+)}_{l} \, t^{(+)}_{l+1} \, 
    \exp{[i k \ell_l]}}
  {1 - r^{(-)}_{l} \, r^{(+)}_{l+1} \, \exp{[2 i k \ell_l]}}.
\label{eq:t_doisv}
\end{equation}
Finally, from $G$ for $x_i$ ($x_f$) in the right (left) lead,
one finds
\begin{equation}
  T^{(-)}_{(l,l+1)} = \frac{t^{(-)}_{l} \, t^{(-)}_{l+1} \, \exp{[i k \ell_l]}}
  {1 - r^{(-)}_{l} \, r^{(+)}_{l+1} \, \exp{[2 i k \ell_l]}}.
\label{eq:t_doisve}
\end{equation}

With proper substitutions, the above Eqs. 
(\ref{eq:r_doisv}), (\ref{eq:r_doisve}), 
(\ref{eq:t_doisv}), and (\ref{eq:t_doisve}) constitute then the 
basic generating expressions to obtain $R$ and $T$ 
for an arbitrary number of vertices in a linear graph.
To exemplify this, let us assume a third vertex $l+2$, as shown in
Fig. \ref{fig:fig19} (b).
For $x_i, x_f$ in the left lead, we can suppose $l$---$(l+1)$ forming a block 
of coefficients $R_{(l,l+1)}^{(\pm)}$ and $T_{(l,l+1)}^{(\pm)}$ (see
Fig. \ref{fig:fig19} (b)).
Hence, by mapping the vertex $l$, the vertex ${l+1}$ and the edge $e_l$ 
of Fig. \ref{fig:fig19} (a) into, respectively, the $l$---$(l+1)$ block, the
vertex $l+2$, and the edge $e_{l+1}$ of Fig. \ref{fig:fig19} (b), we can
directly infer from Eq. \eqref{eq:r_doisv} that
\begin{equation}
\label{eq:r_tresv}
R^{(+)}_{(l,l+2)} =
R^{(+)}_{(l,l+1)} +
\frac{T^{(+)}_{(l,l+1)} \, r^{(+)}_{l+2} \, T^{(-)}_{(l,l+1)} \, 
\exp[2 i k \ell_{l+1}]}
{1 - R^{(-)}_{(l,l+1)} \, r^{(+)}_{l+2} \, \exp[2 i k \ell_{l+1}]}.
\end{equation}

To close, based on the previous examples, one can readily generalize
the above results for a block $(l,l+n)$ of $n+1$ vertices, obtaining the 
following recursive relations
\begin{equation}
R^{(+)}_{(l,l+n)} = R^{(+)}_{(l,l+n-1)} +
\frac{T^{(+)}_{(l,l+n-1)} \, r^{(+)}_{l+n} \, T^{(-)}_{(l,l+n-1)} \, 
\exp[2 i k \ell_{l+n-1}]}
{1 - R^{(-)}_{(l,l+n-1)} \, r^{(+)}_{l+n} \, \exp[2 i k \ell_{l+n-1}]},
\end{equation}
\begin{equation}
R^{(-)}_{(l,l+n)} = r^{(-)}_{l+n} +
\frac{t^{(-)}_{l+n} \, R^{(-)}_{(l,l+n-1)} \, t^{(+)}_{l+n}
\, \exp[2 i k\ell_{l+n-1}]}
{1- R^{(-)}_{(l,l+n-1)} \, r^{(+)}_{l+n} \exp[2 i k \ell_{l+n-1}]},
\end{equation}
\begin{equation}
T^{(\pm)}_{(l,l+n)} =
\frac{T^{(\pm)}_{(l,l+n-1)} \, t^{(\pm)}_{l+n} \exp[i k \ell_{l+n-1}]}
{1- R^{(-)}_{(l,l+n-1)} \, r^{(+)}_{l+n} \exp[2 i k \ell_{l+n-1}]},
\end{equation}

\subsection{Green's function as a transition probability amplitude and
the determination of quasi-bound states}
\label{sec:7-3}

Once we now know the recurrence formulas for the scattering coefficients
of a linear quantum graph, we can return to the Green's function in Eq.
\eqref{eq:green_qe}.
But first we shall recall that $G(x_f, x_i;k)$ can be generally interpreted 
as the transition probability amplitude for a particle (of fixed energy 
$E = \hbar^2 k^2/(2 \mu)$) initially in $x_i$ to get to $x_f$ 
\cite{Book.2005.Schulman}.
Thus, the overall multiplicative term in Eq. \eqref{eq:green_qe},
namely,
\begin{equation}
\mathcal{A}_{i,l}(k) = \frac{T_{1,l}^{(+)}(k)}
  {1 - R_{1,l}^{(-)}(k) \, R_{l+1,N}^{(+)}(k) \, \exp{[2 i k \ell_{j}]}},
\label{eq:amp_quase_estados}
\end{equation}
represents the probability amplitude for a particle (of wavenumber $k$)
to leave the left semi-infinite lead $i$ and to tunnel to the edge $e_l$.

So, if the graph supports a quasi-bound state totally or partially  
localized in $e_l$, an incident wave (from lead $i$) with $k$ close to 
the corresponding quasi-bound state $k^{(\rm{qb})}$ value should have a very 
high probability to be transmitted to the edge $e_l$ region.
In this way, the plot of $|\mathcal{A}_{i,l}|^2$ as function of $k$ (or 
likewise of $E$) should display peaks\footnote{Here we mention a minor 
technical point.
Differently from $|R|^2$ and $|T|^2$, the quantity $|\mathcal{A}|^2$ is
not normalized to one.
However, this is not a problem since we are only concerned with the
quasi-energies locations and their widths.
So, the peaks actual heights are not relevant (unless for comparative 
purposes between distinct $E^{(\rm{qb})}$'s).}
centered at the correct $E^{(\rm qb)}$'s, as schematically depicted in Fig. 
\ref{fig:fig17}.
Moreover, such peaks widths at half height would correspond to the
$\Gamma$'s.

\begin{figure}
  \centering
  \includegraphics*[width=0.45\textwidth]{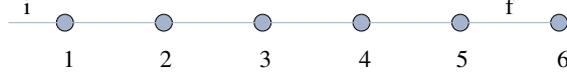}
  \caption{\label{fig:fig20} (Color online).
    The graph of Fig. \ref{fig:fig18} for $N=6$ and all edges of the
    same length $\ell_l = \ell$.
    The vertex 6 is the system `dead end', for which is assumed either 
    Dirichlet or Neumann boundary conditions.
    The other vertices are delta interactions of strength $\gamma$.}
\end{figure}

As an example, consider the linear open graph with six vertices of 
Fig. \ref{fig:fig20}, where the last vertex 6 is a `dead end'.
We suppose for all edges $\ell_l = \ell = 1$ and for the vertices 1 to 5 
generalized $\delta$ interactions of a same strength $\gamma$.
However, for vertex 6 we assume either Dirichlet (so $r_6^{(+)} = -1$)
or Neumann (so $r_6^{(+)} = +1$) boundary conditions.
For two values of the delta intensity, $\gamma = 1$ and $\gamma = 2$,
and for $l$ varying from 1 to 5, we plot in Figs. \ref{fig:fig21} and 
\ref{fig:fig22} the quantity $|\mathcal{A}_{i,l}|^2$ as function of $k$ for,
respectively, the Dirichlet and Neumann boundary conditions at vertex 6.

\begin{figure}
  \centering
    \includegraphics*[width=0.5\textwidth]{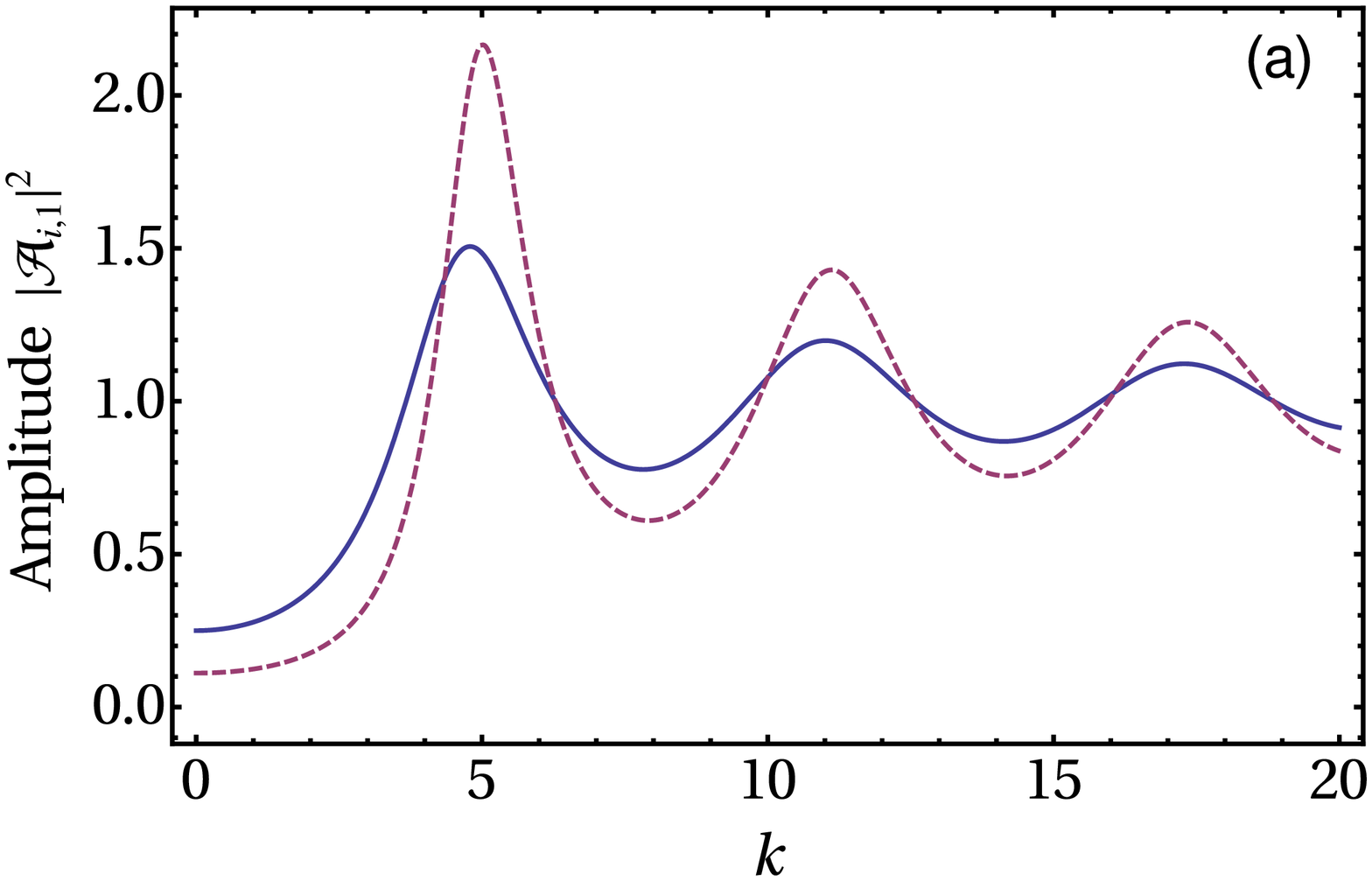}%
    \includegraphics*[width=0.5\textwidth]{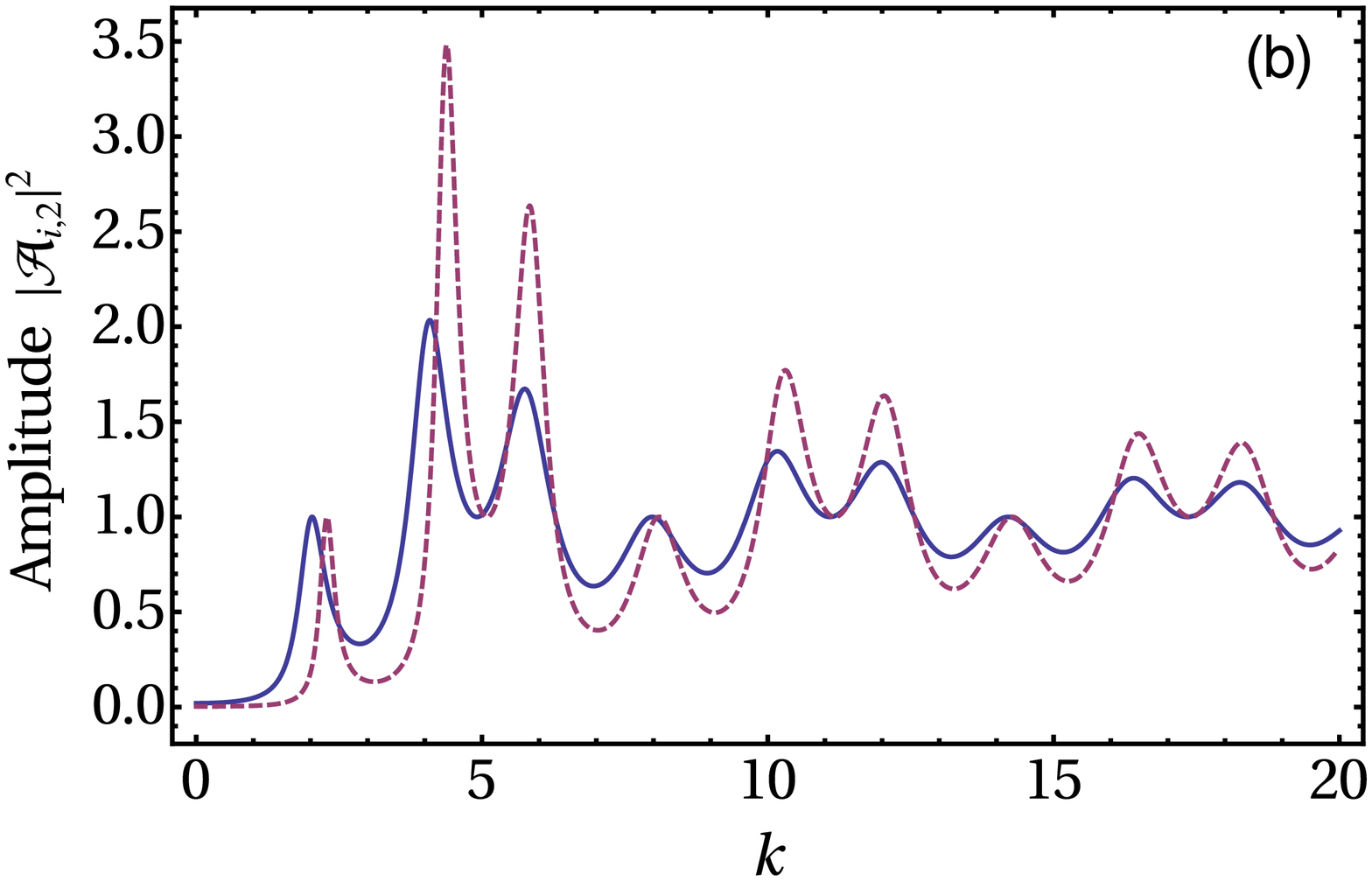}\\
    \includegraphics*[width=0.5\textwidth]{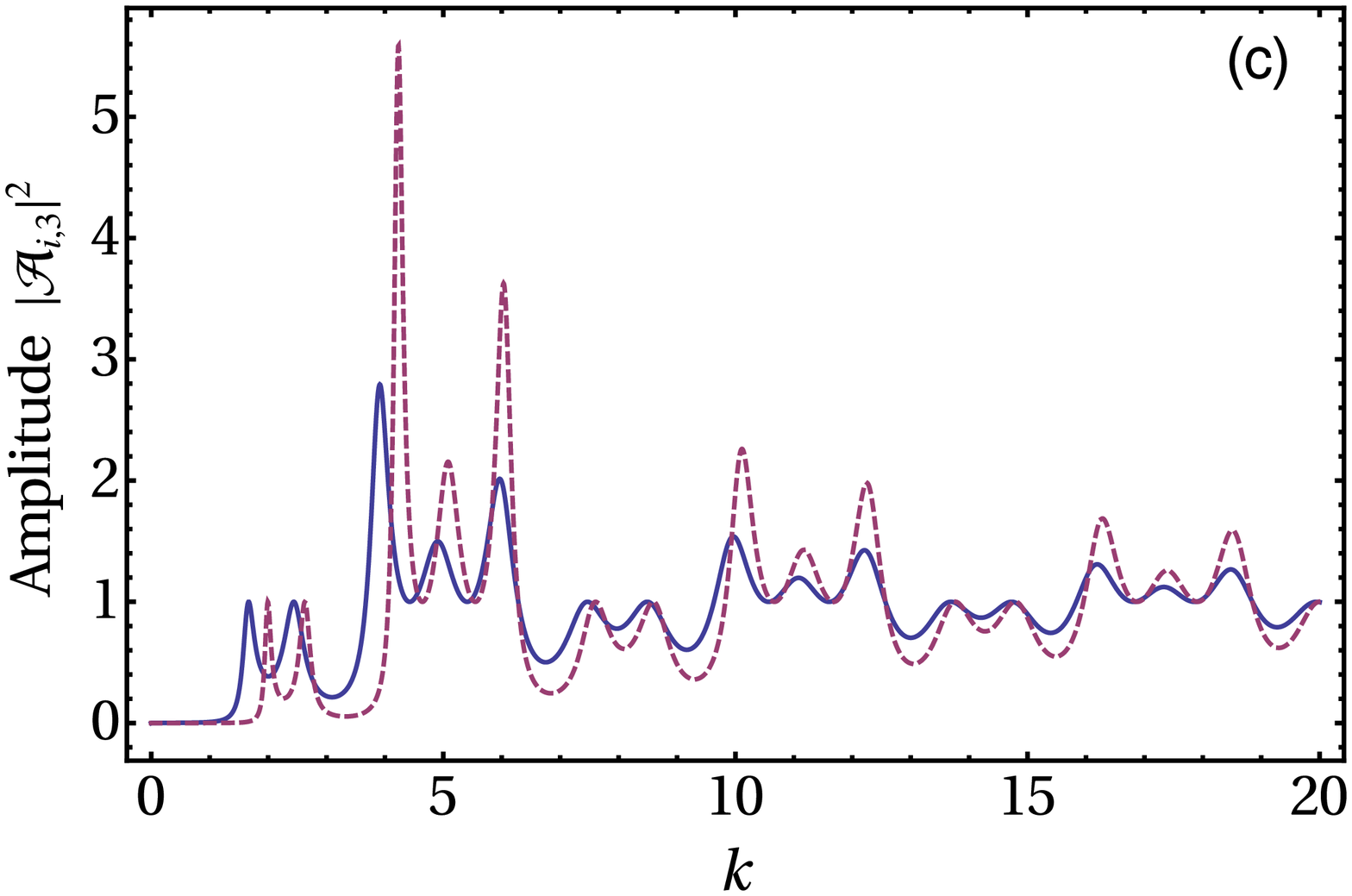}%
    \includegraphics*[width=0.5\textwidth]{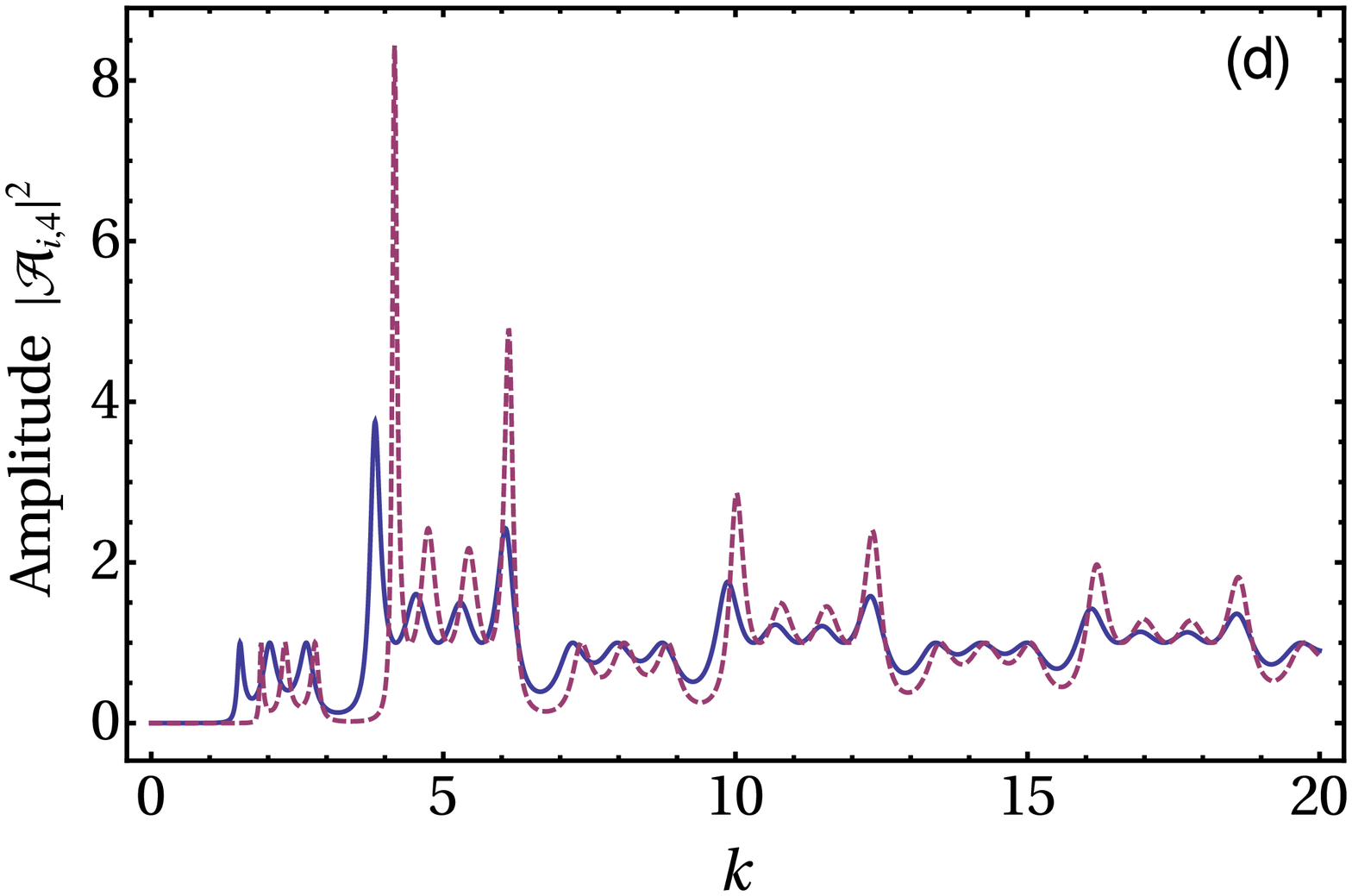}\\
    \includegraphics*[width=0.5\textwidth]{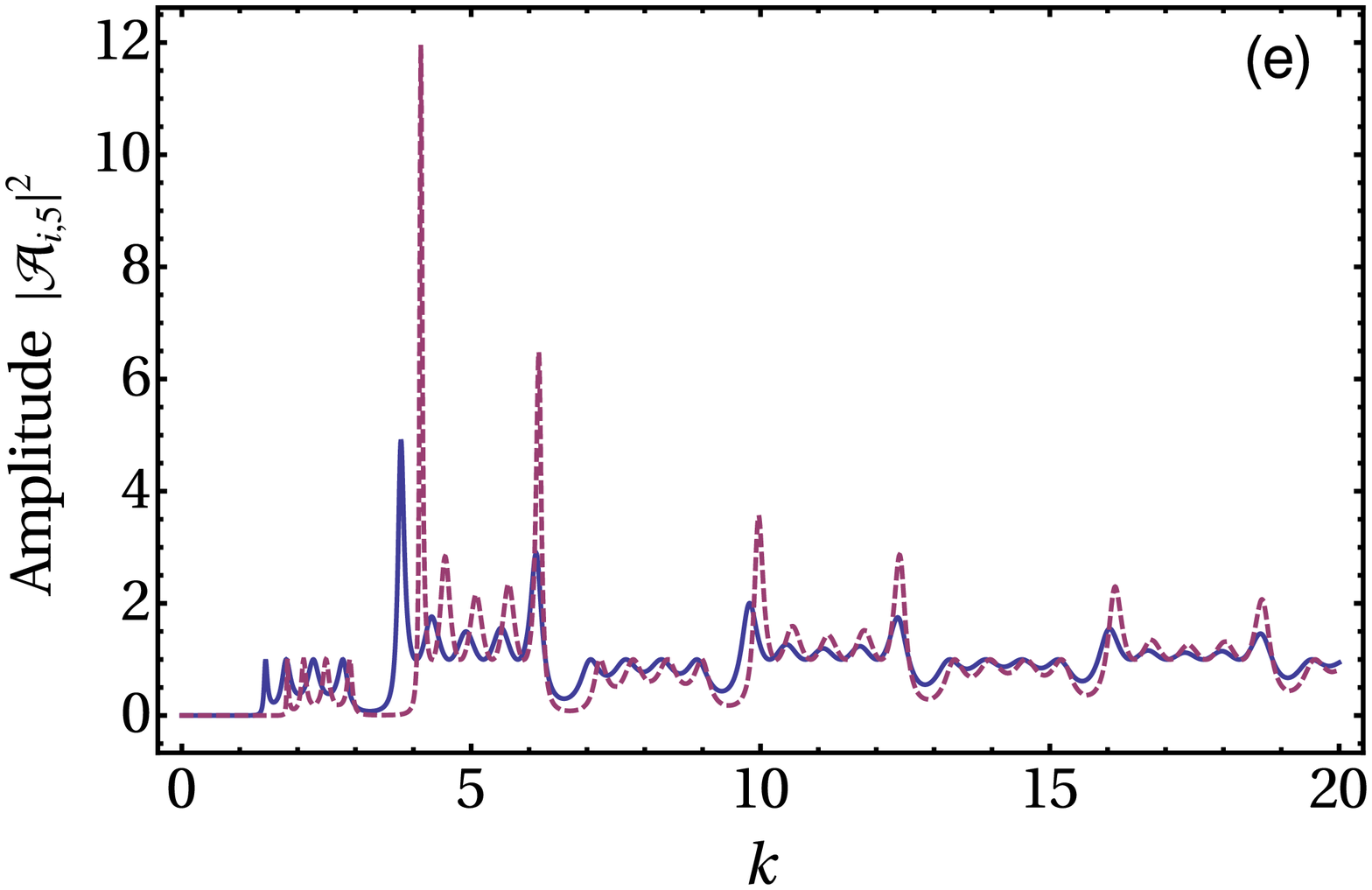}
  \caption{\label{fig:fig21} (Color online).
    The transition probability $|\mathcal{A}_{i,l}|^2$, Eq. 
    (\ref{eq:amp_quase_estados}), as a function of
    $k$ for the graph of Fig. \ref{fig:fig20} with the
    Dirichlet boundary condition at the vertex 6 
    (so, $r_6^{(+)}=-1$).
    The $l$'s are (a) 1, (b) 2, (c) 3, (d) 4, and (e) 5.
    The solid (dashed) line is for $\gamma=1$ ($\gamma=2$).
    Here $\ell = 1$.
}
\end{figure}

\begin{figure}
  \centering
  \includegraphics*[width=0.5\textwidth]{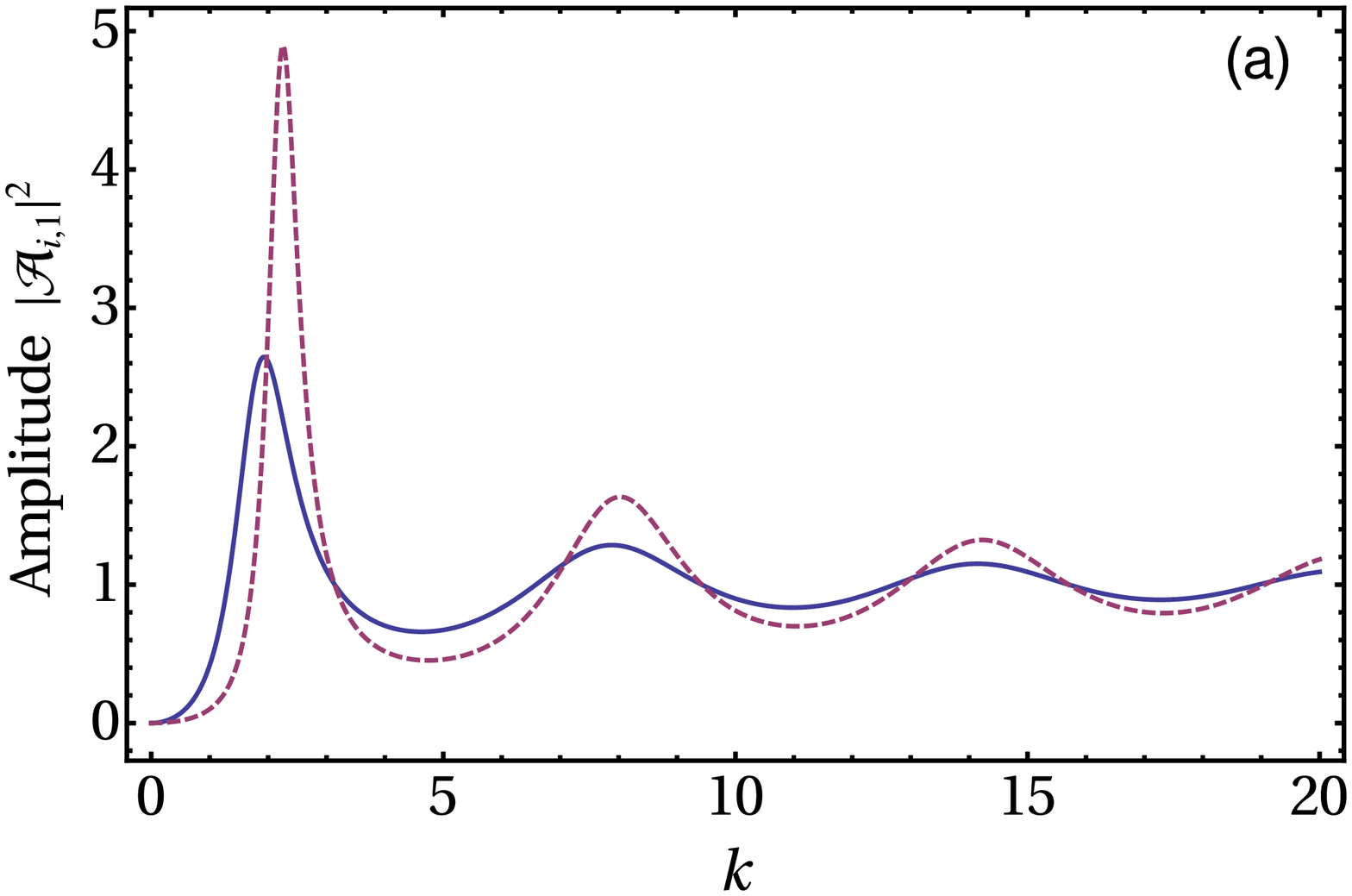}%
  \includegraphics*[width=0.5\textwidth]{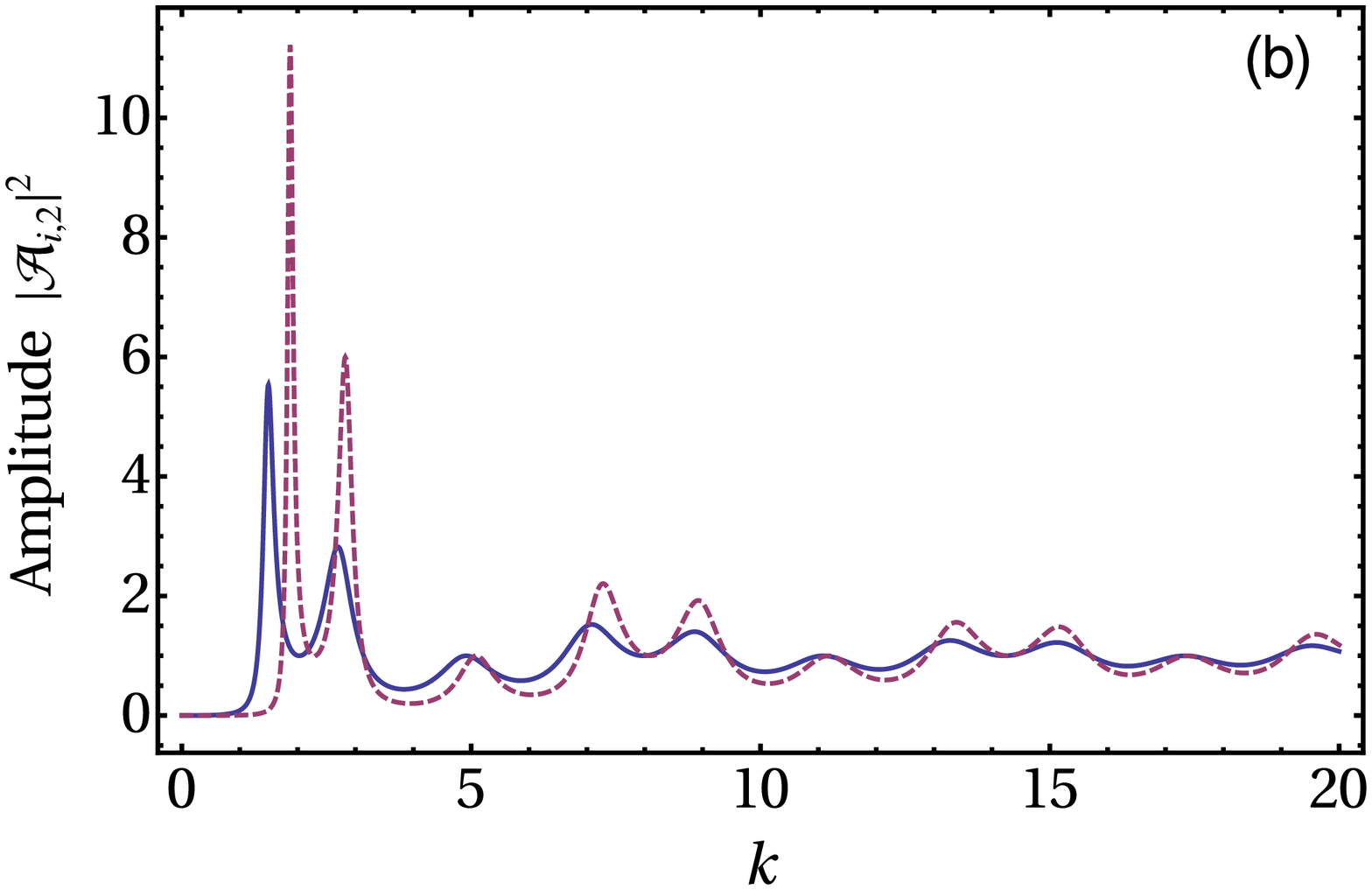}\\
  \includegraphics*[width=0.5\textwidth]{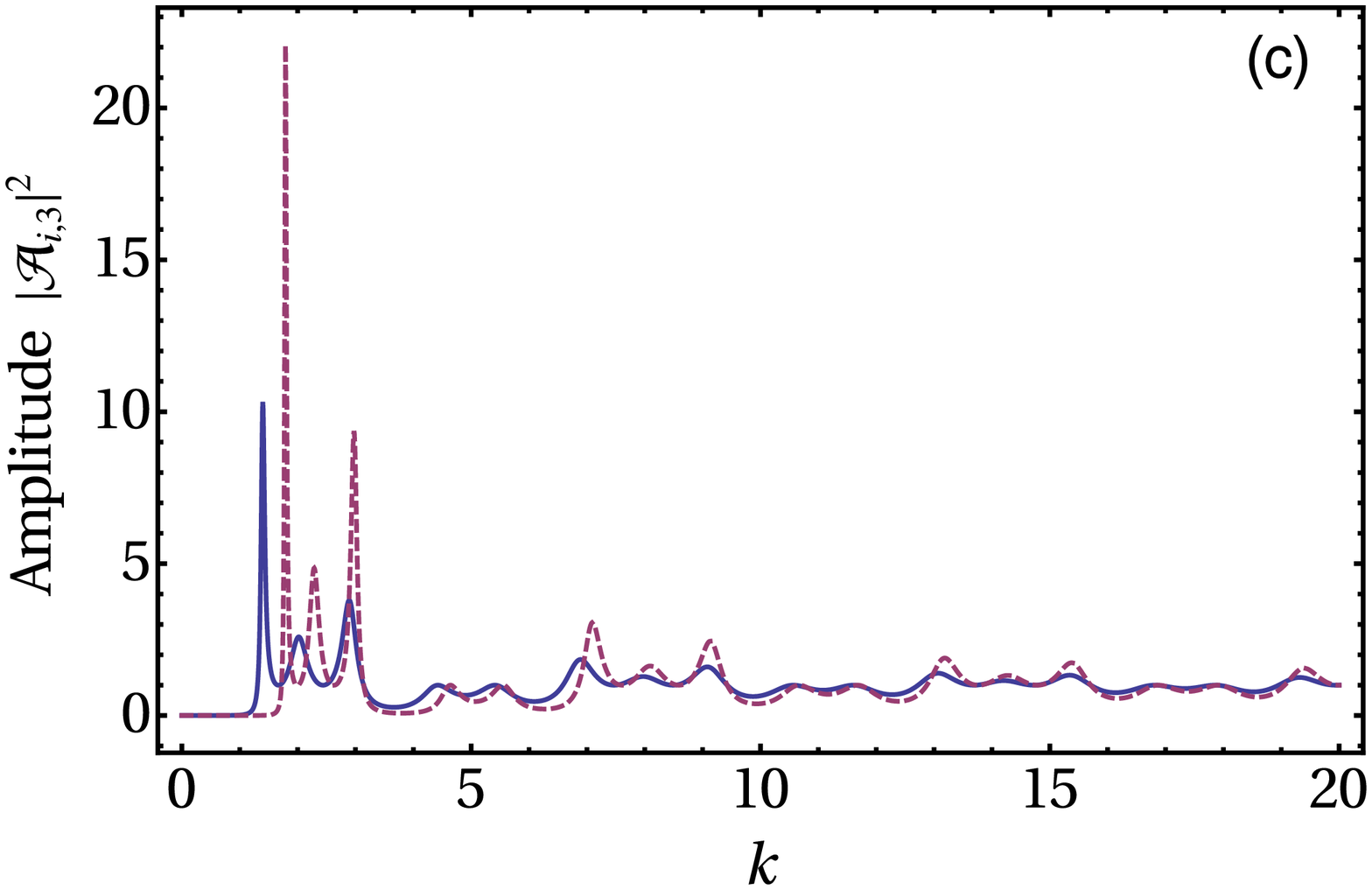}%
  \includegraphics*[width=0.5\textwidth]{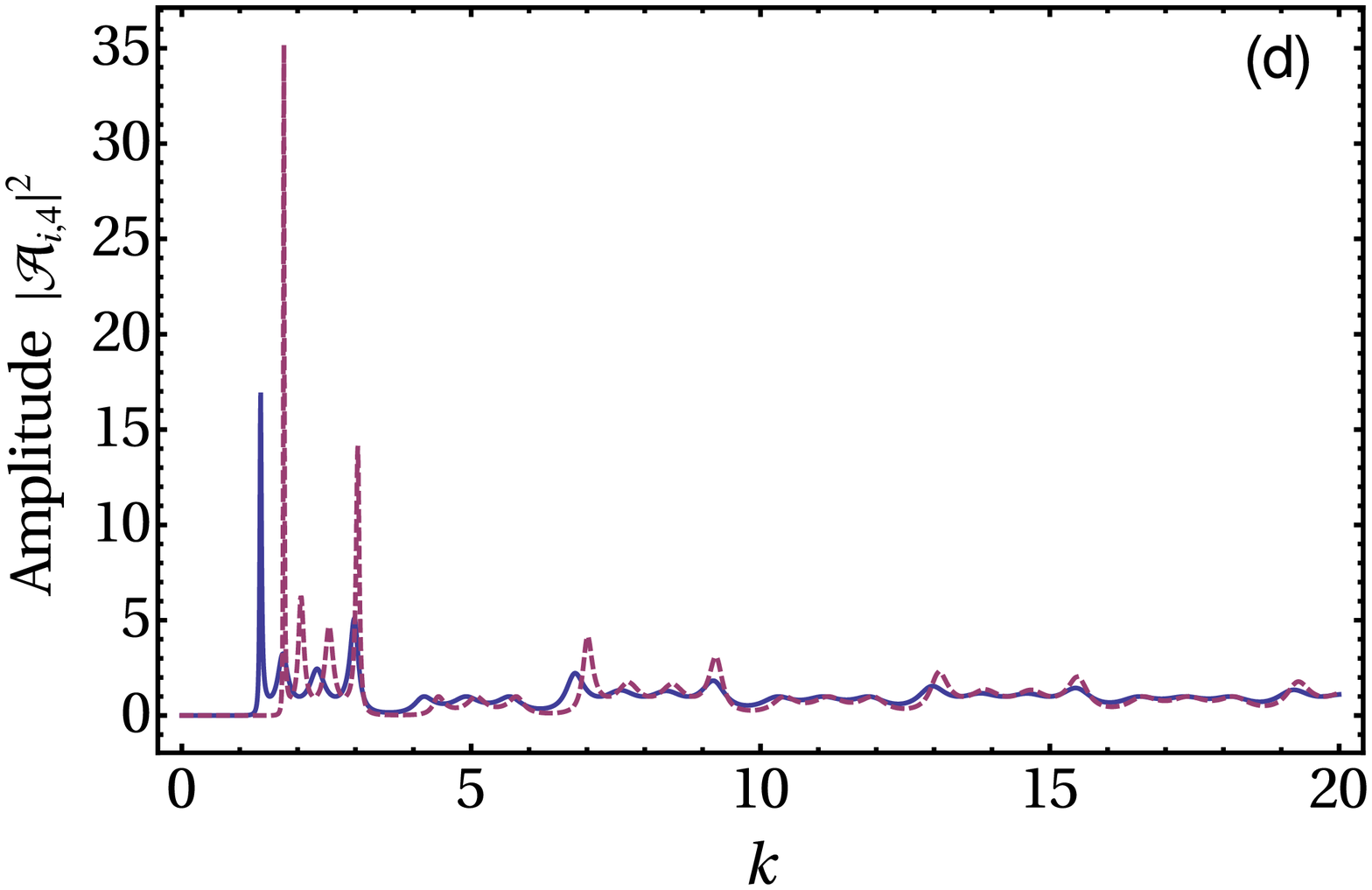}\\
  \includegraphics*[width=0.5\textwidth]{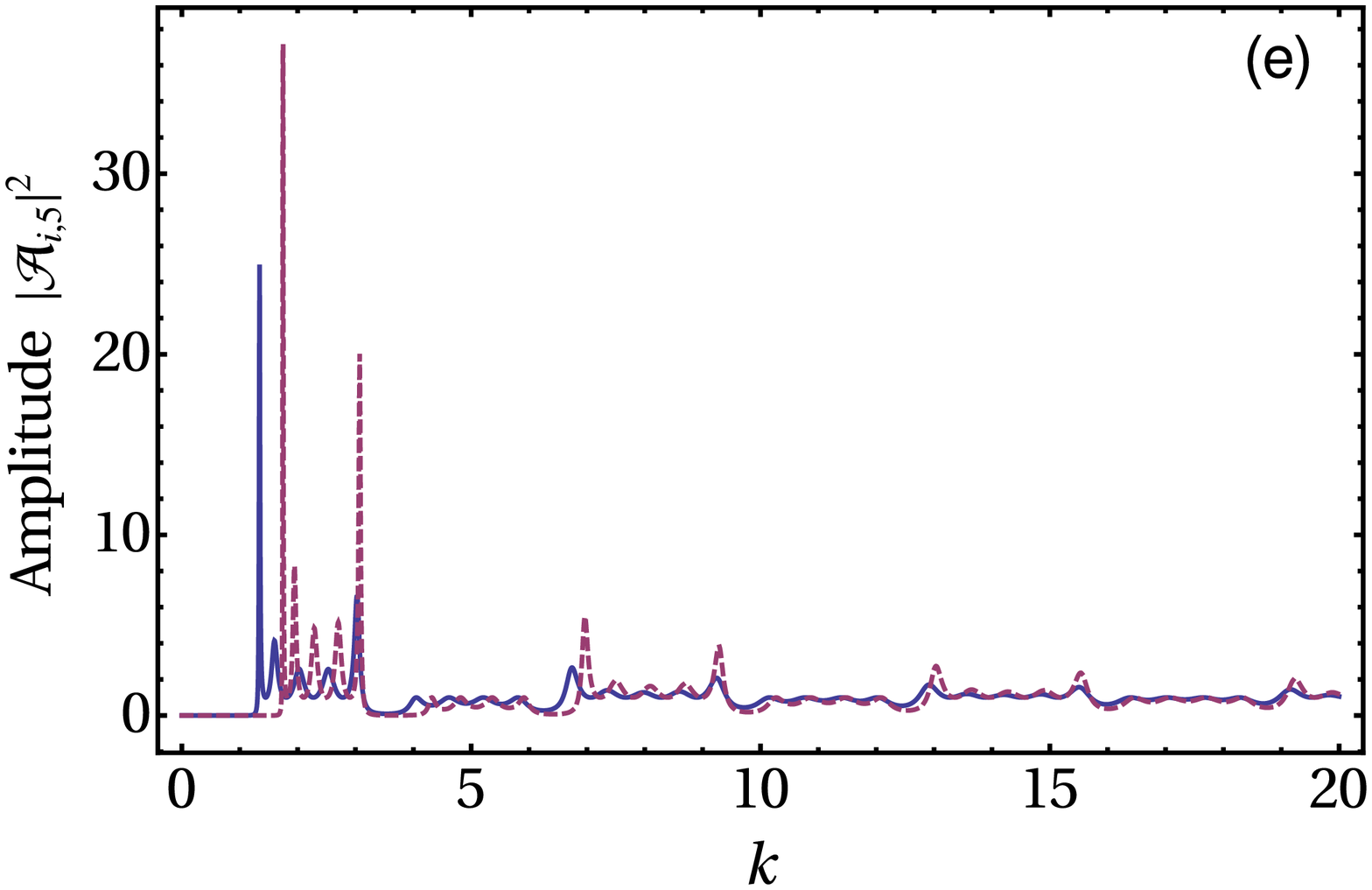}%
  \caption{\label{fig:fig22} (Color online).
    The same as in Fig. (\ref{fig:fig21}), but for the Neumann boundary 
    condition at the vertex 6 (so, $r_6^{(+)} = +1$).}
\end{figure}

From the plots in Figs. \ref{fig:fig21} and \ref{fig:fig22} we see
that the analysis of the $\mathcal{A}_{i,l}$'s for distinct $l$'s renders
a much more detailed information than just to examine the system global 
reflection coefficient $R_{i i} = \exp[i \phi_R(k)]$.
For instance, for the $r_6^{(+)} = -1$ case, Fig. \ref{fig:fig21},
and when $\gamma=2$, it is clear the existence of a 
$k^{(\rm{qb})} \approx 4.2$.
Indeed, we see peaks around this wavenumber value for fairly all the $l$'s.
Nevertheless, they are much higher and narrower for $l = 4, 5$.
Hence, such quasi-bound state must be much more localized in these 
two edges.
Another observed feature is that the quasi-bound states are longer-lived
for $\gamma = 2$ than for $\gamma = 1$ (compare the heights and widths of
the peaks in the two situations).
This is simple to understand: a delta interaction of greater strength 
is more efficient in trapping an initially localized state.
Finally, from the general trends in Figs. \ref{fig:fig21} and \ref{fig:fig22} 
we also can conclude that it is the Neumann boundary condition (at the
`dead end' vertex 6) which is able to create quasi-bound states of
longer $\tau$'s.

\subsection{Quasi-bound state in arbitrary graphs}
\label{sec:7-4}

Inspecting the expression for $\mathcal{A}_{i,l}$ in Eq.
\eqref{eq:amp_quase_estados} (as well as other similar formulas along 
this review), we conclude that typical transitions amplitudes
between parts of a quantum graph -- in which $x_i$ is in a lead $i$
and $x_f$ is in an edge $e_l$ --  are given by
\begin{equation}
\mathcal{A}_{i,l} = \frac{T_{i,l}}{1 - R_{\rm{right}} \, R_{\rm{left}} \,
\exp[2 i k \ell_l]}.
\label{eq:AIJ}
\end{equation}
The numerator is a transmission coefficient, corresponding to the 
graph region between $x_i$ (in the lead $i$) and $x_f$ (in the edge $l$).
In the denominator, $R_{\rm{right}}$ ($R_{\rm{left}}$) is the global 
reflection coefficient for a part of the graph, so to speak,
to the `right' of edge $l+1$ (to the `left' of vertex $l$, between $x_i$
and the vertex $l$).
Note also that the term in the denominator is associated with eventual
energy eigenvalues \cite{Mastherthesis.2001.Andrade,JPA.2003.36.227}, and
in general can be derived from a sum over periodic orbits in the graph
(i.e., scattering paths leaving and arriving at the same edge $l$)
\cite{JPA.2003.36.545,PRL.1997.79.4794,AoP.1999.274.76}.

\begin{figure}
  \centering
  \includegraphics*[width=0.15\textwidth]{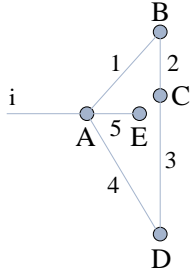}
  \caption{\label{fig:fig23} (Color online).
    Example of an open quantum graph, whose a modified version has 
    been studied in \cite{JPA.2003.36.545}.}
\end{figure}

Therefore, Eq. (\ref{eq:AIJ}) is not restricted to a linear graph, and
in fact should work for any topology (provided one properly defines and
constructs the $R$'s and  $T$).
As an example, consider the structure\footnote{
We should mention that a modified version of this graph, with extra
leads at $B$, $C$, and $D$, has been studied in \cite{JPA.2003.36.545} 
in the context of quantum protocols for transmission of information.}
in Fig \ref{fig:fig23}.
Such graph can display interesting features if one assumes different
boundary conditions at each vertex and distinct lengths for each edge
(see \cite{JPA.2003.36.545}). 
But here we restrict the discussion to generalized delta point interactions
of a same strength $\gamma$ at the vertices
$A$, $B$, $C$, $D$, and either Dirichlet or Neumann boundary conditions
(see previous section) at the vertex $E$.
Also, we suppose all the edges with the length $\ell=1$.
So, due to symmetry, the edges 1 and 4 and 2 and 3 must present similar
scattering properties and we can focus just on the inequivalent
amplitudes
$\mathcal{A}_{i,1}(k)$, $\mathcal{A}_{i,2}(k)$, and $\mathcal{A}_{i,5}(k)$. 
Using Eq. (\ref{eq:AIJ}) and the appropriate corresponding reflection and
transmission quantum amplitudes for the graph of Fig. \ref{fig:fig23},
we show in Fig. \ref{fig:fig24} the behavior of the modulus square of
these three quantities as function of $k$ for the Dirichlet and
Neumann boundary conditions at $E$ and the delta interactions strength
value $\gamma=0.5$ and $\gamma=1$.

Because the graph distinct geometric characteristics, when compared to
the simpler linear case (Fig. \ref{fig:fig20}), we do observe here a
richer profile of quasi-bound states.
Also, the distinct boundary conditions at $E$ considerably change
the positions and sizes of the $E^{(\rm{qb})}$ peaks
(this is a same sort of sensibility also found for the transmission
probabilities for the related graph studied in \cite{JPA.2003.36.545}).
Finally, in general the peaks are higher and narrower, so longer-lived,
for the greater value of $\gamma$ ($\gamma=1$).

\begin{figure}
  \centering
  \includegraphics*[width=0.5\textwidth]{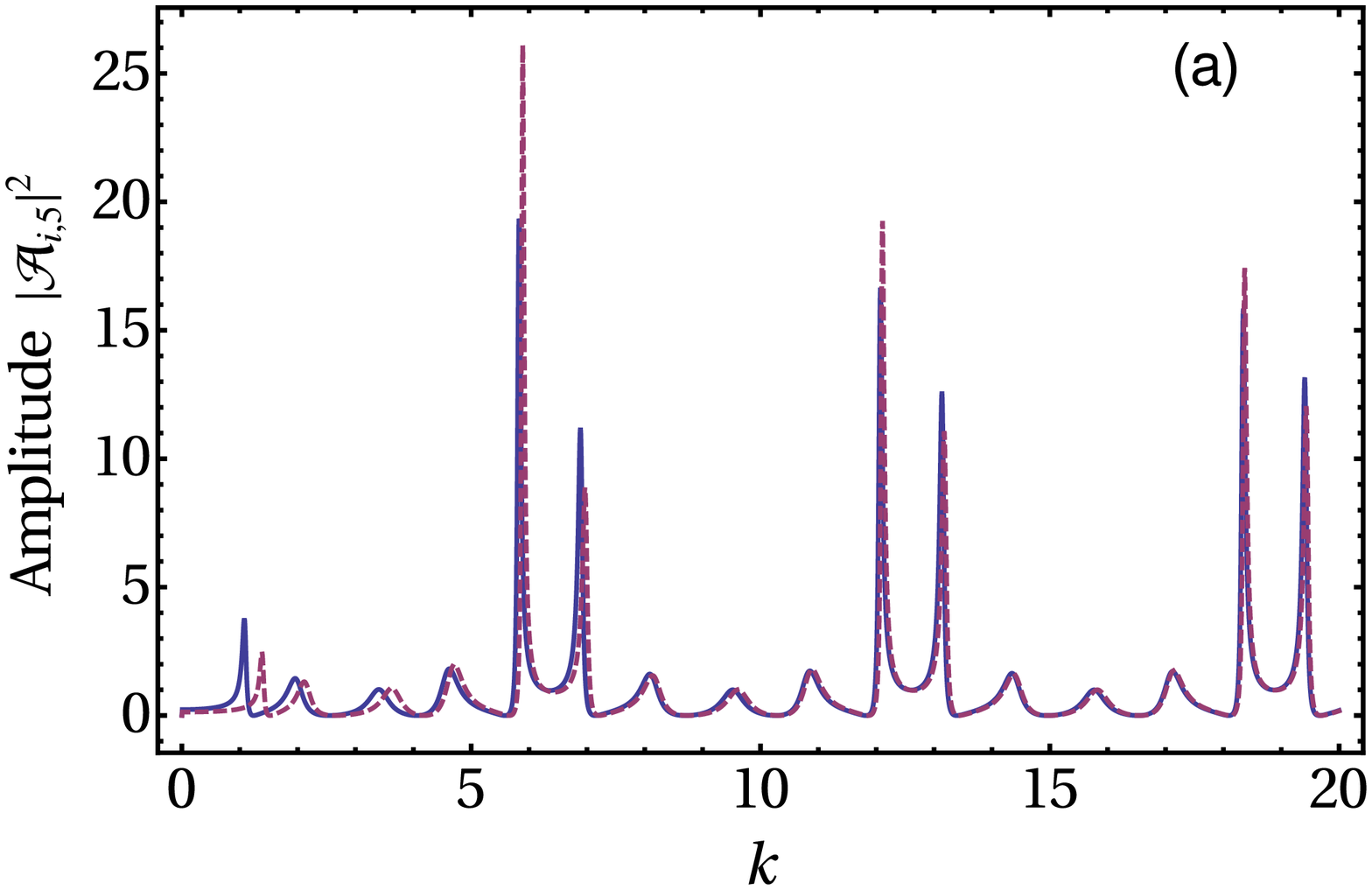}%
  \includegraphics*[width=0.5\textwidth]{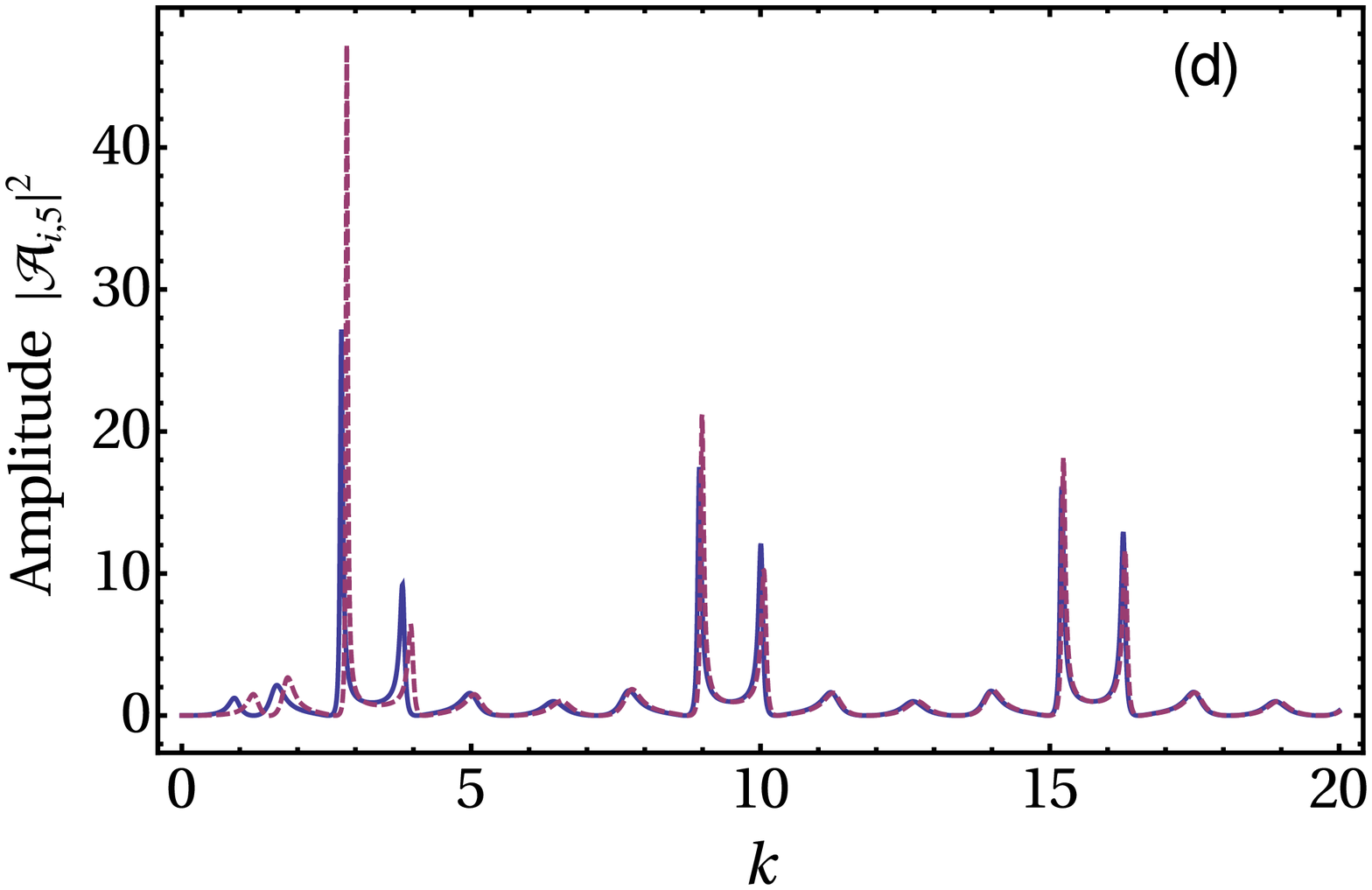}\\
  \includegraphics*[width=0.5\textwidth]{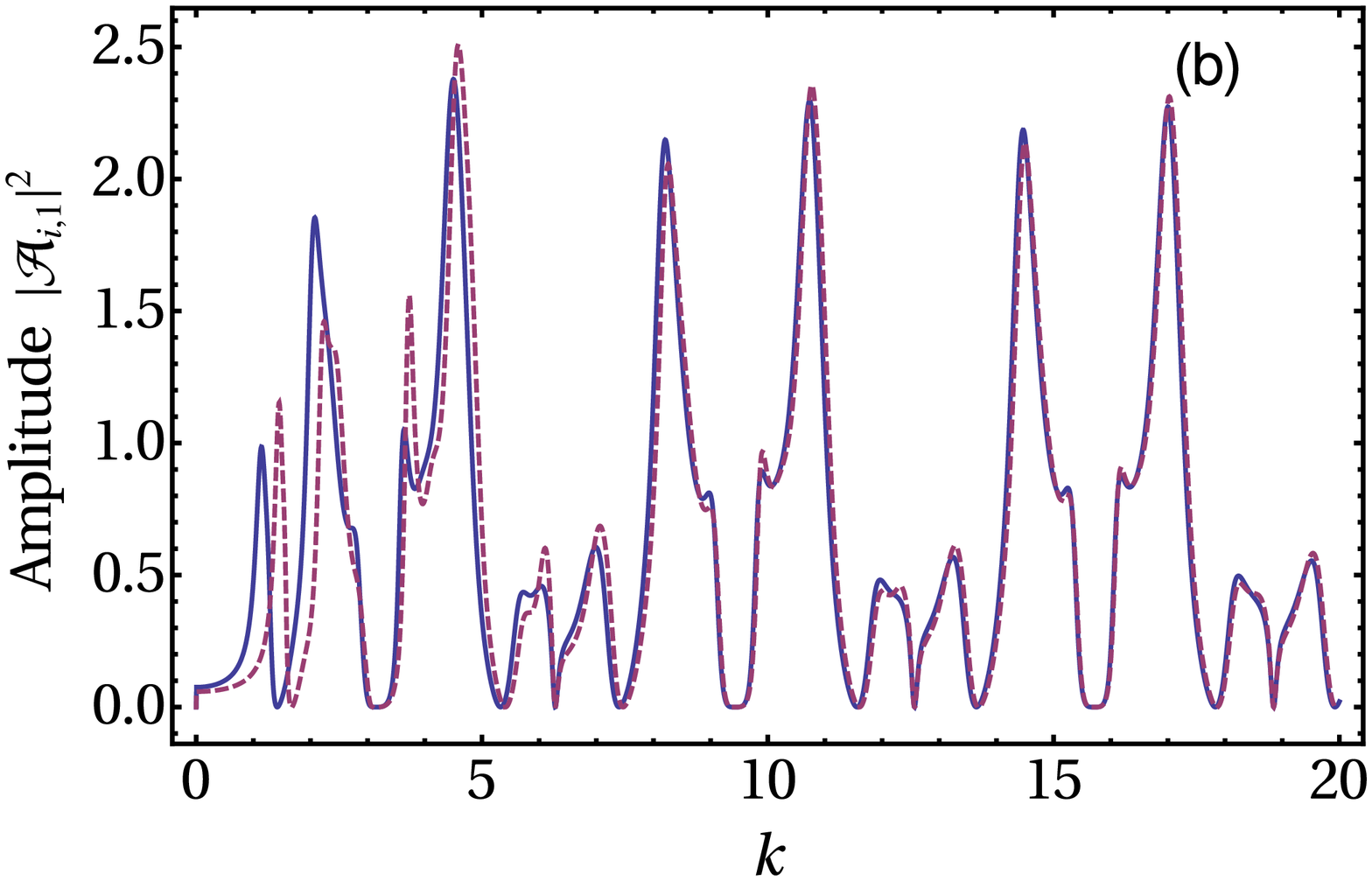}%
  \includegraphics*[width=0.5\textwidth]{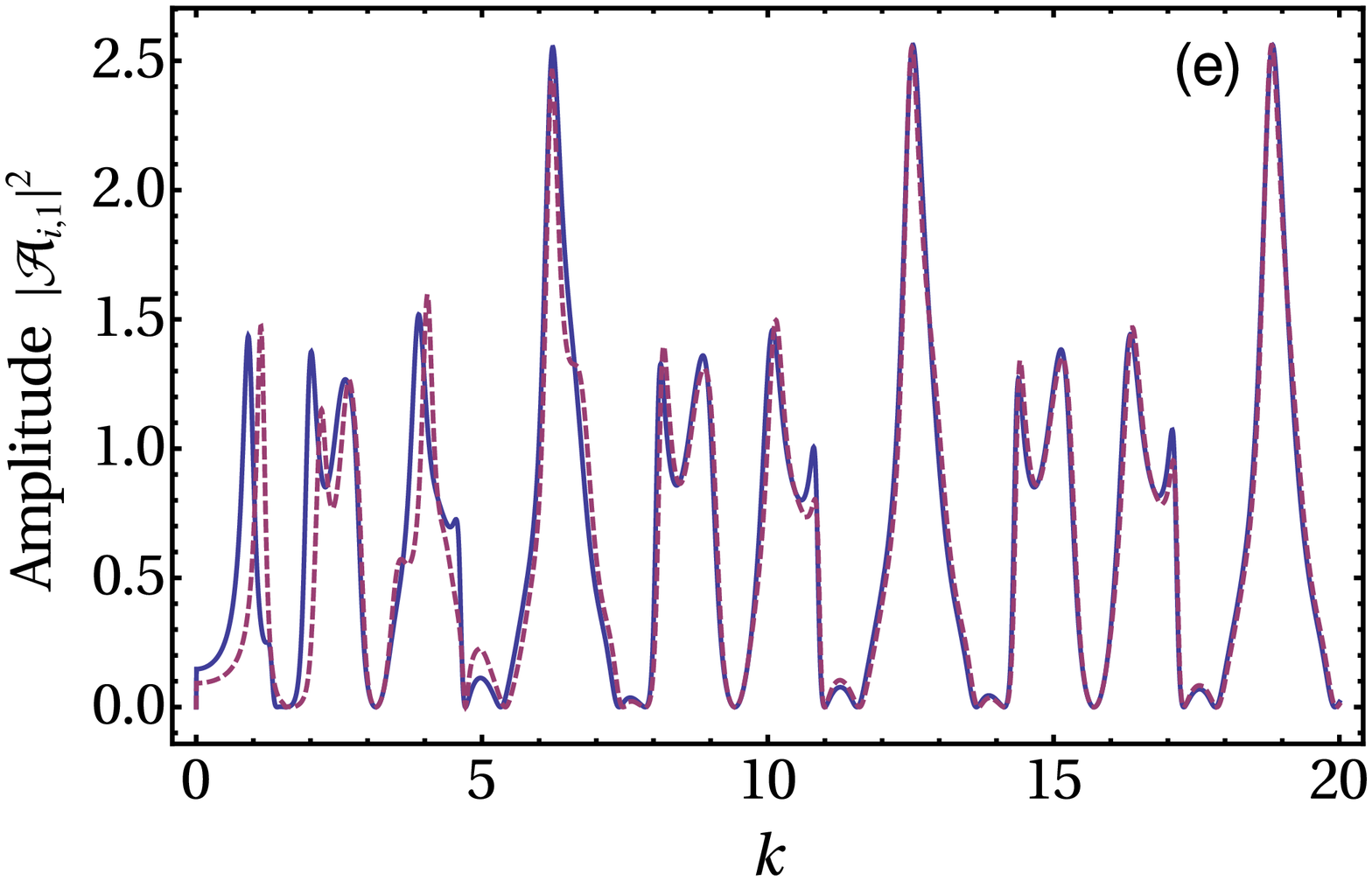}\\
  \includegraphics*[width=0.5\textwidth]{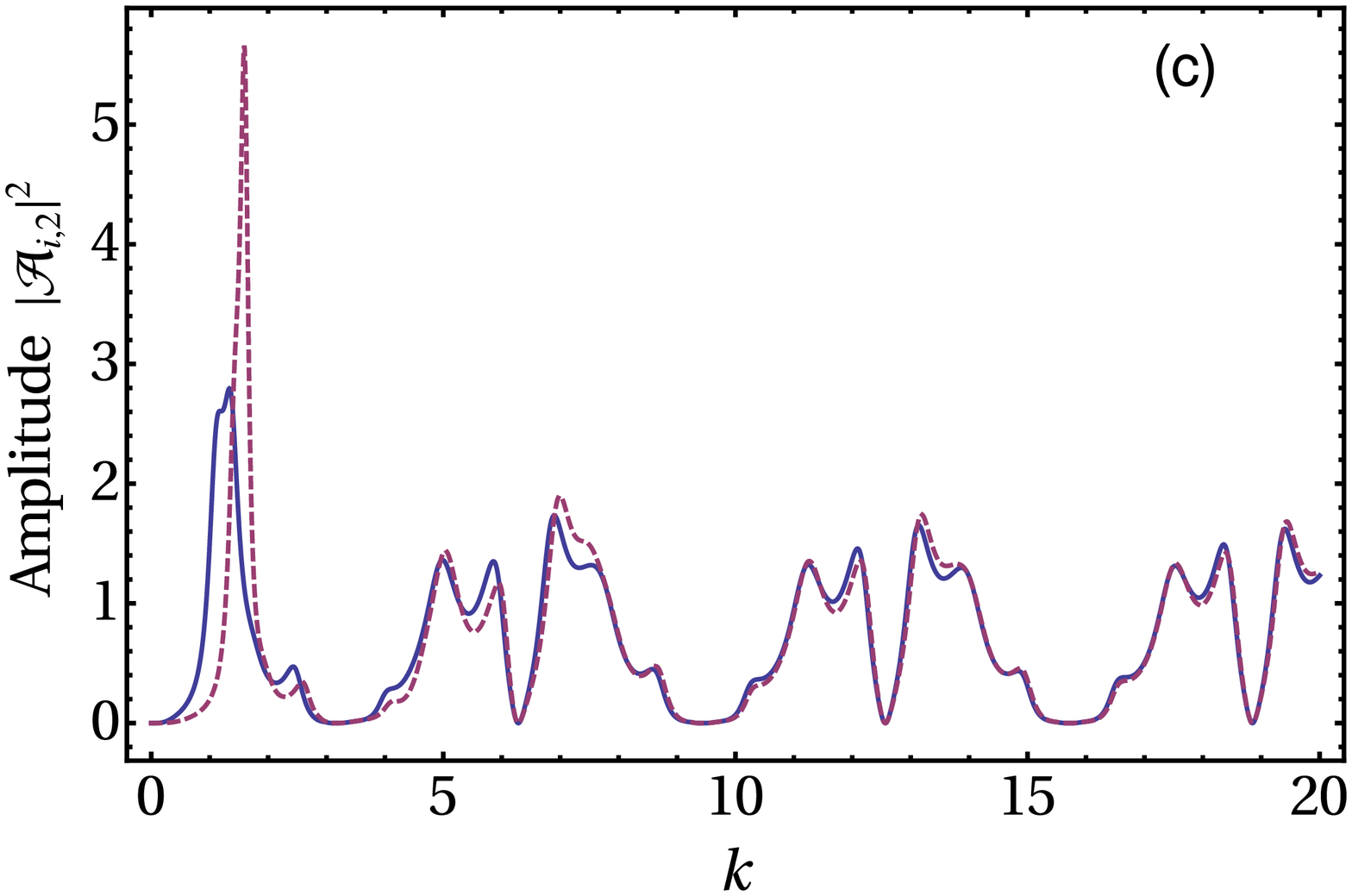}%
  \includegraphics*[width=0.5\textwidth]{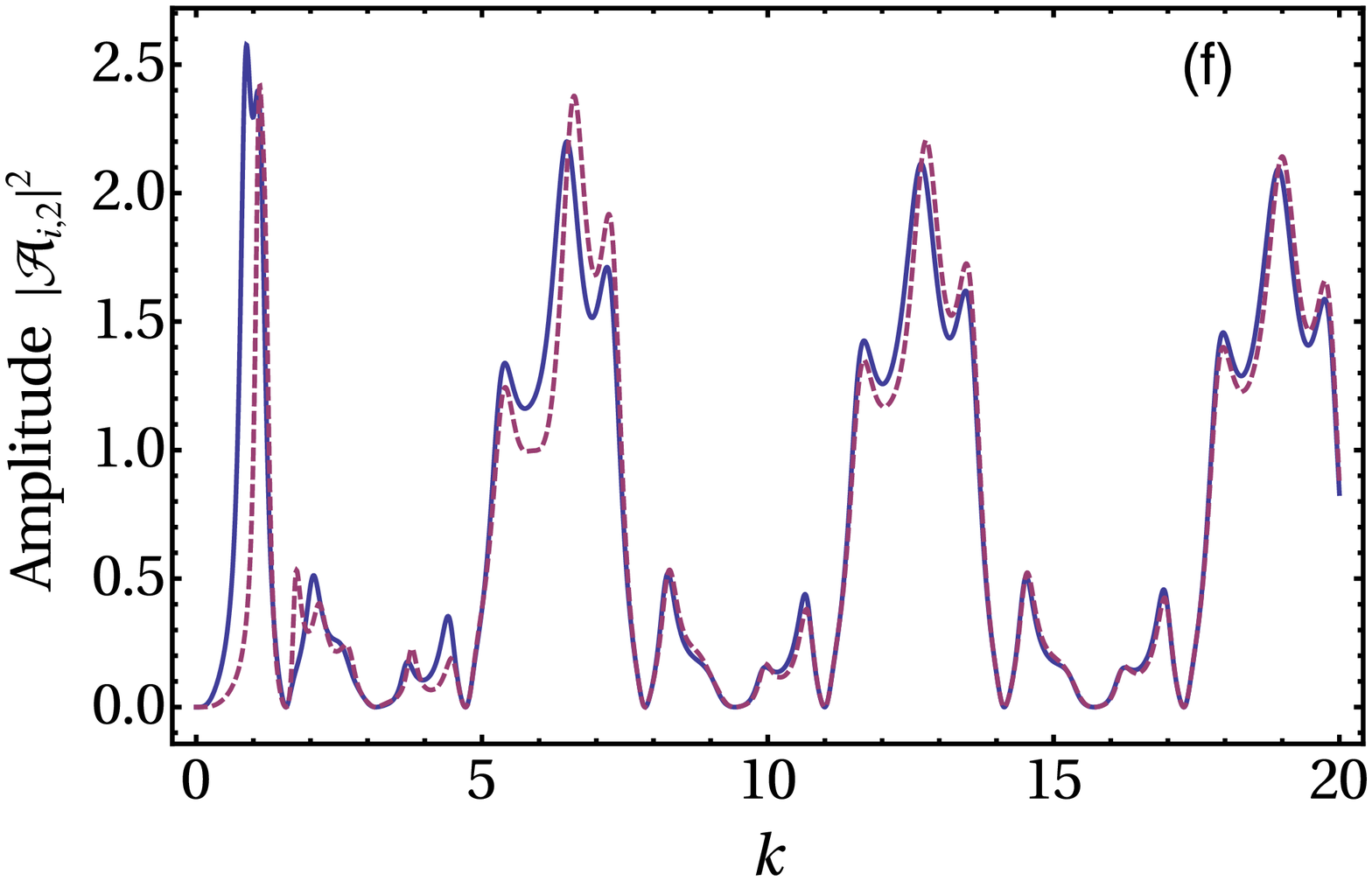}%
  \caption{\label{fig:fig24} (Color online).
    Behavior of $|\mathcal{A}_{i,l}|^2$ as function of $k$ 
    (calculate from Eq. (\ref{eq:AIJ})) for the
    graph of Fig. \ref{fig:fig23} and $l=1,2,5$.
    The vertices $A$, $B$, $C$, and $D$, are generalized delta
    interactions of strength $\gamma$, with $\gamma=0.5$ (solid)
    and $\gamma=1$ (dashed) lines.
    The boundary conditions at vertex $E$ are Dirichlet's (so, $r_E=-1$)
    in (a)--(c) and Neumann's (so, $r_E=1$) in (d)--(f).
    All edges have the same length $\ell=1$.
     }
\end{figure}

%%% Local Variables:
%%% mode: latex
%%% TeX-master: "green-qg-pr"
%%% ispell-local-dictionary: "american"
%%% End:
\section{Conclusion}
\label{sec:conclusion}

The discussions throughout this article have highlighted the usefulness of
graphs to study some fundamental theoretical aspects of quantum mechanics
as well as to model different phenomena associated to quantum wave-like
behavior.
But despite the conceptual simplicity of these systems, the calculation
of their quantum properties might demand sophisticated and involving 
methods \cite{Inproceedings.2006.Kostrykin}.
Further, certain standard mathematical procedures may require
modifications when applied to a graph structure, as to obtain the 
Green's function from the Krein's resolvent formula
\cite{JPA.2005.38.4859}.

Since the Green's function is one of the most powerful techniques to solve
quantum systems \cite{Book.2006.Economou}, in this review we have specifically
considered such approach to address finite open and closed undressed
quantum graphs of any topology.
We have so discussed a physically appealing procedure to construct $G$,
summarized in the Eq. (\ref{eq:gf})\footnote{We should observe that
Eq. (\ref{eq:gf}) is ultimately akin to the type of calculations proposed
in the very interesting work in \cite{PRE.2001.65.016205}, but which is
devoted only to open quantum graphs.}:
the exact Green's function given as a sum over all the possible
scattering ``classical'' paths (sp) along the edges, for which local quantum
effects are taking into account through reflections and transmissions
amplitudes defined at the vertices (constituting thus the scatterers centers).
Then, the present Green's function method somehow generalizes the
Kirchhoff's quantum rules \cite{JPA.1999.32.595} by ascribing a general
scattering matrix to each vertex of the quantum graph.

In particular, we have described in details recursive ways to sum up all
the sp's contributions to $G$.
Basically, they rely on two simplification schemes:
(a) to regroup infinite many paths into a single trajectory family; and 
(b) to divide a larger graph into smaller pieces, to derive for each
piece a global scattering matrix, and finally to compose all the pieces
back together.
As concrete examples, certain representative quantum graphs commonly
found in the literature have been considered, as the cube, binary tree
and Sierpi\'{n}ski-like structures.

The protocols outlined here could likewise be applied to dressed quantum
graphs if the potentials along the edges decay at least exponentially
\cite{JPA.29.2839.1996}.
In fact, in this case very good analytical approximations for the Green's
function can be derived \cite{JPA.2001.34.5041,JPA.2003.36.227}.
But then, besides the vertices quantum amplitudes, it is also necessary
to consider the potentials reflection and transmission coefficients and
to compute the classical actions for a particle under these potentials.
Furthermore, a close related class of systems, namely, scattering quantum 
walks, can be treated exactly in the same fashion.
As shown in \cite{PRA.2011.84.042343,PRA.2012.86.042309}, the exact
Green's function -- written as a sum over sp's -- allows to identify
the precise paths responsible for distinct effects, like the ones
resulting in the super-diffusion observed in quantum walks.

Finally, a very interesting application for Green's functions (and in the
context of open quantum graphs, eventually difficult by other means)
is to search for possible quasi-bound states.
We have illustrated how to do, moreover analyzing the influence of few
different boundary conditions at the vertices in setting the quasi-energies 
and corresponding widths.

We hope that this review, discussing exact closed analytic expressions for 
the Green's functions of quantum graphs, can become a helpful guide to all 
those interested in this diverse and conceptually and phenomenologically 
so rich class of systems.

%%% Local Variables:
%%% mode: latex
%%% TeX-master: "green-qg-pr"
%%% ispell-local-dictionary: "american"
%%% End:

\section{Acknowledgments}
FMA would like to thank Simone Severini and Sougato Bose at University
College London for the hospitality extended during the period this
manuscript was finished.
Support is acknowledge from CNPq for researcher grants.
FMA also thanks CNPq for grants No. 460404/2014-8 (Universal) and
206224/2014-1 (PDE).

\appendix
\section{The most general point interaction conserving
probability flux  as a quantum graph vertex}
\label{app:flux}

\subsection{The usual case: the line}
\label{appendix:a1}

The probability density flux in the usual 1D quantum mechanics reads
(here for $\hbar = \mu = 1$)
\begin{equation}
j(x)=\frac{1}{2i}[\psi^{*}(x)\psi'(x)-\psi(x)\psi'^{*}(x)].
\label{eq:fluxo}
\end{equation}
Thus, if we define ($\psi'(x) \equiv d \psi(x)/dx$)
\begin{equation}
\Phi(x) = \left(
\begin{array}{c}
\psi(x) \\
\psi'(x)
\end{array}
\right),
\end{equation}
and
\begin{equation}
J = \left(
\begin{array}{cc}
0 & 1 \\
-1 & 0
\end{array}
\right),
\end{equation}
$j(x)$ can be written in a complex symplectic-like form as
\begin{equation}
j(x) = \frac{1}{2i} \Phi^{\dagger}(x) \, J \, \Phi(x).
\label{eq:sympletic}
\end{equation}

Now, suppose a free particle of energy $E = k^2/2$ on the line
($-\infty < x < +\infty$), obeying to $-d^2 \psi(x)/dx^2 = k^2 \psi(x)$
for $x\neq 0$.
At $x = 0$ we assume a point interaction.
Since, by definition, the range of action of such kind of potential
is zero, its only effect is to set a specific BC for the wave
function $\psi(x)$ at $x = 0$.
Thus, the most general point potential corresponds to the
most general linear boundary condition, represented by
\begin{equation}
\Phi(0^{+}) = \Gamma \, \Phi(0^{-}),
\label{eq:bc}
\end{equation}
with
\begin{equation}
\Gamma = \omega
\left(
\begin{array}{cc}
a & b \\
c & d
\end{array}
\right).
\label{eq:gamma-matrix}
\end{equation}
For example, for the common delta function potential $\gamma \, \delta(x)$
(so, with $\gamma$ being the strength), the parameters are
$a = d = \omega = 1$, $b = 0$, and $c = \gamma$.

Using the Eqs. \eqref{eq:sympletic} and \eqref{eq:bc},
we have
\begin{equation}
j(0^{+}) = \frac{1}{2i}
\Phi^{\dagger}(0^{-}) \, \Gamma^{\dagger} \,  J \, \Gamma \, \Phi(0^{-}).
\end{equation}
If we impose $j(0^{+}) = j(0^{-})$, it follows that
$\Gamma^{\dagger} \, J \, \Gamma = J$, yielding
\begin{equation}
a d - b c = 1, \ a, b, c, d \ \mbox{real numbers and}
\ |\omega| = 1.
\label{eq:parameters-relations}
\end{equation}
Therefore, the most general point interaction consistent with
flux conservation is characterized by Eq. \eqref{eq:bc}, with
$\Gamma$ given by Eqs. \eqref{eq:gamma-matrix} and
\eqref{eq:parameters-relations}.

Next, to consider a ${\mathcal S}$ matrix formalism 
\cite{Book.1989.Chadan}, suppose typical plane wave scattering solutions 
(of wavenumber $k$).
The incoming and outgoing parts of the state should then be related 
through
\begin{equation}
\left(
\begin{array}{c}
\psi_k^{(\mbox{\scriptsize out})}(0^{-}) \\
\psi_k^{(\mbox{\scriptsize out})}(0^{+})
\end{array}
\right) = {\mathcal S}(k)
\left(
\begin{array}{c}
\psi_k^{(\mbox{\scriptsize in})}(0^{-}) \\
\psi_k^{(\mbox{\scriptsize in})}(0^{+})
\end{array}
\right).
\label{eq:in-out}
\end{equation}
Probability conservation at the origin,
\begin{equation}
|\psi_k^{(\mbox{\scriptsize in})}(0^{-})|^2 +
 |\psi_k^{(\mbox{\scriptsize in})}(0^{+})|^2 =
 |\psi_k^{(\mbox{\scriptsize out})}(0^{-})|^2 +
 |\psi_k^{(\mbox{\scriptsize out})}(0^{+})|^2,
\end{equation}
inserted into Eq. \eqref{eq:in-out} leads to
${\mathcal S}(k) {\mathcal S}^{\dagger}(k) =
{\mathcal S}^{\dagger}(k) {\mathcal S}(k) = \mathbf{1}$, i.e.,
${\mathcal S}$ is unitary.
Furthermore, making in Eq. \eqref{eq:in-out} the substitution
$k \rightarrow -k$, we can write
\begin{equation}
\left(
\begin{array}{c}
\psi_{-k}^{(\mbox{\scriptsize in})}(0^{-}) \\
\psi_{-k}^{(\mbox{\scriptsize in})}(0^{+})
\end{array}
\right) = {\mathcal S}^{\dagger}(-k)
\left(
\begin{array}{c}
\psi_{-k}^{(\mbox{\scriptsize out})}(0^{-}) \\
\psi_{-k}^{(\mbox{\scriptsize out})}(0^{+})
\end{array}
\right).
\label{eq:in-out-inv}
\end{equation}
But $k \rightarrow -k$ inverts the flux direction, physically implying
in
$\psi^{(\mbox{\scriptsize in})} \leftrightarrow \psi^{(\mbox{\scriptsize out})}$.
So, given such in-out exchange in Eq. \eqref{eq:in-out-inv} and once the
relation between incoming and outgoing wave function components
is always set in the form of Eq. \eqref{eq:in-out}, we must have
${\mathcal S}(k) = {\mathcal S}^{\dagger}(-k)$.

For any arbitrary point interaction, we can write the scattering solutions
$\psi_k^{(\pm)}(x)$ assuming a plane wave, of wavenumber $k$,
incident either from the left $(+)$ or right $(-)$, so that
(${\mathcal N} = 1/\sqrt{2 \pi}$)
\begin{equation}
 \psi_k^{(\pm)}(x) = {\mathcal N} \times
 \left\{
   \begin{array}{ll}
     \exp{[\pm ikx]} + R^{(\pm)}(k) \exp{[\mp ikx]}, & x \lessgtr 0 \\
     T^{(\pm)}(k) \exp{[\pm ikx]}, &  x \gtrless  0.
   \end{array}
 \right.
 \label{eq:psi-scattering}
\end{equation}

Observing that $\exp{[\pm ikx]}$ are the incoming and the terms
involving $R$ and $T$ are the outgoing parts of the above full
scattering states, one gets that arbitrary linear combinations
of $\psi_k^{(+)}$ and $\psi_k^{(-)}$ results,
from Eq. \eqref{eq:in-out}, in
\begin{equation}
{\mathcal S}(k) =
\left(
\begin{array}{cc}
R^{(+)}(k) & T^{(-)}(k) \\
T^{(+)}(k) & R^{(-)}(k)
\end{array}
\right).
\label{eq:pi-s-matrix}
\end{equation}
Now, imposing ${\mathcal S} \, {\mathcal S}^{\dagger} =
{\mathcal S}^{\dagger} \, {\mathcal S} = \mathbf{1}$ and
${\mathcal S}(k) = {\mathcal S}^{\dagger}(-k)$ to Eq.
\eqref{eq:pi-s-matrix}, ones finds that
\begin{align}
  |{R}|^2+|{T}|^2=1,
  \qquad
  {{R}^{(+)}}^*{T}^{(\pm)} + {{T}^{(\mp)}}^* {R}^{(-)} = 0,
  \nonumber \\
  {{R}^{(\pm)}}^*(k) = {R}^{(\pm)}(-k),
  \qquad
  {{T}^{(\pm)}}^*(k) = {T}^{(\mp)}(-k).
  \label{eq:rt-relations}
\end{align}
These are the basic conditions to assure proper features for the
scattering solutions in quantum
mechanics \cite{Book.1989.Chadan}, e.g.,
orthonormalization, flux conservation, and the existence of the 
scattering inverse problem.
If, furthermore, one also requires time-reverse invariance --
what we are not imposing in this work -- then ${T}^{(+)} = {T}^{(-)}$.

Finally, to establish a full correspondence between the two
approaches, the boundary condition treatment and the ${\mathcal S}$
matrix formalism, let us assume Eq. \eqref{eq:bc} (with 
Eq. \eqref{eq:parameters-relations}) for the states
in Eq. \eqref{eq:psi-scattering}.
Thus \cite{PRA.2002.66.062712}
\begin{equation}
{R}^{(\pm)}(k) = \frac{c \pm ik(d-a) + bk^2} {-c + i k(d+a) + b k^2},
 \ \
{T}^{(\pm)}(k) = \frac{2 i k \omega^{\pm 1}}
 {-c + i k(d+a) + b k^2 }.
\label{eq:rt-point-interaction}
\end{equation}
It is easy to verify that the quantum amplitudes in Eq.
\eqref{eq:rt-point-interaction} satisfies {\em all} the
fundamental requirements in Eq. \eqref{eq:rt-relations}
\cite{PRA.2002.66.062712}.
Hence, up to a global phase $\omega$, the problem is likewise specified
from the parameters $a, b, c$ and $d$ or from the coefficients
$R^{(\pm)}$ and $T^{(\pm)}$.
Thus, the two approaches are completely equivalent and arbitrary point
interactions can be defined entirely in terms of their $\mathcal{S}$
matrix (for a more detailed analysis, see, e.g., \cite{JPA.2006.39.2493}).

\subsection{A point interaction in 1D for multiple directions: a star
graph topology}
\label{app:generalization}

The above prescription for the line is directly extended to the more
general case.
To see how, first note that in the 1D case, a zero-range potential at the
origin divides the interval $- \infty < x < + \infty$ into two semi-infinite
lines.
Thus, from the identification $x_1 = - x$ and $x_2 = +x$, the left
($-\infty < x < 0$) and right ($0 < x < +\infty$) regions could be
represented by $0 \leq x_1 \leq +\infty$ and
$0 \leq x_2 \leq +\infty$.
Hence, in a quantum graph framework, the system topology is that of a
single vertex joining two leads.
Also, the original nomenclature $0^{+}$ ($0^{-}$) now becomes
$x_2 = 0$ ($x_1 = 0$), indicating that we are considering the vertex but
from the right (left) side, i.e., at the beginning of lead 2 (1).

A zero-range potential located at 0 and attached to $N=|E(\Gamma)|$
semi-infinite lines constitutes a star graph-like topology, depicted in 
Figure
\ref{fig:fig1}(c).
Along each lead $s$ (with $s = 1,2,\ldots,N$) the spatial coordinate 
$x_s$ ranges from 0 to $+\infty$ and $\psi_k^{(\mbox{\scriptsize in})}(x_s)$ and
$\psi_k^{(\mbox{\scriptsize out})}(x_s)$ denote, respectively, incoming and
outgoing $k$ plane wave states.
In this case, the equivalent of Eqs. \eqref{eq:in-out} and 
\eqref{eq:in-out-inv}
read
\begin{equation}
\Psi_k^{(\mbox{\scriptsize out})}(0) =
{\mathcal S}(k) \, \Psi_k^{(\mbox{\scriptsize in})}(0) \ \ \mbox{and} \ \
\Psi_{-k}^{(\mbox{\scriptsize in})}(0)
= {\mathcal S}^{\dagger}(-k) \, \Psi_{-k}^{(\mbox{\scriptsize out})}(0),
\end{equation}
with $\Psi$ a $N$-components column vector (naturally extending the
$2$-components for the line) and ${\mathcal S}(k)$ a $N \times N$
scattering matrix, whose element ${\mathcal S}^{(s r)}(k)$ yield the 
quantum transition amplitude to go from lead $r$ to lead $s$ for a state 
of wave number $k$.
Probability conservation and moment inversion reciprocity, namely,
\begin{equation}
{\Psi_k^{(\mbox{\scriptsize out})}(0)}^{\dagger} \,
\Psi_k^{(\mbox{\scriptsize out})}(0) = {\Psi_k^{(\mbox{\scriptsize in})}(0)}^{\dagger} \,
\Psi_k^{(\mbox{\scriptsize in})}(0) \ \mbox{and} \
k \leftrightarrow -k \Longleftrightarrow \
\Psi^{(\mbox{\scriptsize out})} \leftrightarrow \Psi^{(\mbox{\scriptsize in})},
\end{equation}
demand ${\mathcal S}(k)$ to be unitary and
${\mathcal S}(k) = {\mathcal S}^{\dagger}(-k)$, exactly as in Sec.
\ref{appendix:a1}.
Therefore, any $N \times N$ matrix satisfying these two conditions will
represent a proper zero-range interaction, resulting in a well-behaved
quantum dynamics on a $N$ star graph.
Furthermore, the scattering states follow from a direct generalization
of Eq. \eqref{eq:psi-scattering}, where the amplitudes are given by the
corresponding matrix elements of ${\mathcal S}(k)$ (cf., Sec. 2).

Finally, the BC approach
in \cite{JPA.1999.32.595,JMP.2001.42.1563} can be put
in a direct relation with the above ${\mathcal S}$ formalism through an
one-to-one correspondence between the $N^2$ independent real parameters
defining the BC at the vertex (see Sec. 2.1) and the matrix elements of
${\mathcal S}$, likewise parameterizable by $N^2$ independent real
constants \cite{JPA.1982.15.3465}.

\subsection{A general graph}
\label{app:general-graph}

To conclude the analysis, we note that in an arbitrary undressed graph,
the region around each vertex $j$ is basically a star structure.
The difference is that instead of going from 0 to $+\infty$, some (or all)
edges can be finite, ending up in another vertex $m$.
Due to the superposition principle -- which holds true for any linear
wave-like differential equation (here Helmholtz) -- the global state for
an spatially extended problem can be construct in terms of a multiple
scattering process \cite{Book.1982.Newton}.
In other words, a proper sum of the locally scattered waves (entirely
determined by ${\mathcal S}_j(k)$) results in the full exact solution.
This is the case even if the system is closed (the graph has no
leads)\footnote{
A trivial example is that of an infinite square well (a graph
with two vertices and one edge), whose typical bounded
$\psi_n(x) \propto \sin[k_n x]$ (with $k_n = n \pi/L$) is given
as the linear combination of the plane waves scattered off by each wall
(vertex), at $x=0$ and $x=L$.}.

In this way, a legitimate and univocal quantum dynamics for any open or
closed graph is utterly obtained by associating to each vertex $j$ a
corresponding scattering matrix ${\mathcal S}_j(k)$
(for ${\mathcal S}_j(k)$ as described in Sec. \ref{app:generalization}).
Then, it also directly follows that the BC prescription and the
${\mathcal S}$ scheme are totally equivalent regardless the graph
topology.

%%% Local Variables:
%%% mode: latex
%%% TeX-master: "green-qg-pr"
%%% ispell-local-dictionary: "american"
%%% End:
\section{The exact Green's function for quantum graphs:
the generalized semiclassical formula}
\label{app:green}

Here we shall outline only the main steps necessary to demonstrate that the
exact Green's function for quantum graphs can be written in the same
functional form of Eq. \eqref{eq:gf}, i.e., as generalized semiclassical
formula.

\subsection{Reviewing a simple case, the Green's function for a point
interaction on the line}

Suppose the usual infinite line and an arbitrary point interaction at the
origin ($x=0$), for which the reflection and transmission coefficients are
$R^{(\pm)}$ and $T^{(\pm)}$ (see Appendix \ref{appendix:a1}).
It is worth recalling that this example corresponds to a quantum graph
with one vertex and two leads.
From \cite{JPA.1998.31.2975}, we can readily write down its exact
Green's function.
Defining
$G_{+-}$ for $x_f > 0 > x_i$,
$G_{-+}$ for $x_i > 0 > x_f$,
$G_{++}$ for $x_f, \ x_i > 0$ and
$G_{--}$ for $x_f, \ x_i < 0$,
one finds
\begin{align}
  G_{\pm \mp}(x_f,x_i;k)  ={}
  &
    \frac{\mu}{i\hbar^2k} T^{(\pm)} \exp[ik|x_f - x_i|],
  \nonumber \\
  G_{\pm \pm}(x_f,x_i;k) = {}
  & \frac{\mu}{i\hbar^2k} \left[\exp[ik |x_f - x_i|]
    + R^{(\pm)} \exp[i k( |x_f| +|x_i|)]\right],
\label{eq:gvcompact}
\end{align}
which have the structure of Eq. \eqref{eq:gf}.
In fact, for $\pm \, \mp$ there is only one sp leaving $x_i$, crossing the
origin, and finally arriving at $x_f$.
In this case, the classical-like action reads
$S_{sp} = p L_{sp} /\hbar = k |x_f - x_i|$, whereas the quantum weight is
given by $W_{sp} = T^{(\pm)}$ (just the amplitude gained in this scattering
process, a transmission).
For $\pm \, \pm$, both end points are at the same side of the zero
range potential.
Therefore, we have (i) a direct sp, going straight from $x_i$ to $x_f$,
so with $W_{sp} = 1$ and $S_{sp} =k |x_f - x_i|$, and (ii) an indirect sp,
along which there is a single reflection (at $x=0$), thus
$W_{sp} = R^{(\pm)}$ and $S_{sp} =k (|x_f| + |x_i|)$.

\subsection{Green's function for a star graph}

Similarly to which has been done in the Appendix \ref{app:generalization},
to see why $G$ for quantum graphs can be written in the general
form of Eq. \eqref{eq:gf}, we can start considering the basic (building
block) star shape depicted in Figure \ref{fig:fig1}(c).
The sole vertex (assumed to be at the origin of all leads, in a total
of $N$) is interpreted as an arbitrary scattering center, so a general
point interaction.

Suppose $\{ \Psi^{(\kappa)}, \Psi^{(\sigma)}(k) \}$ to represent the complete
full set of solutions for the Schr\"odinger equation for this graph, where
$\Psi^{(\sigma)}(k) = (\psi_1^{(\sigma)}(x_1;k), \ldots, \psi_N^{(\sigma)}(x_N;k))^{T}$
and
$\Psi^{(\kappa)} = (\psi_1^{(\kappa)}(x_1), \ldots, \psi_N^{(\kappa)}(x_N))^{T}$
are, respectively, the scattering and bound states with energy
$E = \hbar^2 k^2 / 2 \mu$ and $E_{\kappa}$.
We also observe that for each wavenumber $k$, we have a scattering state
$\sigma$ (here, $\sigma$ labels through which initial lead $\sigma$ the 
plane wave is incident to the vertex).
This is equivalent to the 1D problem where one has two leads and so two
solutions ($\sigma = \pm$), one incoming from the left and other from the
right of the origin \cite{JPA.1998.31.2975,JPA.2001.34.5041,JPA.2003.36.227}
(cf, Eq. \eqref{eq:psi-scattering} in Appendix \ref{appendix:a1}).

From the Green's function spectral decomposition property, we
can write \cite{Book.2006.Economou} (for $x_f$ and $x_i$ in the
edges $l$ and $n$, respectively)
\begin{align}
  G_{l n}(x_f,x_i;E) = {}
  &
    G_{l n}^{\mbox{\scriptsize (b.s.)}}(x_{f}, x_{i}; E) +
    G_{l n}^{\mbox{\scriptsize (s.s.)}}(x_{f}, x_{i}; E), \\
  G_{l n}^{\mbox{\scriptsize (b.s.)}}(x_{f}, x_{i}; E) = {}
  &
    \sum_{\kappa} \frac{\psi_{l}^{(\kappa)}(x_{f}) \,
    {\psi_{n}^{(\kappa)}}^*(x_{i})}
    {E - E_{\kappa}}, \\
    G_{l n}^{\mbox{\scriptsize (s.s.)}}(x_{f}, x_{i}; E) =
  &
  \int_{0}^{\infty} dk \sum_{\sigma=1}^{N}
  \frac{\psi_{l}^{(\sigma)}(x_f;k) \, {\psi_{n}^{(\sigma)}}^{*}(x_i;k)}
  {E - \hbar^2 k^2/(2 \mu)}.
  \label{eq:greenexp}
\end{align}
The scattering solution for a plane wave of energy $E=\hbar^2k^2/2\mu$,
incoming from lead $\sigma$ towards the vertex, is given by
(with $x$ in $l$, for $l=1,\ldots,N$)
\begin{equation}
  \psi_l^{(\sigma)}(x;k)=\frac{1}{\sqrt{2\pi}}
  \Big(\delta_{l \sigma}\exp[-i k x]+S^{(l \sigma)}(k)\exp[ikx]\Big),
  \label{eq:psigrafo}
\end{equation}
By inserting \eqref{eq:psigrafo} into \eqref{eq:greenexp}, then
$(E=\hbar^2\lambda^2/(2\mu))$
\begin{align}
  G_{l n}(x_f,x_i;\lambda) = {}
  & G_{l n}^{\mbox{\scriptsize (b.s.)}}(x_{f},x_{i}; E)
    + \frac{2\mu}{\hbar^2}\frac{1}{2\pi}
    \int_{0}^{\infty}\frac{dk}{\lambda^2-k^2}
    \nonumber \\
  &  \times \Big\{
    \delta_{nl}\exp[-ik(x_f-x_i)] +
    S^{(l n)}(k)\exp[ik(x_f+x_i)]
    \nonumber \\
  &
    + {S^{(l n)}}^{*}(k)\exp[-ik(x_f+x_i)]
  \nonumber \\
  &
    +\sum_{\sigma=1}^{N}S^{(l \sigma)}(k) \,
    {S^{(n \sigma)}}^{*}(k)\exp[ik(x_f-x_i)] \Big\}.
\end{align}
Using the relations in Eq. \eqref{eq:sqg}, the above equation can be
written as
\begin{align}
 G_{ln}(x_f,x_i;\lambda) = {}
  & G_{ln}^{\mbox{\scriptsize (b.s.)}}(x_{f},x_{i}; E) +
\frac{2\mu}{\hbar^2} \frac{1}{2\pi}
\int_{-\infty}^{\infty}\frac{dk}{\lambda^2-k^2}
\Big\{\delta_{nl}\exp[-ik(x_f-x_i)]
    \nonumber \\
  &
+S^{(l n)}(k)\exp[ik(x_f+x_i)]\Big\}.
\label{eq:ggraphaux}
\end{align}

Above, the integral involving \mbox{$\exp[-ik(x_f-x_i)]$} leads to the
free particle Green's function.
For the other integral, we consider a contour integration along the real
axis closed by a infinite semicircle in the upper half of the complex
plane.
The pole contributions are due the denominator $\lambda^2-k^2$ and
possible singularities of $S^{(l n)}(k)$.
If the single vertex (a zero range potential) does not allow bounded states,
$G^{\mbox{\scriptsize (b.s.)}}=0$ and $S^{(l n)}(k)$ does not have poles.
On the other hand, for a very large number of situations the terms in the
integration resulting from the bound energy poles exactly cancel out
with $G^{\mbox{\scriptsize (b.s.)}}$ \cite{JMP.1964.5.591,JMP.1981.22.306,
PRA.1988.37.973}.
This is precisely which takes place for general point
interactions \cite{JPA.2006.39.2493}.
Putting all this together, the remaining steps in evaluating
Eq \eqref{eq:ggraphaux} are straightforward.
Thus, reverting to the notation $k$ for the wave number variable, we
finally get
\begin{equation}
  G_{ln}(x_f,x_i;k)=\frac{\mu}{i\hbar^2 k}\Big\{
  \delta_{nl}\exp[ik|x_f-x_i|]+S^{(l n)}(k)\exp[ik(x_f+x_i)]\Big\}.
\label{eq:fgge}
\end{equation}

Now, notice that Eq. (\ref{eq:fgge}) would readily follow from the sum over
scattering paths prescription.
In fact, for a particle with $x_i$ in lead $n$, arriving at $x_f$ in
lead $l$, we have two possibilities.
(i) The leads $n$ and $l$ are the same, so there are two scattering
paths:
straight propagation from $x_i$ to $x_f$, corresponding to
$\exp[ik|x_f-x_i|]$ and $W =1$; and
propagation from $x_i$ to the vertex, reflection (gaining a factor
$S^{(n n)}(k)$) and then propagation to $x_f$,
in this case yielding $\exp[ik(x_f+x_i)]$ and an amplitude $S^{(n n)}(k)$
(i.e., the reflection coefficient from $n$ to $n$).
These contributions result in
$G_{n n}^{(\mbox{\scriptsize semicl gen})}(x_f,x_i;k) = (\mu/(i\hbar^2 k))
\big\{\exp[-ik|x_f-x_i|] + S^{(n n)}(k)\exp[ik(x_f+x_i)]\big\}$.
(ii) The leads are distinct, thus there is only one scattering path:
propagation from $x_i$ to the vertex, a transmission through it
(gaining a factor $S^{(l n)}(k)$), and finally propagation to $x_f$.
So,
$G_{l n}^{(\mbox{\scriptsize semicl gen})}(x_f,x_i;k)= (\mu/( i\hbar^2 k))
\big\{S^{(l n)}(k) \exp[ik(x_f+x_i)]\big\}$.
These two possibilities are exactly summarized by Eq. \eqref{eq:fgge}.

\subsection{The Green's function for an arbitrary graph}

Last, for an arbitrary case the reasoning resembles that in the Appendix
\ref{app:general-graph}.
For the star graph, the exact $G$ is written in terms of a (finite) sum of
scattering paths.
Extending for any topology (as considered in this work), the local scattering
-- around each vertex, so in a star-like configuration -- can be associated
to a stretch of a much larger sp, leaving from $x_i$, traveling across the
totality or parts of the whole graph, and finally arriving at $x_f$.
This is just the usual multiple scattering process, valid to describe any
wave propagation in the linear context.
Along the way, the $W_{sp}$ are built from the quantum amplitudes gained
through the successive scattering at the vertices.
On its turn $S_{sp} = k L_{sp}$, for $L_{sp}$ the sp total classical
distance traveled between the end points.
Of course, generally the number of sp can be infinite (thus
demanding the techniques of Sec. 4 for explicit calculations).
But the main point is that Eq. \eqref{eq:gf} represents the exact
construction for the Green's function of any quantum graph.

%%% Local Variables:
%%% mode: latex
%%% TeX-master: "green-qg-pr"
%%% ispell-local-dictionary: "american"
%%% End:
\section{Certain common boundary conditions for quantum graphs and
the wave function solution for the example of Sec. \ref{sec:cesog}}
\label{app:boundary-conditions}

The purpose here is twofold.
To discuss some of the more common boundary conditions (BCs) for
quantum graphs and to illustrate their usage considering the Schr\"odinger 
equation solution for the example of Sec. \ref{sec:cesog}.

\subsection{Few usual boundary conditions for quantum graphs}
\label{app:boundary-conditions-1}

Consider the set of edges attached to a certain vertex $V$ of an arbitrary
quantum graph.
Locally (i.e., around $V$) the topology is that seen in Fig. \ref{fig:fig1} 
(c).
So, to define the BCs and the scattering amplitudes for such particular 
vertex, without loss of generality we always can treat $V$ and its edges 
as a star graph.

Now, let us depart a little bit from the previous notation and for
simplicity to label the unique vertex in Fig. \ref{fig:fig1} (c) by $V$ and
the leads by $n = 1, 2, \ldots, N$.
To each lead we can associate the coordinate $x_n$, whose origin is at
$V \equiv 0$ and prolongs to $\sigma_n \times \infty$.
As already mentioned (see footnote 5), usually one takes $\sigma_n = +1$
for any $n$.
But here we shall discuss the most general case, since it is just a 
matter of convenience (according to each specific situation) to set 
$\sigma_n = \pm 1$.
Further, we denote the wave function at lead $n$ by $\psi_n(x_n)$.
Usually, the spatial derivatives of $\psi$ along any edge or lead (with 
respect to a reference vertex $V$) are taken in the outgoing direction
from $V$.
Hence, a simple way to assure that for the star graph is to define
$D^{\mbox{\scriptsize out}}_x \psi(x) \equiv \sigma \, d\psi(x)/dx$.
Hereafter we set $\hbar = \mu = 1$.

First, assume the following BCs at $V$
(with $\gamma_V$ any real number)
\begin{equation}
\psi_1(V) = \psi_2(V) = \ldots = \psi_N(V) = \psi(V), \qquad
\sum_{n=1}^{n=N} \, D^{\mbox{\scriptsize out}}_{x_n} \, \psi_n(x_n)\big|_{x_n = V} =
\sum_{n=1}^{n=N} \, \sigma_n \, d\psi_n(x_n)/dx_n\big|_{x_n = V} = 
2 \gamma_V \, \psi(V).
\label{eq:gen-delta}
\end{equation}
These BCs correspond to the generalized $\delta$ interaction of strength
$\gamma_V$ (see, e.g., \cite{PRL.1995.74.3503}).
To understand why, suppose an initial plane wave (of wave number $k$)
incoming from lead $m$ and then being scattered off at $V$.
The system full scattering state (satisfying to the Schr\"odinger
equation) reads
\begin{align}
  \psi_m(x_m) = {} & {\mathcal C} \, \Big(\exp[- i \sigma_m k x_m] +
  r_v^{(m)} \exp[+ i \sigma_m k x_m]\Big),
  \nonumber \\
  \psi_{n}(x_n) = {} & {\mathcal C} \, t_V^{(n,m)} \,
  \exp[+i \sigma_n k x_n], \ \ n \neq m.
\end{align}
Applying the BCs in Eq. (\ref{eq:gen-delta}) to the above expressions, we
get (recalling that $x|_V = 0$)
\begin{equation}
  t_V^{(n,m)} = t_V^{(m)}, \ \ \forall n \neq m, \qquad
  1 + r_V^{(m)} = t_V^{(m)}, \qquad
  i k \, (-1 + r_V^{(m)}) + i k \, (N-1) \, t_V^{(m)} = 2 \gamma_V \, t_V^{(m)}.
\end{equation}
Solving for $r$ and $t$ (where we can drop the superscript indices),
we find
\begin{equation}
  r_V = \frac{2 \gamma_V - (N-2) \, i k}{N i k - 2 \gamma_V},
  \qquad
  t_V = \frac{2 i k}{N i k - 2 \gamma_V}.
\label{eq:rtgendelta}
\end{equation}
Note that when $N=2$, such expressions do reduce to the usual reflection and 
transmission coefficients for the $\delta$ function potential on the line,
explaining the nomenclature ``generalized delta'' for $N > 2$.

Second, it is very common to set $\gamma_V = 0$ in Eq. (\ref{eq:gen-delta}),
resulting in the so called Neumann-Kirchhoff BCs 
\cite{Book.2006.Berkolaiko,PRL.2013.110.094101}.
One of their notable characteristics is that the corresponding reflection
and transmission coefficients are $k$-independent, since in this case
$r_V = 2/N - 1$ and $t_V = 2/N$.
Moreover, these $r$'s and $t$'s displays another interesting feature, but
which is barely explored in the literature.
Although trivial when $N=2$ (for which $r_V = 0$ and $t_V = 1$, i.e., the
vertex $V$ is eliminate with the two edges becoming merged) the
Neumann-Kirchhoff quantum amplitudes are exactly the matrix elements of a
$N \geq 2$ dimensional Grover operator
\cite{PRA.2010.81.042330,NJP.2005.7.156,NJP.2003.5.83}, an essential gate
in quantum computation.
So, quantum graphs with generalized $\delta$ functions of vanishing
strengths at the vertices have a close relation with quantum walks
driving by Grover `coins' \cite{PRA.2010.81.042330}.

Lastly, assume that the vertex $V$ is a `dead end', with $N=1$.
This means $V$ is joined only to one lead, $m$.
Defining $\lambda = 2 \gamma_V$, we have from the wave function in the
lead $m$ and from the delta BC that
$-i k + i k \, r_V = \lambda \, (1 + r_V)$, so
\begin{equation}
r_V = \frac{i k + \lambda}{i k - \lambda}.
\end{equation}
This corresponds to the most general possible BC (consistent with flux
conservation) for a quantum particle interacting with an infinite wall in 
the half-line \cite{PRD.22.3012.1980,JOB.7.S77.2005}.

\subsection{The wave function solution for the graph of Sec. 
\ref{sec:cesog}: the bound state case}
\label{app:boundary-conditions-2}

Now, consider the system of Fig. \ref{fig:fig6} (a).
Denoting $\gamma_O = \gamma$ and $2 \gamma_A = \lambda$, with at least one
of these parameters negative, we can have bound state.
For $k = i \kappa$ with $\kappa > 0$, and once for the leads
$i$ and $f$ and the edge $1$ it holds, respectively, that
$0 \leq x_i, x_f < + \infty$ (so, in both $i$ and $f$ cases $\sigma = +1$), 
and $0 < x_1 < \ell_1$, we can write (dropping the subscript for $x$)
\begin{equation}
\psi_i(x) = {\mathcal C} \, \exp[- \kappa x],
\qquad 
\psi_{f}(x) = {\mathcal C} \, \exp[- \kappa x],
\qquad
\psi_1(x) = {\mathcal C} \, \Big(A \, \exp[- \kappa x] + 
                                 B \, \exp[+ \kappa x]\Big).
\end{equation}
Applying the BCs in Eq. (\ref{eq:gen-delta}) to the above wave functions,
namely,
\begin{equation}
-\frac{d\psi_1(x)}{dx}\Big|_{x=\ell_1} = \lambda \, \psi_1(\ell_1),
\qquad
\psi_i(0) = \psi_f(0) = \psi_1(0), 
\qquad
\Big(\frac{d\psi_i(x)}{dx} + \frac{d\psi_f(x)}{dx} + 
\frac{d\psi_1(x)}{dx}\Big)\Big|_{x=0} 
= 2 \gamma \, \psi_f(0),
\end{equation}
we get for $\kappa$ (with $r_O(k)$ and $r_A(k)$ the coefficients given 
in Sec. \ref{sec:cesog})
\begin{equation}
g(i \kappa) = 
1 - r_O(i \kappa) \, r_A(i \kappa) \, \exp[-2 \kappa \ell_1] = 0.
\label{eq:ev-wave}
\end{equation}
Note that Eq. (\ref{eq:ev-wave}) is the same than Eq. (\ref{eq:ev-green})
with $k = i \kappa$.
Hence, the eigenvalues derived from the Schr\"odinger equation are exactly
those calculated from the Green's function approach in Sec.
\ref{sec:cesog}.
We also obtain (using Eq. (\ref{eq:ev-wave}) as well as the fact that for
any $k$, $1 + r_O(k) = t_O(k)$)
\begin{equation}
A = \frac{1}{1+r_A(i \kappa) \, \exp[-2 \kappa \ell_1]} =
\frac{r_O(i \kappa)}{t_O(i \kappa)}, 
\qquad
B = \frac{r_A(i \kappa) \, \exp[-2 \kappa \ell_1]}{1+r_A(i \kappa) \, 
\exp[-2 \kappa \ell_1]} = \frac{1}{t_O(i \kappa)}.
\end{equation}
In this way (also redefining ${\mathcal C} \equiv t_O(i \kappa) \,
{\mathcal N}_S(i \kappa)$)
\begin{equation}
\psi_i(x) = {\mathcal N}_S(i \kappa) \, t_O(i \kappa) \, \exp[- \kappa x],
\ \ \ \
\psi_f(x) = {\mathcal N}_S(i \kappa) \, t_O(i \kappa) \, \exp[- \kappa x],
\ \ \ \
\psi_1(x) = {\mathcal N}_S(i \kappa) \, \Big(\exp[+ \kappa x] + r_O(i \kappa) \,
                                   \exp[- \kappa x]\Big),
\end{equation}
which agree with the wave functions in Eqs. (\ref{eq:wave-if}) and 
(\ref{eq:wave-11}) in Sec. \ref{sec:cesog}.

Finally, the normalization constant ${\mathcal N}_S(i \kappa)$ follows from
\begin{align}
{{\mathcal N}_S}(i \kappa)  = {} & \left\{ 2 \, t_O(i \kappa)^2
\int_{0}^{\infty} dx \, \exp[-2 \kappa x] +
\int_{0}^{\ell_1} dx \,
\Big(\exp[+\kappa x] + r_O(i \kappa) \, \exp[-\kappa x]\Big)^2 
\right\}^{-1/2}
\nonumber \\
 = {} & \sqrt{2 \kappa} \ \left\{ 
2 \, (1 + r_O(i \kappa))^2 + (\exp[2 i \kappa \ell_1] - 1) + 
4 \kappa \ell_1 r_O(i \kappa)
+ r_O(i \kappa)^2 (1-\exp[-2 \kappa \ell_1])\right\}^{-1/2}.
\end{align}
Although a somehow trick exercise, one should be able to show that 
${\mathcal N}_S$ yields ${\mathcal N}_G$ of Eq. 
(\ref{eq:norm-green}).

%%% Local Variables:
%%% mode: latex
%%% TeX-master: "green-qg-pr"
%%% ispell-local-dictionary: "american"
%%% End:

\section{References}
\bibliographystyle{elsarticle-num}

\end{document}